\documentclass[12pt,a4paper]{article}

\usepackage{ifthen}
\newboolean{pdflatex}
\setboolean{pdflatex}{true}

\newboolean{articletitles}
\setboolean{articletitles}{true}

\newboolean{uprightparticles}
\setboolean{uprightparticles}{false}

\newboolean{inbibliography}
\setboolean{inbibliography}{false}

\newcommand{\mpr}        {\mbox{$m^\prime$}}
\newcommand{\thpr}       {\mbox{$\theta^\prime$}}

\usepackage{rotating}

\usepackage{microtype}
\usepackage{lineno}
\usepackage{xspace}

\usepackage{multirow}
\usepackage{caption}

\usepackage{graphicx}
\usepackage{color}
\usepackage{colortbl}
\graphicspath{{./figs/}}

\usepackage{amsmath}
\usepackage{amssymb}
\usepackage{amsfonts}
\usepackage{upgreek}

\newcommand*\patchAmsMathEnvironmentForLineno[1]{
\expandafter\let\csname old#1\expandafter\endcsname\csname #1\endcsname
\expandafter\let\csname oldend#1\expandafter\endcsname\csname
end#1\endcsname
 \renewenvironment{#1}
   {\linenomath\csname old#1\endcsname}
   {\csname oldend#1\endcsname\endlinenomath}
}
\newcommand*\patchBothAmsMathEnvironmentsForLineno[1]{
  \patchAmsMathEnvironmentForLineno{#1}
  \patchAmsMathEnvironmentForLineno{#1*}
}
\AtBeginDocument{
\patchBothAmsMathEnvironmentsForLineno{equation}
\patchBothAmsMathEnvironmentsForLineno{align}
\patchBothAmsMathEnvironmentsForLineno{flalign}
\patchBothAmsMathEnvironmentsForLineno{alignat}
\patchBothAmsMathEnvironmentsForLineno{gather}
\patchBothAmsMathEnvironmentsForLineno{multline}
\patchBothAmsMathEnvironmentsForLineno{eqnarray}
}

\allowdisplaybreaks

\usepackage{hyperref}
\usepackage[all]{hypcap}

\def\lhcb {\mbox{LHCb}\xspace}

\def\babar  {\mbox{BaBar}\xspace}

\ifthenelse{\boolean{uprightparticles}}
{

 \def\Pmu         {\ensuremath{\upmu}\xspace}

 \def\Ppi         {\ensuremath{\uppi}\xspace}

 \def\Ppsi        {\ensuremath{\uppsi}\xspace}

 \def\PDelta      {\ensuremath{\Delta}\xspace}
 \def\PXi      {\ensuremath{\Xi}\xspace}
 \def\PLambda      {\ensuremath{\Lambda}\xspace}
 \def\PSigma      {\ensuremath{\Sigma}\xspace}
 \def\POmega      {\ensuremath{\Omega}\xspace}
 \def\PUpsilon      {\ensuremath{\Upsilon}\xspace}

 \def\PB      {\ensuremath{\mathrm{B}}\xspace}
 
 \def\PD      {\ensuremath{\mathrm{D}}\xspace}

 \def\PJ      {\ensuremath{\mathrm{J}}\xspace}
 \def\PK      {\ensuremath{\mathrm{K}}\xspace}

 \def\Pb      {\ensuremath{\mathrm{b}}\xspace}
 \def\Pc      {\ensuremath{\mathrm{c}}\xspace}
 
 \def\Pe      {\ensuremath{\mathrm{e}}\xspace}

 \def\Pi      {\ensuremath{\mathrm{i}}\xspace}

 \def\Pp      {\ensuremath{\mathrm{p}}\xspace}

 \def\Ps      {\ensuremath{\mathrm{s}}\xspace}

}
{

 \def\Pmu         {\ensuremath{\mu}\xspace}

 \def\Ppi         {\ensuremath{\pi}\xspace}

 \def\Ppsi        {\ensuremath{\psi}\xspace}
 
 \mathchardef\PDelta="7101
 \mathchardef\PXi="7104
 \mathchardef\PLambda="7103
 \mathchardef\PSigma="7106
 \mathchardef\POmega="710A
 \mathchardef\PUpsilon="7107
 
 \def\PB      {\ensuremath{B}\xspace}
 
 \def\PD      {\ensuremath{D}\xspace}

 \def\PJ      {\ensuremath{J}\xspace}
 \def\PK      {\ensuremath{K}\xspace}

 \def\Pb      {\ensuremath{b}\xspace}
 \def\Pc      {\ensuremath{c}\xspace}
 
 \def\Pe      {\ensuremath{e}\xspace}

 \def\Pi      {\ensuremath{i}\xspace}

 \def\Pp      {\ensuremath{p}\xspace}

 \def\Ps      {\ensuremath{s}\xspace}

}

\makeatletter
\ifcase \@ptsize \relax
  \newcommand{\miniscule}{\@setfontsize\miniscule{4}{5}}
\or
  \newcommand{\miniscule}{\@setfontsize\miniscule{5}{6}}
\or
  \newcommand{\miniscule}{\@setfontsize\miniscule{5}{6}}
\fi
\makeatother

\DeclareRobustCommand{\optbar}[1]{\shortstack{{\miniscule (\rule[.5ex]{1.25em}{.18mm})}
  \\ [-.7ex] $#1$}}

\def\epem       {{\ensuremath{\Pe^+\Pe^-}}\xspace}

\def\mumu       {{\ensuremath{\Pmu^+\Pmu^-}}\xspace}

\def\squark    {{\ensuremath{\Ps}}\xspace}

\def\cquark    {{\ensuremath{\Pc}}\xspace}

\def\bquark    {{\ensuremath{\Pb}}\xspace}

\def\pion   {{\ensuremath{\Ppi}}\xspace}
\def\piz    {{\ensuremath{\pion^0}}\xspace}

\def\pip    {{\ensuremath{\pion^+}}\xspace}
\def\pim    {{\ensuremath{\pion^-}}\xspace}
\def\pipm   {{\ensuremath{\pion^\pm}}\xspace}

\def\kaon    {\ensuremath{\PK}\xspace}
  \def\Kbar    {{\kern 0.2em\overline{\kern -0.2em \PK}{}}\xspace}

\def\KorKbar    {\kern 0.18em\optbar{\kern -0.18em K}{}\xspace}

\def\Kp      {\ensuremath{\kaon^+}\xspace}
\def\Km      {\ensuremath{\kaon^-}\xspace}
\def\Kpm     {\ensuremath{\kaon^\pm}\xspace}

\def\Kstarz  {\ensuremath{\kaon^{*0}}\xspace}
\def\Kstarzb {\ensuremath{\Kbar{}^{*0}}\xspace}

\def\Kstarb  {\ensuremath{\Kbar{}^*}\xspace}

  \def\Dbar    {{\kern 0.2em\overline{\kern -0.2em \PD}{}}\xspace}
\def\D       {{\ensuremath{\PD}}\xspace}

\def\DorDbar    {\kern 0.18em\optbar{\kern -0.18em D}{}\xspace}
\def\Dz      {{\ensuremath{\D^0}}\xspace}
\def\Dzb     {{\ensuremath{\Dbar{}^0}}\xspace}

\def\Dstar   {{\ensuremath{\D^*}}\xspace}

\def\Dstarzb {{\ensuremath{\Dbar{}^{*0}}}\xspace}
\def\Dstarp  {{\ensuremath{\D^{*+}}}\xspace}

\def\Dstarpm {{\ensuremath{\D^{*\pm}}}\xspace}

\def\Dsm     {{\ensuremath{\D^-_\squark}}\xspace}

\def\B       {{\ensuremath{\PB}}\xspace}
\def\Bbar    {{\ensuremath{\kern 0.18em\overline{\kern -0.18em \PB}{}}}\xspace}

\def\BorBbar    {\kern 0.18em\optbar{\kern -0.18em B}{}\xspace}
\def\Bz      {{\ensuremath{\B^0}}\xspace}

\def\Bu      {{\ensuremath{\B^+}}\xspace}

\def\Bp      {{\ensuremath{\Bu}}\xspace}

\def\Bd      {{\ensuremath{\B^0}}\xspace}
\def\Bs      {{\ensuremath{\B^0_\squark}}\xspace}
\def\Bsb     {{\ensuremath{\Bbar{}^0_\squark}}\xspace}
\def\Bdb     {{\ensuremath{\Bbar{}^0}}\xspace}

\def\jpsi     {{\ensuremath{{\PJ\mskip -3mu/\mskip -2mu\Ppsi\mskip 2mu}}}\xspace}

  \def\Y#1S{\ensuremath{\PUpsilon{(#1S)}}\xspace}

\def\proton      {{\ensuremath{\Pp}}\xspace}
\def\antiproton  {{\ensuremath{\overline \proton}}\xspace}

\def\Lz          {{\ensuremath{\PLambda}}\xspace}
\def\Lbar        {{\ensuremath{\kern 0.1em\overline{\kern -0.1em\PLambda}}}\xspace}
\def\LorLbar    {\kern 0.18em\optbar{\kern -0.18em \PLambda}{}\xspace}

\def\Lbbar   {{\ensuremath{\Lbar{}^0_\bquark}}\xspace}

\def\to                 {\ensuremath{\rightarrow}\xspace}

\def\CP                {{\ensuremath{C\!P}}\xspace}

\def\AT#1     {\ensuremath{A_{\mathrm{T}}^{#1}}\xspace}

\def\C#1      {\ensuremath{\mathcal{C}_{#1}}\xspace}
\def\Cp#1     {\ensuremath{\mathcal{C}_{#1}^{'}}\xspace}
\def\Ceff#1   {\ensuremath{\mathcal{C}_{#1}^{\mathrm{(eff)}}}\xspace}
\def\Cpeff#1  {\ensuremath{\mathcal{C}_{#1}^{'\mathrm{(eff)}}}\xspace}
\def\Ope#1    {\ensuremath{\mathcal{O}_{#1}}\xspace}
\def\Opep#1   {\ensuremath{\mathcal{O}_{#1}^{'}}\xspace}

\newcommand{\tev}{\ifthenelse{\boolean{inbibliography}}{\ensuremath{~T\kern -0.05em eV}\xspace}{\ensuremath{\mathrm{\,Te\kern -0.1em V}}}\xspace}
\newcommand{\gev}{\ensuremath{\mathrm{\,Ge\kern -0.1em V}}\xspace}
\newcommand{\mev}{\ensuremath{\mathrm{\,Me\kern -0.1em V}}\xspace}
\newcommand{\kev}{\ensuremath{\mathrm{\,ke\kern -0.1em V}}\xspace}
\newcommand{\ev}{\ensuremath{\mathrm{\,e\kern -0.1em V}}\xspace}
\newcommand{\gevc}{\ensuremath{{\mathrm{\,Ge\kern -0.1em V\!/}c}}\xspace}
\newcommand{\mevc}{\ensuremath{{\mathrm{\,Me\kern -0.1em V\!/}c}}\xspace}
\newcommand{\gevcc}{\ensuremath{{\mathrm{\,Ge\kern -0.1em V\!/}c^2}}\xspace}
\newcommand{\gevgevcccc}{\ensuremath{{\mathrm{\,Ge\kern -0.1em V^2\!/}c^4}}\xspace}
\newcommand{\mevcc}{\ensuremath{{\mathrm{\,Me\kern -0.1em V\!/}c^2}}\xspace}

\def\mm   {\ensuremath{\rm \,mm}\xspace}

\def\mum  {\ensuremath{{\,\upmu\rm m}}\xspace}

\def\fm   {\ensuremath{\rm \,fm}\xspace}

\newcommand{\stat}{\ensuremath{\mathrm{\,(stat)}}\xspace}
\newcommand{\syst}{\ensuremath{\mathrm{\,(syst)}}\xspace}

\newcommand{\chisq}{\ensuremath{\chi^2}\xspace}

\newcommand{\chisqip}{\ensuremath{\chi^2_{\rm IP}}\xspace}

\def\gsim{{~\raise.15em\hbox{$>$}\kern-.85em
          \lower.35em\hbox{$\sim$}~}\xspace}
\def\lsim{{~\raise.15em\hbox{$<$}\kern-.85em
          \lower.35em\hbox{$\sim$}~}\xspace}

\def\sPlot{\mbox{\em sPlot}}

\def\ptot       {\mbox{$p$}\xspace}
\def\pt         {\mbox{$p_{\rm T}$}\xspace}

\def\evtgen     {\mbox{\textsc{EvtGen}}\xspace}

\def\gauss      {\mbox{\textsc{Gauss}}\xspace}
\def\geant      {\mbox{\textsc{Geant4}}\xspace}

\def\photos     {\mbox{\textsc{Photos}}\xspace}

\def\pythia     {\mbox{\textsc{Pythia}}\xspace}

\def\tell1  {TELL1\xspace}
\def\ukl1   {UKL1\xspace}

\newcommand{\eg}{\mbox{\itshape e.g.}\xspace}
\newcommand{\ie}{\mbox{\itshape i.e.}\xspace}

\def\phani{\phantom{1}}

\def\Bds     {{\ensuremath{\B^0_{(\squark)}}}\xspace}

\def\Kstarbsubz  {{\ensuremath{\Kbar{}^*_0}}\xspace}

\def\Kstarbsubt  {{\ensuremath{\Kbar{}^*_2}}\xspace}

\def\DorDstarzb {{\ensuremath{\Dbar{}^{(*)0}}}\xspace}

\usepackage{cite}
\usepackage{mciteplus}

\begin{document}

\renewcommand{\thefootnote}{\fnsymbol{footnote}}
\setcounter{footnote}{1}

\begin{titlepage}
\pagenumbering{roman}

\vspace*{-1.5cm}
\centerline{\large EUROPEAN ORGANIZATION FOR NUCLEAR RESEARCH (CERN)}
\vspace*{1.5cm}
\hspace*{-0.5cm}
\begin{tabular*}{\linewidth}{lc@{\extracolsep{\fill}}r}
\ifthenelse{\boolean{pdflatex}}
{\vspace*{-2.7cm}\mbox{\!\!\!\includegraphics[width=.14\textwidth]{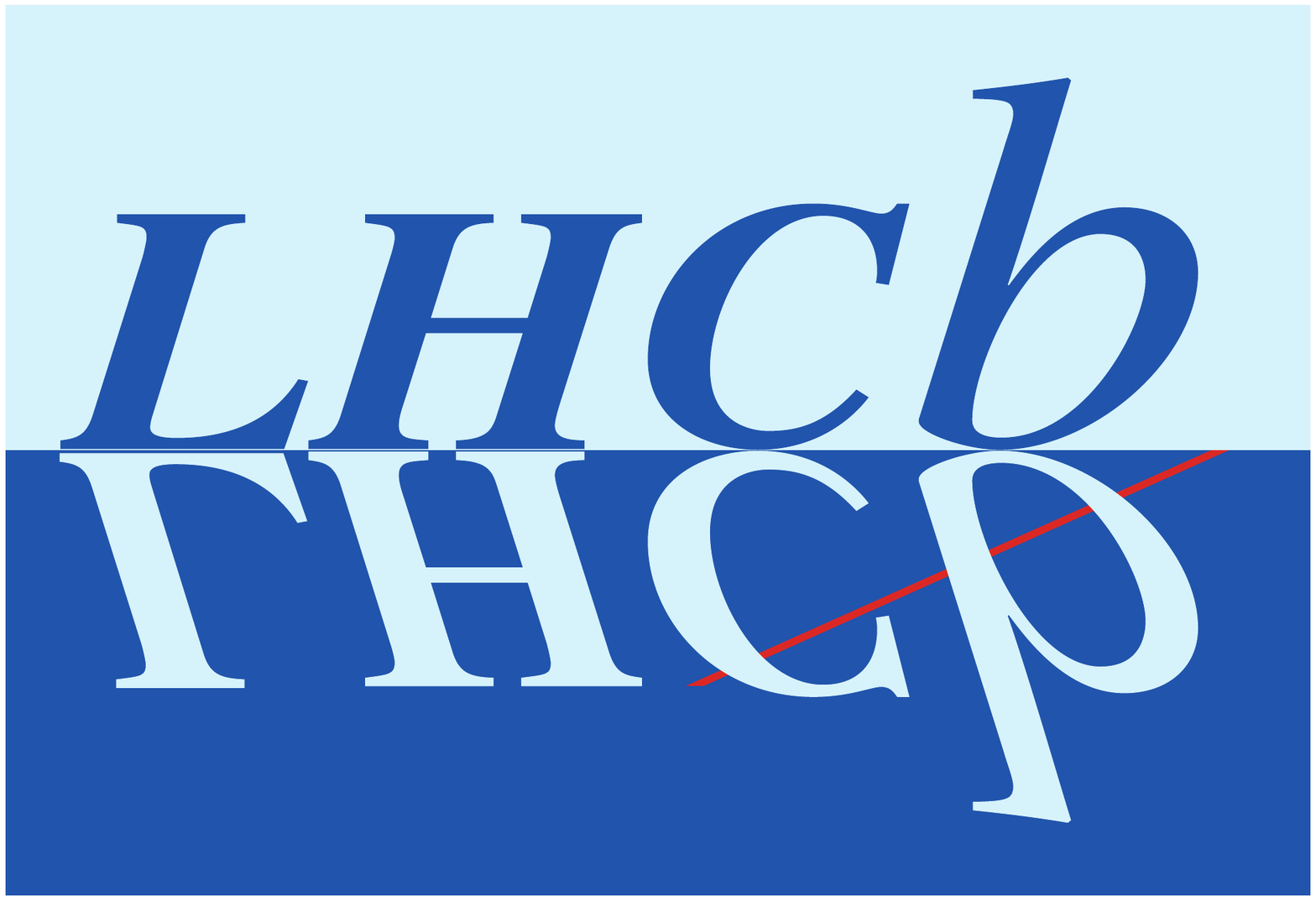}} & &}
{\vspace*{-1.2cm}\mbox{\!\!\!\includegraphics[width=.12\textwidth]{lhcb-logo.eps}} & &}
\\
 & & CERN-PH-EP-2014-184 \\
 & & LHCb-PAPER-2014-036 \\
 & & 20 October 2014 \\
 & & \\
\end{tabular*}

\vspace*{2.0cm}

{\bf\boldmath\huge
\begin{center}
  Dalitz plot analysis of $\Bs \to \Dzb \Km \pip$ decays
\end{center}
}

\vspace*{1.0cm}

\begin{center}
  The LHCb collaboration\footnote{Authors are listed at the end of this paper.}
\end{center}

\vspace{\fill}

\begin{abstract}
  \noindent
  The resonant substructure of $\Bs \to \Dzb\Km\pip$ decays is studied with
  the Dalitz plot analysis technique.
  The study is based on a data sample corresponding to an integrated
  luminosity of $3.0  \,{\rm fb}^{-1}$ of $pp$ collision data recorded by LHCb.
  A structure at $m(\Dzb\Km) \approx 2.86 \gevcc$ is found to be an admixture of spin-1 and spin-3 resonances.
  The masses and widths of these states and of the $D^{*}_{s2}(2573)^{-}$
  meson are measured, as are the complex amplitudes and fit fractions for all
  the $\Dzb\Km$ and $\Km\pip$ components included in the amplitude model.
  In addition, the $D^{*}_{s2}(2573)^{-}$ resonance is confirmed to be spin-2.
\end{abstract}

\vspace*{2.0cm}

\begin{center}
  Published in Phys.~Rev.~D
\end{center}

\vspace{\fill}

{\footnotesize
\centerline{\copyright~CERN on behalf of the \lhcb collaboration, license \href{http://creativecommons.org/licenses/by/4.0/}{CC-BY-4.0}.}}
\vspace*{2mm}

\end{titlepage}

\newpage
\setcounter{page}{2}
\mbox{~}

\cleardoublepage

\renewcommand{\thefootnote}{\arabic{footnote}}
\setcounter{footnote}{0}

\pagestyle{plain}
\setcounter{page}{1}
\pagenumbering{arabic}

\section{Introduction}
\label{sec:Introduction}

Several recent experimental discoveries have reinvigorated the
field of heavy meson spectroscopy.
Among the most interesting are the observations of the
$D_{s0}^*(2317)^-$~\cite{Aubert:2003fg} and $D_{s1}(2460)^-$~\cite{Besson:2003cp}
states. In contrast to prior predictions, these are below the $DK$ and $D^*K$ thresholds, respectively, and hence are narrow.
The $D_{s0}^*(2317)^-$ and $D_{s1}(2460)^-$ states are usually interpreted~\cite{PDG2012} as two of the orbitally excited (1P) states,
the other two being the long-established $D_{s1}(2536)^-$ and $D_{s2}^*(2573)^-$
resonances, though the reason for the large mass splitting between the mesons
below and above the $D^{(*)}K$ thresholds is not fully understood.
Further interest in the field has been generated by the discovery of several $D_{sJ}^-$ states with masses above that of the $D_{s2}^*(2573)^-$ resonance through production in $e^+e^-$~\cite{Aubert:2006mh,Aubert:2009ah} or $pp$~\cite{LHCb-PAPER-2012-016} collisions.
A summary is given in Table~\ref{tab:BaBarLHCb}.

The $D_{s1}^*(2700)^-$ and $D_{sJ}^*(2860)^-$ states are usually interpreted as members of the 2S or 1D families.
The 2S family is a doublet with spin-parity quantum numbers $J^P = 0^-, 1^-$,
while there are four 1D states with $J^P = 1^-, 2^-, 2^-, 3^-$.
Among these, only resonances with natural spin-parity ($0^+, 1^-, 2^+, 3^-, ...$) can decay to two pseudoscalar mesons.
If the 2S and 1D $J^P = 1^-$ states are close in mass they may mix.
In the literature, the $D_{s1}^*(2700)^-$ is usually interpreted as being the $1^-$ 2S state, while the $D_{sJ}^*(2860)^-$ is a candidate to be the $3^-$ 1D state~\cite{Colangelo:2006rq,Zhang:2006yj,Li:2007px,Zhong:2008kd,Shan:2008ga,Chen:2009zt,Ebert:2009ua,Badalian:2011tb,Godfrey:2013aaa}.
However, several papers (\eg~Ref.~\cite{Li:2009qu}) point out that the $D_{sJ}^*(2860)^-$ could be the 1D $1^-$ state or, more generally, if the $D_{s1}^*(2700)^-$ is interpreted as an admixture of 2S and 1D $1^-$ states, the $D_{sJ}^*(2860)^-$ could be its orthogonal partner.
Several authors (\eg~Ref.\cite{Zhong:2009sk}) point out that the observed relative rates of $D_{sJ}^*(2860)^- \to D^*K$ and $D_{sJ}^*(2860)^- \to DK$ decays suggest that the observed signal for the former may include additional contributions from states with unnatural parity such as the $2^-$ 1D states.
Other authors have considered the possibility that the observed states may have a significant component from multiquark states (tetraquarks or molecules)~\cite{Vijande:2008zn,Ebert:2010af,Guo:2011dd}.
For detailed reviews, see Refs.~\cite{Swanson:2006st,Rosner:2006jz,Klempt:2007cp,Colangelo:2012xi}.

An observation of a state with $J^P = 3^-$ would be a clear signature of that state being a member of the 1D family.
Although candidates for spin-1 and spin-2 1D $c\bar{c}$ and $b\bar{b}$ states have been reported~\cite{PDG2012,Bonvicini:2004yj,delAmoSanchez:2010kz}, no spin-3 meson involving a $c$ or $b$ quark has previously been observed.
Production of high-spin states is expected to be suppressed in \B meson decay due to the angular momentum barrier~\cite{blatt-weisskopf}, and indeed has never yet been observed.
However, as the decays of high-spin resonances are suppressed for the same reason, they are expected to have relatively small widths, potentially enhancing their observability.

\begin{table}[!tb]
\centering
\caption{\small
  Excited charm-strange states above the $D_{s2}^*(2573)^-$ seen in $\D^{(*)}\kaon$ spectra by BaBar~\cite{Aubert:2009ah} in \epem collisions and by LHCb~\cite{LHCb-PAPER-2012-016} in $pp$ collisions.
  Units of $\mevcc$ are implied.
  The first source of uncertainty is statistical and the second is systematic.
}
\begin{tabular}{cccc}
  \hline
  State & Mass & Width & Comment \\
  \hline
  \multicolumn{4}{c}{BaBar} \\
  \hline
  $D_{s1}^*(2700)^-$ & $2710 \pm 2 \,^{+12}_{-7}$ & $149 \pm \phani7 ~^{+39}_{-52}$ &
  Seen in $DK$ and $D^*K$ \\
  $D_{sJ}^*(2860)^-$ & $2862 \pm 2 \,^{+5\phani}_{-2}$  & $\phani48 \pm \phani3 \pm 6$ & Seen in
  $DK$ and $D^*K$ \\
  $D_{sJ}(3040)^-$   & $3044 \pm 8 \,^{+30}_{-5}$ & $239 \pm 35 ~^{+46}_{-42}$ & Seen in $D^*K$ only \\
  \hline
  \multicolumn{4}{c}{LHCb} \\
  \hline
  $D_{s1}^*(2700)^-$ & $2709.2 \pm 1.9 \pm 4.5$ & $115.8 \pm 7.3 \pm 12.1$ &
  \multirow{2}{*}{Only $DK$ studied} \\
  $D_{sJ}^*(2860)^-$ & $2866.1 \pm 1.0 \pm 6.3$ & $ 69.9 \pm 3.2 \pm  6.6$ & \\
  \hline
\end{tabular}
\label{tab:BaBarLHCb}
\end{table}

The Dalitz plot~\cite{Dalitz:1953cp} analysis technique has proven to be a
powerful tool for studies of charm meson spectroscopy.
Analyses by the Belle~\cite{Abe:2003zm,Kuzmin:2006mw} and
BaBar~\cite{Aubert:2009wg} collaborations of $B\to D\pi\pi$ decays have
provided insight into the orbitally excited charm mesons.
Such analyses complement those on inclusive production of charm
mesons~\cite{Abulencia:2005ry,delAmoSanchez:2010vq,LHCb-PAPER-2013-026} as the
lower background allows broader states to be distinguished and the
well-defined initial state allows the quantum numbers to be unambiguously
determined.
These advantages compensate to some extent for the smaller samples that are
available from \B meson decay compared to inclusive production.

Until now, few results on charm-strange meson spectroscopy have become
available from Dalitz plot analyses, because the available samples of such
mesons from \Bp and \Bz decays are much smaller than those of non-strange charm mesons.
An exception is a study of $\Bp \to \Dz\Dzb\Kp$ decays by Belle~\cite{Brodzicka:2007aa}, which produced the first observation of the $D_{s1}^*(2700)^-$ meson and showed that it has $J^P = 1^-$.
Copious samples of charm-strange mesons are, however, available from decays of \Bs mesons produced at high energy hadron colliders.
These have been exploited to study the properties of the
$D_{s1}(2536)^-$~\cite{Abazov:2007wg} and
$D_{s2}^*(2573)^-$~\cite{LHCb-PAPER-2011-001} states produced in semileptonic \Bs
decays.
Production of orbitally excited charm-strange mesons has also been seen in
hadronic \Bs decays~\cite{LHCb-PAPER-2012-033}.

In this paper, the first Dalitz plot analysis of the $\Bs \to \Dzb \Km \pip$
decay is presented.
The \Dzb meson is reconstructed through the $\Kp\pim$ decay mode, which is
treated as flavour-specific \ie\ the heavily suppressed $\Bs \to \Dz \Km \pip, \Dz \to \Kp\pim$ contribution is neglected.
The inclusion of charge conjugated processes is implied throughout the paper.
Previously the resonant contribution from $\Bs \to \Dzb \Kstarb(892)^0$ has
been observed~\cite{LHCb-PAPER-2011-008} and the inclusive three-body
branching fraction has been measured~\cite{LHCb-PAPER-2013-022}.
In this work the contributions from excited charm-strange mesons and excited kaon states are separated from each other with the amplitude analysis technique.
The results are important not only from the point-of-view of spectroscopy, but
also as they will provide input to future studies of \CP violation.
In particular, the angle $\gamma$ of the Cabibbo-Kobayashi-Maskawa Unitarity Triangle~\cite{PhysRevLett.10.531,PTP.49.652} can be determined from studies of \CP violation in $\Bd \to \Dzb \Kp\pim$ decays~\cite{Gronau:2002mu,Gershon:2008pe,Gershon:2009qc}.
In such analyses, \Bs decays provide both an important control channel and a potential source of background (see, \eg, Ref.~\cite{LHCb-PAPER-2012-042,LHCb-PAPER-2014-028}).

The analysis is based on a data sample corresponding to an integrated luminosity of $3.0  \,{\rm fb}^{-1}$ of $pp$ collision data collected with the LHCb detector, approximately one third of which was collected during 2011 when the collision centre-of-mass energy was $\sqrt{s} = 7 \tev$ and the rest during 2012 with $\sqrt{s} = 8 \tev$.
Amplitude analysis techniques have previously been used by LHCb to study $\Bd$
and $\Bs$ meson decays to
$\jpsi\Kp\Km$~\cite{LHCb-PAPER-2012-040,LHCb-PAPER-2013-045} and
$\jpsi\pip\pim$~\cite{LHCb-PAPER-2012-005,LHCb-PAPER-2012-045,LHCb-PAPER-2013-069,LHCb-PAPER-2014-012}
final states, and to determine the quantum numbers of the
$X(3872)$~\cite{LHCb-PAPER-2013-001} and $Z(4430)$~\cite{LHCb-PAPER-2014-014}
resonances.
This is, however, the first time that such an analysis has been performed by
LHCb with a decay into a fully hadronic final state (\ie\ without muons).

The paper is organised as follows.
A brief description of the LHCb detector as well as reconstruction and simulation software is given in Sec.~\ref{sec:Detector}.
The selection of signal candidates and the fit to the \Bs candidate invariant mass distribution used to separate signal and background are described in Secs.~\ref{sec:Selection} and~\ref{sec:MassFit}, respectively.
An overview of the Dalitz plot analysis formalism and a definition of the square Dalitz plot (SDP) are given in Sec.~\ref{sec:DalitzGeneralities}, and details of the implementation of the amplitude analysis are presented in Sec.~\ref{sec:Dalitz}.
The evaluation of systematic uncertainties is described in Sec.~\ref{sec:Systematics}.
The results are given in Sec.~\ref{sec:Results}, and a summary concludes the paper in Sec.~\ref{sec:Summary}.
The highlights of the analysis are described in a shorter companion paper~\cite{LHCb-PAPER-2014-035}.

\section{LHCb detector}
\label{sec:Detector}

The \lhcb detector~\cite{Alves:2008zz} is a single-arm forward
spectrometer covering the \mbox{pseudorapidity} range $2<\eta <5$,
designed for the study of particles containing \bquark or \cquark
quarks. The detector includes a high-precision tracking system
consisting of a silicon-strip vertex detector~\cite{LHCb-DP-2014-001}
surrounding the $pp$ interaction region, a large-area silicon-strip detector
located upstream of a dipole magnet with a bending power of about
$4{\rm\,Tm}$, and three stations of silicon-strip detectors and straw
drift tubes~\cite{LHCb-DP-2013-003} placed downstream of the magnet.
The combined tracking system provides a momentum measurement with
a relative uncertainty that varies from 0.4\% at low momentum, \ptot, to 0.6\% at 100\gevc,
and an impact parameter (IP) measurement with a resolution of 20\mum for
charged particles with large momentum transverse to the beamline, \pt~\cite{LHCb-DP-2014-002}.
Different types of charged hadrons are distinguished using information
from two ring-imaging Cherenkov detectors~\cite{LHCb-DP-2012-003}.
Photon, electron and
hadron candidates are identified by a calorimeter system consisting of
scintillating-pad and preshower detectors, an electromagnetic
calorimeter and a hadronic calorimeter. Muons are identified by a
system composed of alternating layers of iron and multiwire
proportional chambers~\cite{LHCb-DP-2012-002}.

The trigger~\cite{LHCb-DP-2012-004} consists of a
hardware stage, based on information from the calorimeter and muon
systems, followed by a software stage, in which all tracks
with a transverse momentum above a threshold of 500 (300)\mevc during 2011 (2012) data-taking are reconstructed.
In the offline selection, the objects that fired the trigger are associated with reconstructed particles.
Selection requirements can therefore be made not only on the trigger line that fired, but on whether the decision was due to the signal candidate, other particles produced in the $pp$ collision, or a combination of both.
Signal candidates are accepted offline if one of the final state particles created a cluster in the hadronic calorimeter with sufficient transverse energy to fire the hardware trigger.
Events that are triggered at the hardware level by another particle in the event are also retained.
After all selection requirements are imposed, 62\,\% of events in the sample were triggered by the signal candidate and 58\,\% were triggered by another particle in the event including 20\,\% that were triggered independently by both by the signal candidate and by another particle.
The software trigger requires a two-, three- or four-track secondary vertex
with a large sum of the \pt of the tracks and a
significant displacement from any of the primary $pp$ interaction vertices~(PVs). At
least one track should have $\pt > 1.7\gevc$ and \chisqip with respect to any
primary interaction greater than 16, where \chisqip is defined as the
difference in \chisq of a given PV reconstructed with and without the
considered particle.

Simulated events are used to characterise the detector response to signal and
certain types of background events.
In the simulation, $pp$ collisions are generated using
\pythia~\cite{Sjostrand:2006za,*Sjostrand:2007gs} with a specific \lhcb
configuration~\cite{LHCb-PROC-2010-056}.  Decays of hadronic particles
are described by \evtgen~\cite{Lange:2001uf}, in which final state
radiation is generated using \photos~\cite{Golonka:2005pn}. The
interaction of the generated particles with the detector and its
response are implemented using the \geant
toolkit~\cite{Allison:2006ve, *Agostinelli:2002hh} as described in
Ref.~\cite{LHCb-PROC-2011-006}.
\section{Selection requirements}
\label{sec:Selection}

The selection requirements are similar to those used in
Refs.~\cite{LHCb-PAPER-2012-056,LHCb-PAPER-2013-022}.
The $\Bd\to\Dzb\pip\pim$ decay, which is topologically and kinematically similar to the signal mode, is used as a control channel to optimise the requirements and is not otherwise used in the analysis.
A set of loose initial requirements is imposed to obtain a visible signal peak in the $\Dzb\pip\pim$ candidates.
The tracks are required to be of good quality and to be above thresholds in $p$, \pt and \chisqip, while the $\Dzb\to\Kp\pim$ candidate must satisfy criteria on its vertex quality ($\chisq_{\rm vtx}$) and flight distance from any PV and from the $B$ candidate vertex.
Only candidates with $1814 < m(\Kp\pim) < 1914 \mevcc$ are retained.
A requirement is also imposed on the output of a boosted decision tree that identifies \Dzb mesons (with the appropriate final state) produced in \bquark hadron decays ($\Dzb$ BDT)~\cite{LHCb-PAPER-2012-025,LHCb-PAPER-2012-050}.
The $B$ candidate must satisfy requirements on its invariant mass, $\chisqip$ and on the cosine of the angle between the momentum vector and the line from the PV under consideration to the $\B$ vertex ($\cos \theta_{\rm dir}$).
A requirement is placed on the $\chisq$ of a kinematic fit~\cite{Hulsbergen:2005pu}, in which the $\Dzb$ mass is constrained to its nominal value, to the $\B$ decay hypothesis of the final state tracks.
The four final state tracks are also required to satisfy pion and kaon identification (PID) requirements.

Further discrimination between signal and combinatorial background is achieved with a neural network~\cite{Feindt:2006pm}.
The \sPlot\ technique~\cite{Pivk:2004ty}, with the \B candidate mass as discriminating variable, is used to statistically separate $\Bd\to\Dzb\pip\pim$ decays from background among the remaining $\Dzb\pip\pim$ candidates.
Signal and background weights obtained from this procedure are applied to the candidates, which are then used to train the network.
A total of 16 variables is used in the network.
They include the $\chisqip$ of the four final state tracks and the following variables associated to the $\Dzb$ candidate:
$\chisqip$; $\chisq_{\rm vtx}$; the square of the flight distance from the PV divided by its uncertainty ($\chisq_{\rm flight}$); $\cos \theta_{\rm dir}$; the output of the $\Dzb$ BDT.
In addition, the following variables associated to the $\B$ candidate are included:
$\pt$; $\chisqip$; $\chisq_{\rm vtx}$; $\chisq_{\rm flight}$; $\cos \theta_{\rm dir}$.
Information from the rest of the event is also included through variables that describe the $\pt$ asymmetry, $A_{\pt}$, and track multiplicity in a cone with half-angle of 1.5 units in the plane of pseudorapidity and azimuthal angle (measured in radians)~\cite{LHCB-PAPER-2012-001} around the $\B$ candidate flight direction, with
\begin{equation}
  A_{\pt} = \frac{\pt(\B) - \sum_n \pt(n)}{\pt(\B) + \sum_n \pt(n)}\, ,
\end{equation}
where the scalar sum is over the tracks contained in the cone excluding those associated with the signal $\B$ candidate.
The input quantities to the neural network depend only weakly on position in
the $\B$ decay Dalitz plot.
A requirement imposed on the network output reduces the combinatorial background remaining after the initial selection by a factor of five while retaining more than $90\,\%$ of the signal.

The $\Bs\to\Dzb\Km\pip$ candidates must satisfy all criteria applied to the $\Dzb\pip\pim$ sample with the exception of the PID requirement on the negatively charged ``bachelor'' track, \ie\ the negatively charged track coming directly from the \Bs decay, which is replaced with a requirement that preferentially selects kaons.
The combined efficiency of the PID requirements on the four tracks in the
final state is around $50\,\%$ and varies depending on the kinematics of the
tracks, as described in detail in Sec.~\ref{sec:DalitzEfficiency}.
The PID efficiency is determined using samples of $\Dz \to \Km\pip$ decays
selected in data by exploiting the kinematics of the $\Dstarp\to\Dz\pip$
decay chain to obtain clean samples without using the PID information~\cite{LHCb-DP-2012-003}.

Track momenta are scaled~\cite{LHCb-PAPER-2012-048,LHCB-PAPER-2013-011} with calibration parameters determined by matching the measured peak of the $\jpsi\to\mumu$ decay to the known $\jpsi$ mass~\cite{PDG2012}.
To improve further the $\Bs$ candidate invariant mass resolution, a kinematic fit~\cite{Hulsbergen:2005pu} is used to adjust the four-momenta of the tracks from the $\Dzb$ candidate so that their combined invariant mass matches the world average value for the $\Dzb$ meson~\cite{PDG2012}.
An additional $\Bs$ mass constraint is applied in the calculation of the variables that are used in the Dalitz plot fit.

\begin{figure}
\centering
\includegraphics[scale=0.50]{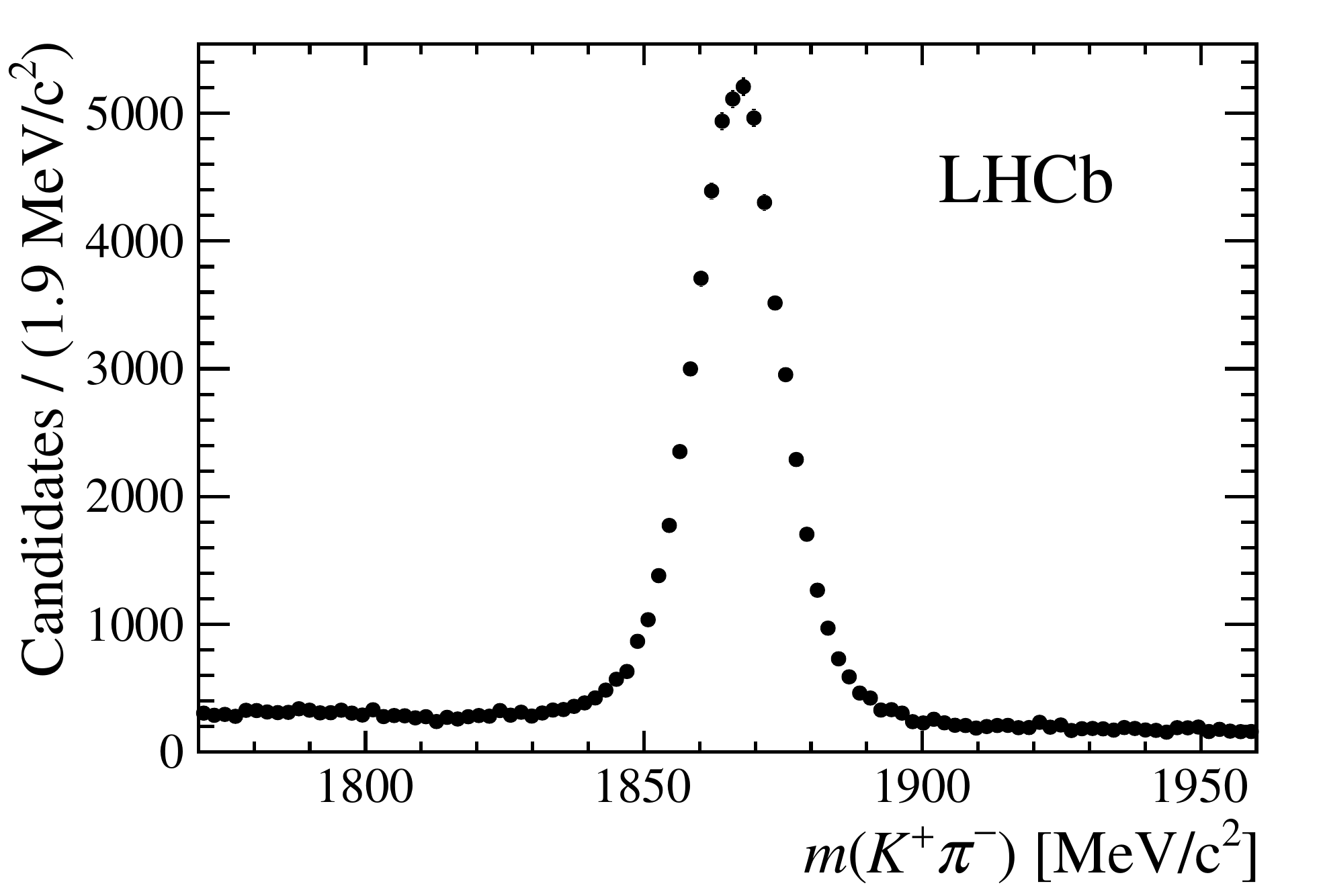}
\caption{\small
  Distribution of \Dzb candidate invariant mass for \Bs candidates in the signal region defined in Sec.~\ref{sec:MassFit}.
  Here the selection criteria have been modified to avoid biasing the distribution: the $\Dzb$ candidate invariant mass requirement has been removed, and the $\chisq$ of the kinematic fit is calculated without applying the $\Dzb$ mass constraint.
}
\label{fig:dmass}
\end{figure}

To remove potential background from $\Dstarpm$ decays, candidates are rejected if the difference between the invariant mass of the combination of the $\Dzb$ candidate and the $\pip$ bachelor and that of the $\Dzb$ candidate itself lies within $\pm 2.5 \mevcc$ of the nominal $\Dstarp$--$\Dz$ mass difference~\cite{PDG2012}.
(This veto removes $\Dstarp \to \Dz\pip$ decays followed by the suppressed $\Dz \to \Kp\pim$ decay; since the the $D$ meson decays is treated as flavour-specific, the final state contains what is referred to as a $\Dzb$ candidate.)
Candidates are also rejected if a similar mass difference calculated with the pion mass hypothesis applied to the kaon bachelor, satisfies the same criterion.
Furthermore, it is required that the kaon from the $\Dzb$ candidate together with the bachelor kaon and the bachelor pion do not form an invariant mass in the range $1955$--$1980 \mevcc$ to remove potential background from $\Bs\to\Dsm\pip$ decays.
Potential background from $\Bs \to \Dz\Dzb$ decays~\cite{LHCb-PAPER-2012-050} is removed by requiring that the pion and kaon originating directly from the \Bs decay give an invariant mass outside the range $1835$--$1880 \mevcc$.
At least one of the pion candidates is required to have no associated hits in the muon counters to remove potential background from $\Bd\to\jpsi\Kstarz$ decays.
Decays of \Bs mesons to the same final state but without an intermediate charm meson are suppressed by the $\Dzb$ BDT criteria, and any surviving background from this source is removed by requiring that the $\Dzb$ candidate vertex is displaced by at least $1\mm$ from the $\Bs$ decay vertex.
Figure~\ref{fig:dmass} shows the \Dzb candidate mass after the selection criteria are applied.

Signal candidates are retained for further analysis if they have an invariant mass in the range $5200$--$5900 \mevcc$.
After all selection requirements are applied, fewer than 1\,\% of events with one candidate also contain a second candidate.
Such multiple candidates are retained and treated in the same manner as other candidates; the associated systematic uncertainty is negligible.
\section{Determination of signal and background yields}
\label{sec:MassFit}

The signal and background yields are obtained from an extended unbinned maximum likelihood fit to the three-body invariant mass distribution of $\Bs \to \Dzb \Km\pip$ candidates.
In addition to signal decays and combinatorial background, the fit allows
background contributions from other \bquark hadron decays.
The decay $\Bs \to \Dstarzb \Km\pip$, with $\Dstarzb \to \Dzb\piz$ or
$\Dzb\gamma$ forms a partially reconstructed background that peaks at values
below the \Bs mass since the $\piz$ or $\gamma$ is missed.
Decays of \Bdb mesons to the $\Dzb \Km\pip$ final state are Cabibbo-suppressed, but may contribute a non-negligible background.
Decays with similar topology and misidentified final state particles can also
populate the mass region used in the fit.
Studies using simulated background events show that contributions from
$\Bd \to \DorDstarzb \pip\pim$ and $\Lbbar \to \DorDstarzb \antiproton
\pip$~\cite{LHCb-PAPER-2013-056} are expected, while background from
$\Bds\to \DorDstarzb\Kp\Km$~\cite{LHCb-PAPER-2012-018,LHCb-PAPER-2013-035} and $\Lbbar \to \DorDstarzb \antiproton \Kp$ is negligible.

The signal and $\Bdb \to \Dzb \Km\pip$ shapes are each modelled with the sum
of two Crystal Ball~\cite{Skwarnicki:1986xj} functions which share a common mean and have tails on opposite sides.
Studies using simulated events and the $\Bd \to \Dzb\pip\pim$ control channel in data verify that this function gives an excellent description of the signal shape.
All tail parameters are fixed to values determined from a fit to simulated
signal decays.
The mass difference between the peaks corresponding to \Bd and \Bs decays is
fixed to its known value~\cite{PDG2012}.
The combinatorial background is modelled using a linear shape.

Smoothed histograms are used to describe the shapes of $\Bs \to \Dstarzb
\Km\pip$, $\Bd \to \DorDstarzb \pip\pim$ and $\Lbbar \to \DorDstarzb
\antiproton \pip$ decays.
The shape for $\Bs \to \Dstarzb\Km\pip$ decays is determined from
simulated events, including contributions from both $\Dstarzb\to\Dzb\gamma$
and $\Dstarzb\to\Dzb\piz$ final states in the correct proportion~\cite{PDG2012}.
The shapes for $\Lbbar \to \DorDstarzb \antiproton\pip$ and
$\Bd\to\DorDstarzb \pip\pim$ decays are derived from simulated samples:
the $\Bd\to\Dzb \pip\pim$ and $\Bd\to\Dstarzb \pip\pim$ samples are combined
in proportion to their branching fractions~\cite{PDG2012}, while the
corresponding \Lbbar decays are combined assuming equal branching fractions since that for the $\Lbbar \to \Dstarzb \antiproton\pip$ decay has not yet been measured.
The shapes of the misidentified backgrounds are reweighted according to: (i) the
known Dalitz plot distributions for the decay modes with $\Dzb$
mesons~\cite{LHCb-PAPER-2013-022,LHCb-PAPER-2013-056}; (ii) the particle identification and misidentification probabilities, accounting for kinematic dependence.
The $K$ and $\pi$ (mis)identification probabilities are obtained from the $\Dstarp\to\Dz\pip, \Dz \to \Km\pip$ samples described in Sec.~\ref{sec:Selection}, while those for (anti)protons are obtained from samples of $\Lz \to p\pim$ decays.

\begin{figure}[!tb]
  \centering
  \includegraphics[scale=0.38]{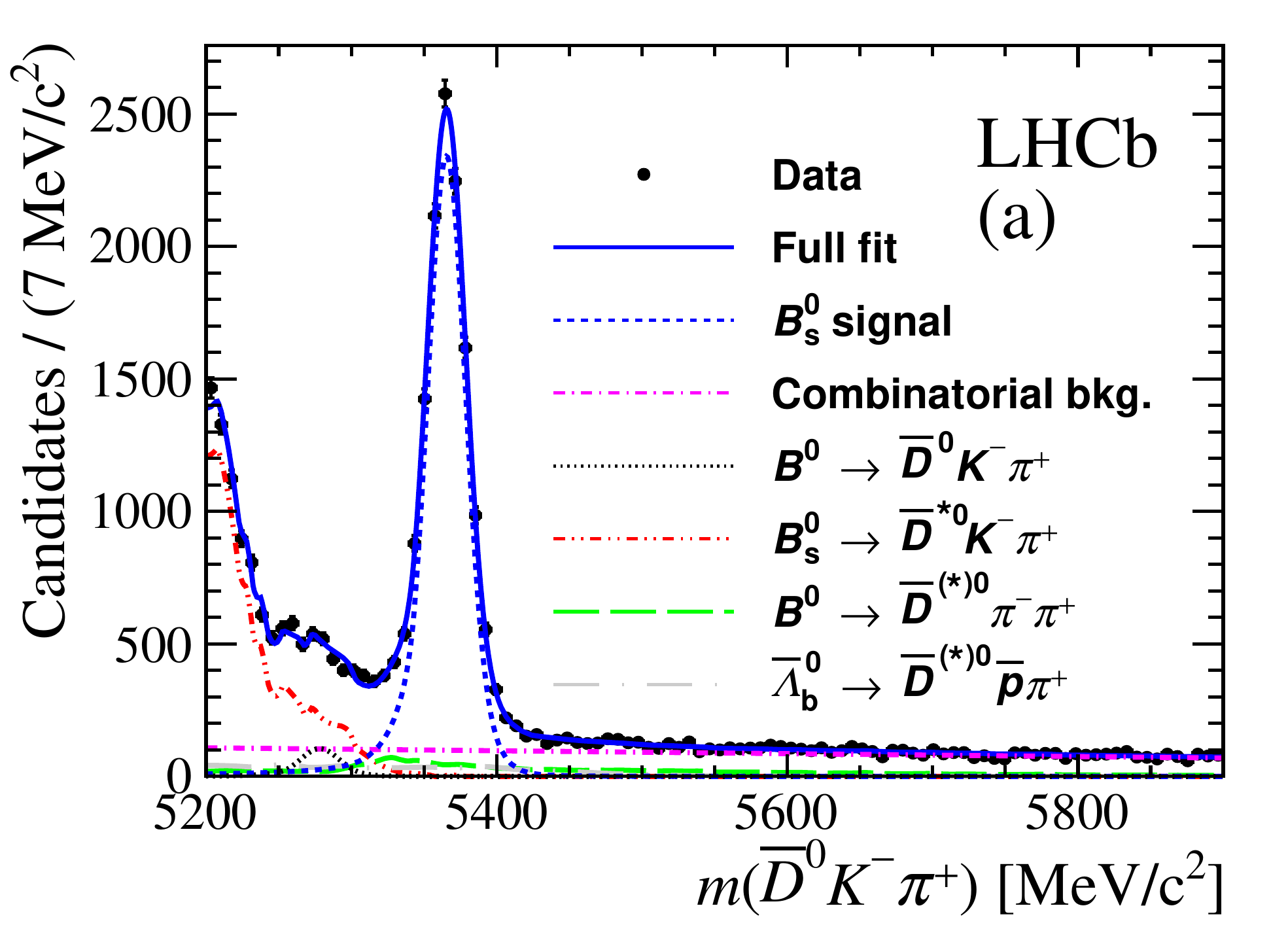}
  \includegraphics[scale=0.38]{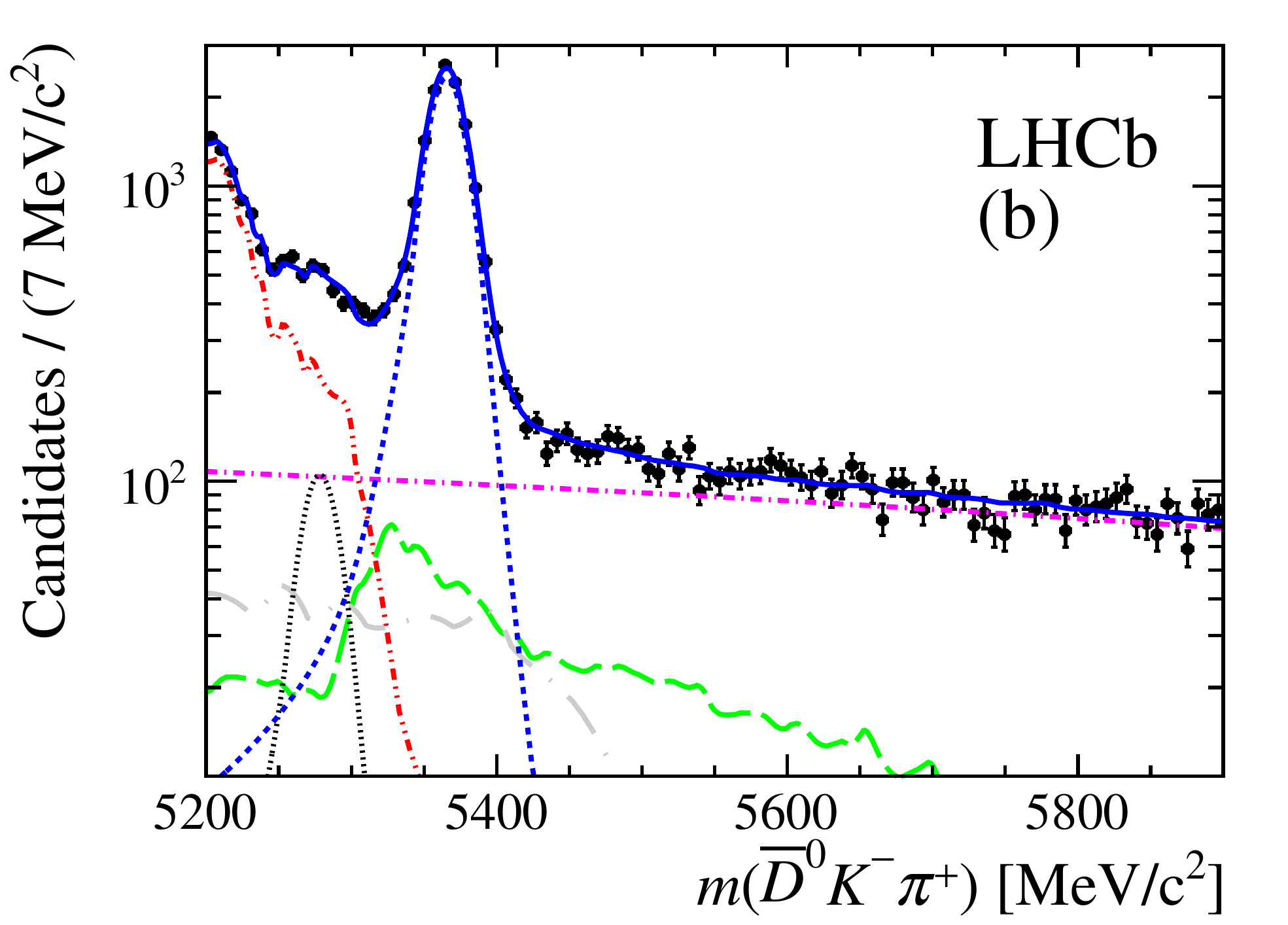}
  \caption{\small
    Result of the fit to the $\Bs\to \Dzb\Km\pip$ candidates invariant mass distribution shown with (a) linear and (b) logarithmic $y$-axis scales.
    Data points are shown in black, the total fit as a solid blue line and the components as detailed in the legend.
  }
  \label{fig:fits}
\end{figure}

There are in total eleven free parameters determined by the fit:
the peak position and the widths of the signal shape, the fraction of the shape contained within the narrower of the two Crystal Ball functions, the linear slope of the combinatorial background, and the yields of the six categories defined above.
The results of the fit are shown in Fig.~\ref{fig:fits} and listed in Table~\ref{tab:massfit}. The fit gives a reduced $\chi^2$ of $98.6/88 = 1.12$.
All yields are consistent with their expectations, based on measured or predicted production rates and branching fractions, and efficiencies or background rejection factors determined from simulations.

\begin{table}[!tb]
  \centering
  \caption{\small
    Results of the $\Bs\to \Dzb\Km\pip$ candidate invariant mass fit.
    Uncertainties are statistical only.
  }
  \label{tab:massfit}
  \vspace{1ex}
  \begin{tabular}{cc}
    \hline
    Parameter & Value \\
    \hline
    $\mu_{\Bs}$                        &            $5365.5    \pm 0.2 \mevcc$ \\
    $\sigma_{1}$                       &            $12.7      \pm 0.2 \mevcc$ \\
    $\sigma_{2}/\sigma_{1}$            &            $1.76 \pm 0.05$ \\
    Relative fraction                  &            $0.797     \pm 0.017$ \\
    Linear slope                       &            $-0.144 \pm 0.006 \ (\!\gev/c^2)^{-1}$ \\
    $N(\Bs \to \Dzb\Km\pip)$          &            $12\,450  \pm 180$  \\
    $N(\Bdb\to \Dzb\Km\pip)$          &            $\phantom{12\,}550      \pm \phani80$  \\
    $N(\text{comb. bkg.})$            &            $\phantom{1\,}9200     \pm 600$  \\
    $N(\Bs\to \Dstarzb \Km\pip)$      &            $\phantom{1\,}7590     \pm 140$  \\
    $N(\Bz\to \DorDstarzb\pip\pim)$   &            $\phantom{1\,}1700     \pm 600$  \\
    $N(\Lbbar\to \DorDstarzb\antiproton\pip)$ &    $\phantom{1\,}1270     \pm 350$  \\
    \hline
  \end{tabular}
\end{table}

\begin{figure}
\centering
\includegraphics[scale=0.38]{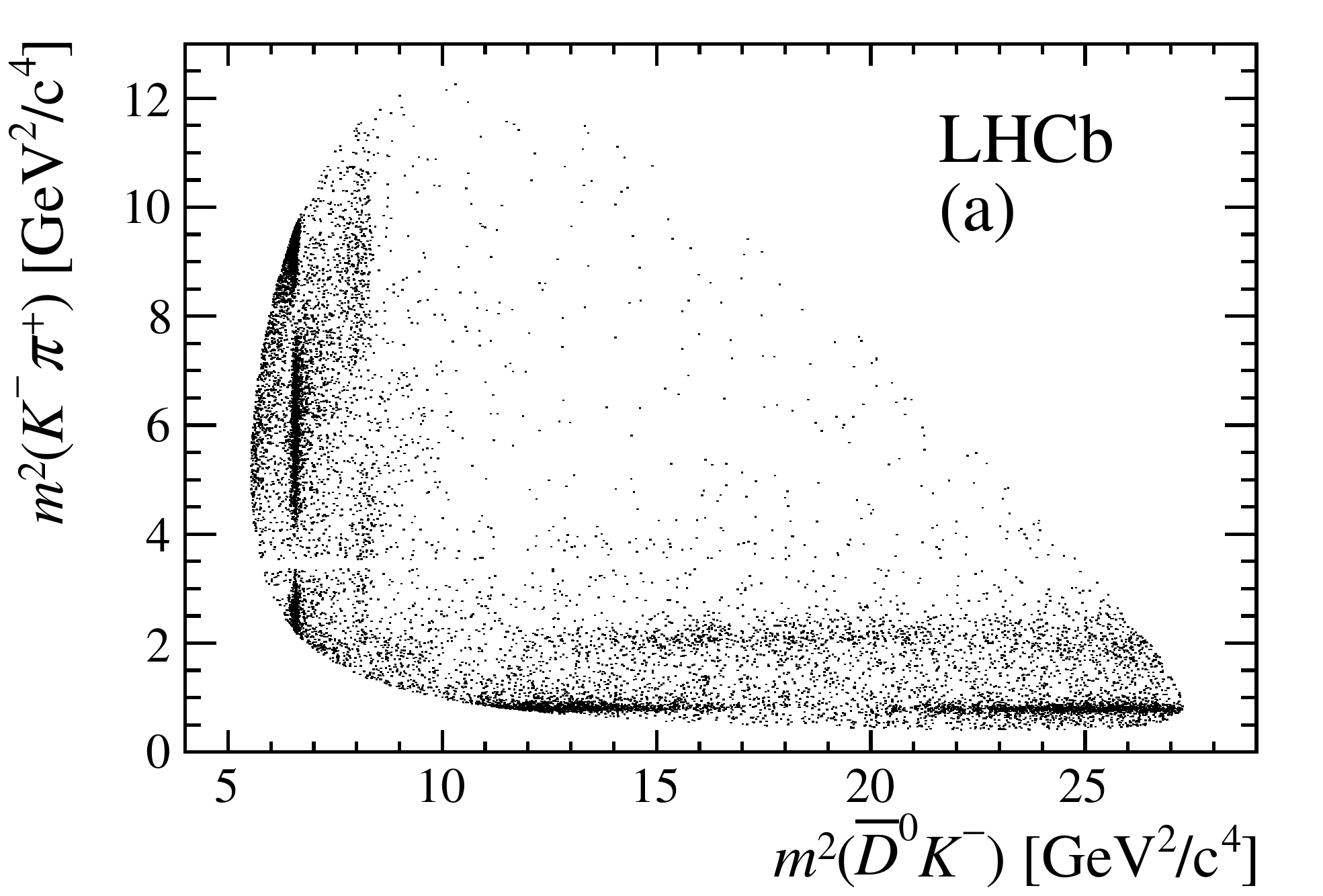}
\includegraphics[scale=0.38]{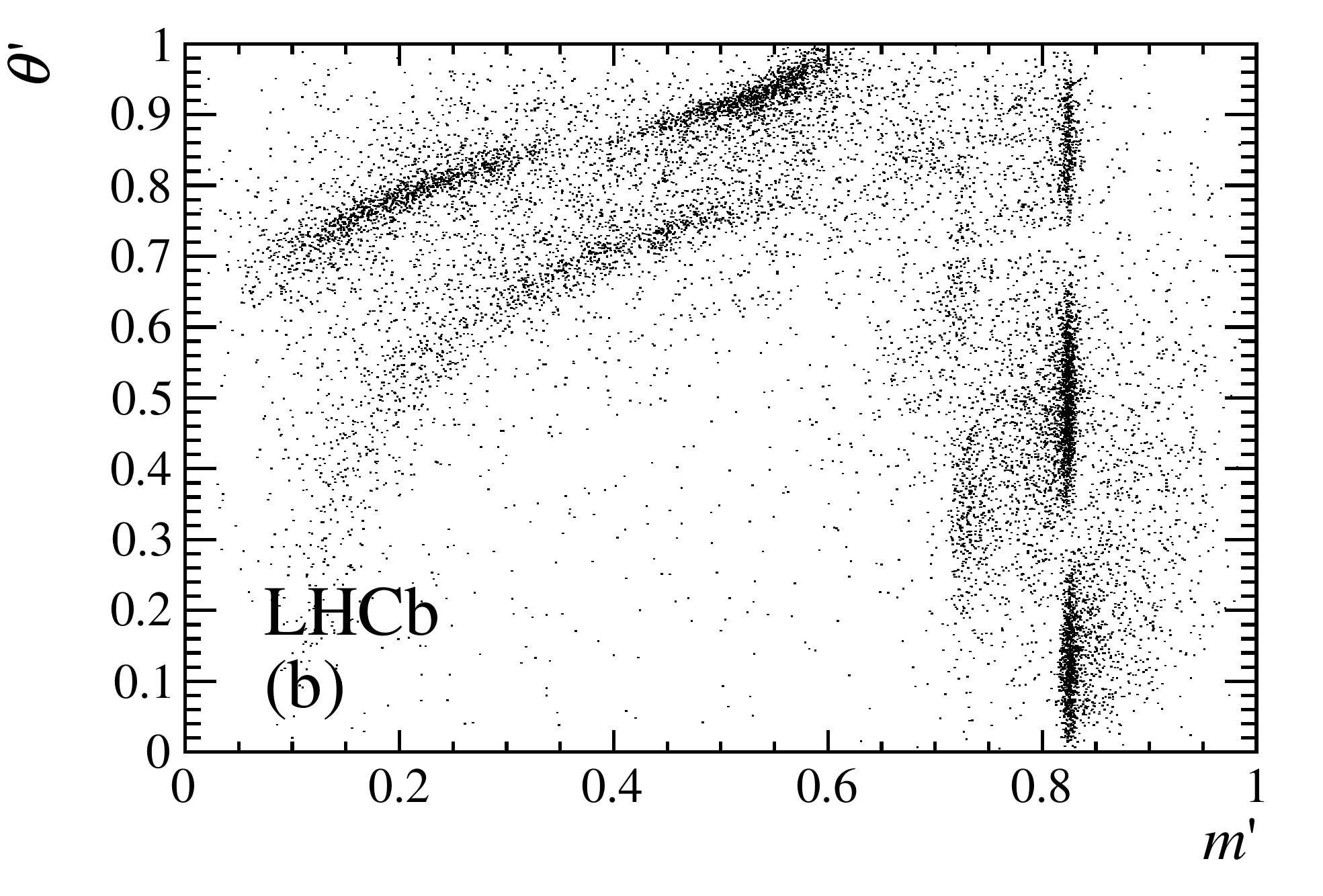}
\caption{\small
  Distribution of $\Bs\to\Dzb\Km\pip$ candidates in the signal region over (a) the Dalitz plot and (b) the square Dalitz plot defined in Eq.~(\ref{eq:sqdp-vars}).
  The effect of the \Dz veto can be seen as an unpopulated horizontal (curved) band in the (square) Dalitz plot.
}
\label{fig:signalevents}
\end{figure}

For the Dalitz plot analysis a signal region is defined as $\mu_\Bs\pm2.5\sigma_1$, where $\mu_\Bs$ and $\sigma_{1}$ are the peak position and core width of the signal shape, respectively, and are taken from the results of the mass fit.
The signal region is then $5333.75$--$5397.25\mevcc$.
The yields in this region are summarised in Table~\ref{tab:yields}.
The distributions of candidates in the signal region over both the Dalitz plot
and the square Dalitz plot defined in the next section are shown in Fig.~\ref{fig:signalevents}.

\begin{table}[!tb]
  \centering
  \caption{\small
    Yields of the fit components within the signal region used for the Dalitz
    plot analysis.
     }
  \label{tab:yields}
  \vspace{1ex}
  \begin{tabular}{cr@{$\,\pm\,$}l}
    \hline
    Component & \multicolumn{2}{c}{Yield} \\
    \hline \\ [-2.5ex]
    $\Bs \to \Dzb \Km\pip$                   & $11\,300$ & $160$ \\
    $\Bdb\to \Dzb \Km\pip$                   &       $2$ & $1$ \\
    comb. bkg.                               &     $950$ & $60$ \\
    $\Bs\to \Dstarzb \Km\pip$                &      $40$ & $1$ \\
    $\Bz\to \DorDstarzb\pip\pim$             &     $360$ & $130$ \\
    $\Lbbar\to \DorDstarzb\antiproton\pip$   &     $300$ & $80$ \\
    \hline
  \end{tabular}
\end{table}

\section{Dalitz plot analysis formalism}
\label{sec:DalitzGeneralities}

The Dalitz plot~\cite{Dalitz:1953cp} describes the phase-space of the three-body decay in terms of two of the three possible two-body invariant mass squared combinations.
In $\Bs \to \Dzb\Km\pip$ decays, resonances are expected in the $m^2(\Dzb\Km)$ and $m^2(\Km\pip)$ combinations, and therefore this pair is a suitable choice to define the Dalitz plot axes.
Given these two invariant mass squared combinations all other kinematic quantities can be uniquely determined for a fixed $\Bs$ mass.

The description of the complex amplitude is based on the isobar model~\cite{Fleming:1964zz,Morgan:1968zza,Herndon:1973yn}, which describes the total amplitude as a coherent sum of amplitudes from resonant or nonresonant intermediate processes.
As such the total amplitude is given by
\begin{equation}\label{eqn:amp}
  {\cal A}\left(m^2(\Dzb\Km), m^2(\Km\pip)\right) = \sum_{j=1}^{N} c_j F_j\left(m^2(\Dzb\Km), m^2(\Km\pip)\right) \,,
\end{equation}
where $c_j$ are complex coefficients giving the relative contribution of each different decay channel.
The resonance dynamics are contained within the $F_j\left(m^2(\Dzb\Km),m^2(\Km\pip)\right)$ terms, which are
composed of invariant mass and angular distributions and
are normalised such that the integral over the Dalitz plot of the squared magnitude of each term is unity.
For example, for a $\Dzb\Km$ resonance
\begin{equation}
  \label{eq:ResDynEqn}
  F\left(m^2(\Dzb\Km), m^2(\Km\pip)\right) =
  R\left(m(\Dzb\Km)\right) \times X(|\vec{p}\,|\,r_{\rm BW}) \times X(|\vec{q}\,|\,r_{\rm BW})
  \times T(\vec{p},\vec{q}\,) \, ,
\end{equation}
where the functions $R$, $X$ and $T$ described below depend on parameters of the resonance such as its spin $L$, pole mass $m_0$ and width $\Gamma_0$.
In the case of a $\Dzb\Km$ resonance, the \pip is referred to as the ``bachelor'' particle. Since the \Bs meson has zero spin, $L$ is equivalently the orbital angular momentum between the resonance and the bachelor.

In Eq.~(\ref{eq:ResDynEqn}), the function $R\left(m(\Dzb\Km)\right)$ is the resonance mass term
(given \eg\ by a Breit--Wigner shape --- the detailed forms for each of the resonance shapes used in the model are described below),
while $\vec{p}$ and $\vec{q}$ are the momenta of the bachelor particle and one of the resonance daughters, respectively, both evaluated in the rest frame of the resonance.
The terms $X(z)$, where $z=|\vec{q}\,|\,r_{\rm BW}$ or $|\vec{p}\,|\,r_{\rm BW}$, are Blatt--Weisskopf barrier form factors~\cite{blatt-weisskopf}, and are given by
\begin{eqnarray}
\label{eq:BWFormFactors}
L = 0 \ : \ X(z) & = & 1\,, \\
L = 1 \ : \ X(z) & = & \sqrt{\frac{1 + z_0^2}{1 + z^2}}\,, \\
L = 2 \ : \ X(z) & = & \sqrt{\frac{z_0^4 + 3z_0^2 + 9}{z^4 + 3z^2 + 9}}\,,\\
L = 3 \ : \ X(z) & = & \sqrt{\frac{z_0^6 + 6z_0^4 + 45z_0^2 + 225}{z^6 + 6z^4 + 45z^2 + 225}}\,,
\label{eq:BWFormFactors-end}
\end{eqnarray}
where $z_0$ represents the value of $z$ when the invariant mass is equal to
the pole mass of the resonance.
The radius of the barrier, $r_{\rm BW}$, is taken to be $4.0\gev^{-1} \approx 0.8\fm$~\cite{Aubert:2005ce} for all resonances.
The angular probability distribution terms, $T(\vec{p},\vec{q})$, are
given in the Zemach tensor formalism~\cite{Zemach:1963bc,Zemach:1968zz} by
\begin{eqnarray}
\label{eq:ZTFactors}
L = 0 \ : \ T(\vec{p},\vec{q}) & = & 1\,,\\
L = 1 \ : \ T(\vec{p},\vec{q}) & = & -\,2\,\vec{p}\cdot\vec{q}\,,\\
L = 2 \ : \ T(\vec{p},\vec{q}) & = & \frac{4}{3} \left[3(\vec{p}\cdot\vec{q}\,)^2 - (|\vec{p}\,||\vec{q}\,|)^2\right]\,,\\
L = 3 \ : \ T(\vec{p},\vec{q}) & = & -\,\frac{24}{15} \left[5(\vec{p}\cdot\vec{q}\,)^3 - 3(\vec{p}\cdot\vec{q}\,)(|\vec{p}\,||\vec{q}\,|)^2\right]\,,
\label{eq:ZTFactors-end}
\end{eqnarray}
which can be seen to have similar forms to the Legendre polynomials, $P_L(x)$,
where $x$ is the cosine of the angle between $\vec{p}$ and $\vec{q}$ (referred
to as the ``helicity angle'').

The majority of the resonant contributions in the decay can have their mass
terms described by the relativistic Breit--Wigner (RBW) function
\begin{equation}
\label{eq:RelBWEqn}
R(m) = \frac{1}{(m_0^2 - m^2) - i\, m_0 \Gamma(m)} \,,
\end{equation}
where the dependence of the decay width of the resonance on $m$ is given by
\begin{equation}
\label{eq:GammaEqn}
\Gamma(m) = \Gamma_0 \left(\frac{q}{q_0}\right)^{2L+1}
\left(\frac{m_0}{m}\right) X^2(q\,r_{\rm BW}) \,,
\end{equation}
where the symbol $q_0$ denotes the value of $q = |\vec{q}\,|$ when $m = m_0$.
This shape can also describe so-called virtual contributions, from resonances
with pole masses outside the kinematically accessible region of the Dalitz plot, with one modification:
in the calculation of the parameter $q_{0}$ the pole mass, $m_0$, is set to a value, $m_0^{\rm{eff}}$, within the kinematically allowed range.
This is accomplished with the {\it ad-hoc} formula
\begin{equation}\label{eqn:effmass}
  m_0^{\rm{eff}}(m_0) = m^{\rm{min}} + (m^{\rm{max}} - m^{\rm{min}}) \left( 1 + \tanh\left( \frac{m_0 - \frac{m^{\rm{min}}+m^{\rm{max}}}{2}}{m^{\rm{max}}-m^{\rm{min}}} \right) \right)\, ,
\end{equation}
where $m^{\rm{max}}$ and $m^{\rm{min}}$ are the upper and lower limits, respectively, of the kinematically allowed mass range.
For virtual contributions, only the tail of the RBW function enters the Dalitz plot.

Because of the large phase-space available in three-body \B meson decays, it is possible to have nonresonant amplitudes (\ie\ contributions that are not associated with any known resonance, including virtual states) that are not, however, constant across the Dalitz plot.
A common approach to model nonresonant terms is to use an exponential form
factor (EFF)~\cite{Garmash:2004wa},
\begin{equation}
  R(m) = e^{-\alpha m^2} \, ,
  \label{eq:nonres}
\end{equation}
where $\alpha$ is a shape parameter that must be determined from the data.

The RBW function is a very good approximation for
narrow resonances well separated from any other resonant or nonresonant
contribution in the same partial wave.
This approximation is known to be invalid in the $K\pi$ S-wave, since the $\Kstarbsubz(1430)$ resonance interferes strongly with a slowly varying nonresonant term (see, for example, Ref.~\cite{Meadows:2007jm}).
The so-called LASS lineshape~\cite{lass} has been developed to combine these amplitudes,
\begin{eqnarray}
  \label{eq:LASSEqn}
  R(m) & = & \frac{m}{q \cot{\delta_B} - iq} + e^{2i \delta_B}
  \frac{m_0 \Gamma_0 \frac{m_0}{q_0}}
       {(m_0^2 - m^2) - i m_0 \Gamma_0 \frac{q}{m} \frac{m_0}{q_0}}\, , \\
{\rm where} \ \cot{\delta_B} & = & \frac{1}{aq} + \frac{1}{2} r q \, ,
\end{eqnarray}
and where $m_0$ and $\Gamma_0$ are now the pole mass and width of the $\Kstarbsubz(1430)$, and $a$ and $r$ are parameters that describe the shape.
Most implementations of the LASS shape in amplitude analyses of \B meson
decays (\eg\ Refs.~\cite{Aubert:2004cp,Aubert:2005ce}) have applied a cut-off
to the slowly varying part close to the charm hadron mass.
The value of the cut-off used in this analysis is $1.7\gevcc$.

In the absence of any reconstruction effects, the Dalitz plot probability
density function would be
\begin{equation}
\label{eq:SigDPLike}
{\cal{P}}_{\rm phys}\left(m^2(\Dzb\Km), m^2(\Km\pip)\right) =
\frac
{|{\cal A}\left(m^2(\Dzb\Km), m^2(\Km\pip)\right)|^2}
{\int\!\!\int_{\rm DP}~{|{\cal A}|^2}~dm^2(\Dzb\Km)\,dm^2(\Km\pip)} \, ,
\end{equation}
where the dependence of ${\cal A}$ on the Dalitz plot position has been suppressed in the denominator for brevity.
In a real experiment, the variation of the efficiency across the Dalitz plot
and the contamination from background processes must be taken into account.
Since signal and background events tend to populate regions close to the
kinematic boundaries of the conventional Dalitz plot, it is convenient to
model the efficiencies and backgrounds using the so-called square Dalitz plot (SDP)
defined by variables \mpr\ and \thpr\ that have validity ranges between 0 and 1 and are given by
\begin{equation}
\label{eq:sqdp-vars}
\mpr \equiv \frac{1}{\pi} \arccos\left(2\frac{m(\Dzb\Km) - m^{\rm min}_{\Dzb\Km}}{m^{\rm max}_{\Dzb\Km} - m^{\rm min}_{\Dzb\Km}} - 1 \right)
\hspace{10mm}{\rm and}\hspace{10mm}
\thpr \equiv \frac{1}{\pi}\theta(\Dzb\Km)\,,
\end{equation}
where $m^{\rm max}_{\Dzb\Km} = m_{\Bs} - m_{\pip}$ and $m^{\rm min}_{\Dzb\Km} = m_{\Dzb} +
m_{\Km}$ are the kinematic boundaries of $m(\Dzb\Km)$ allowed in the $\Bs \to
\Dzb\Km\pip$ decay and $\theta(\Dzb\Km)$ is the helicity angle of the $\Dzb\Km$ system
(the angle between the $\pi$ and the $D$ meson in the $\Dzb\Km$ rest frame).

The primary results of a Dalitz plot analysis are the complex amplitudes given by $c_j$ in Eq.~(\ref{eqn:amp}) that describe the relative contributions of each resonant component.
However, the choice of normalisation, phase convention and amplitude formalism may not be the same for different implementations.
Fit fractions and interference fit fractions provide a convenient convention-independent method to allow meaningful comparisons of results.
The fit fraction is defined as the integral of a single decay amplitude squared divided by that of the coherent matrix element squared for the complete Dalitz plot,
\begin{equation}
{\it FF}_j =
\frac
{\int\!\!\int_{\rm DP}\left|c_j F_j\left(m^2(\Dzb\Km), m^2(\Km\pip)\right)\right|^2~dm^2(\Dzb\Km)\,dm^2(\Km\pip)}
{\int\!\!\int_{\rm DP}\left|{\cal A}\right|^2~dm^2(\Dzb\Km)\,dm^2(\Km\pip)} \, .
\label{eq:fitfraction}
\end{equation}
The sum of these fit fractions is not necessarily unity due to the potential presence of net constructive or destructive interference quantified by interference fit fractions defined for $i<j$ only by
\begin{equation}
  {\it FF}_{ij} =
  \frac
  {\int\!\!\int_{\rm DP} 2 \, {\rm Re}\left[c_ic_j^* F_iF_j^*\right]~dm^2(\Dzb\Km)\,dm^2(\Km\pip)}
  {\int\!\!\int_{\rm DP}\left|{\cal A}\right|^2~dm^2(\Dzb\Km)\,dm^2(\Km\pip)} \, ,
  \label{eq:intfitfraction}
\end{equation}
where the dependence of $F_i^{(*)}$ and ${\cal A}$ on the Dalitz plot position has been omitted.
\section{Dalitz plot fit}
\label{sec:Dalitz}

\subsection{Square Dalitz plot distributions for backgrounds}
\label{sec:DalitzBackground}

\begin{figure}[!tb]
 \centering
 \includegraphics[scale=0.38]{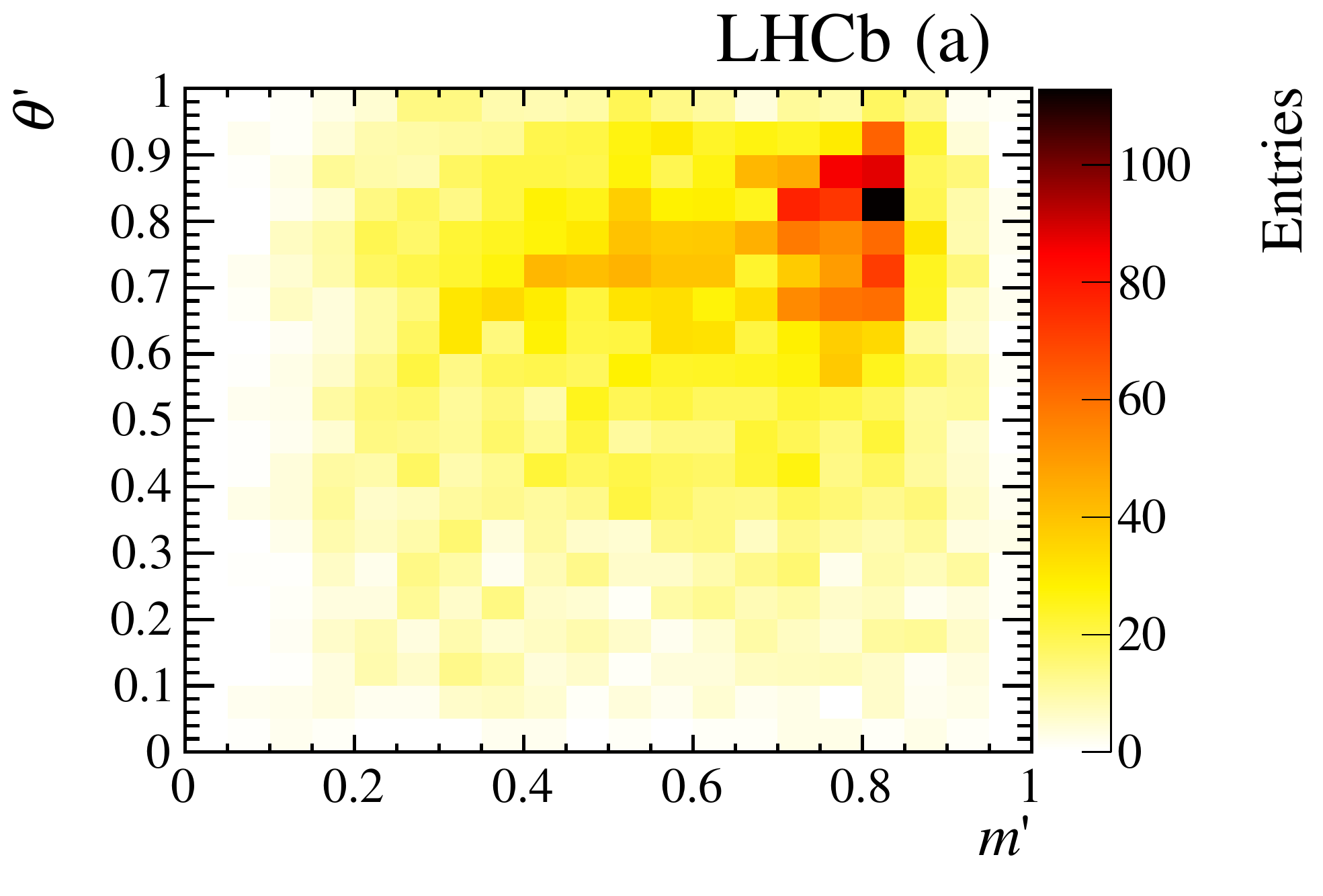}
 \includegraphics[scale=0.38]{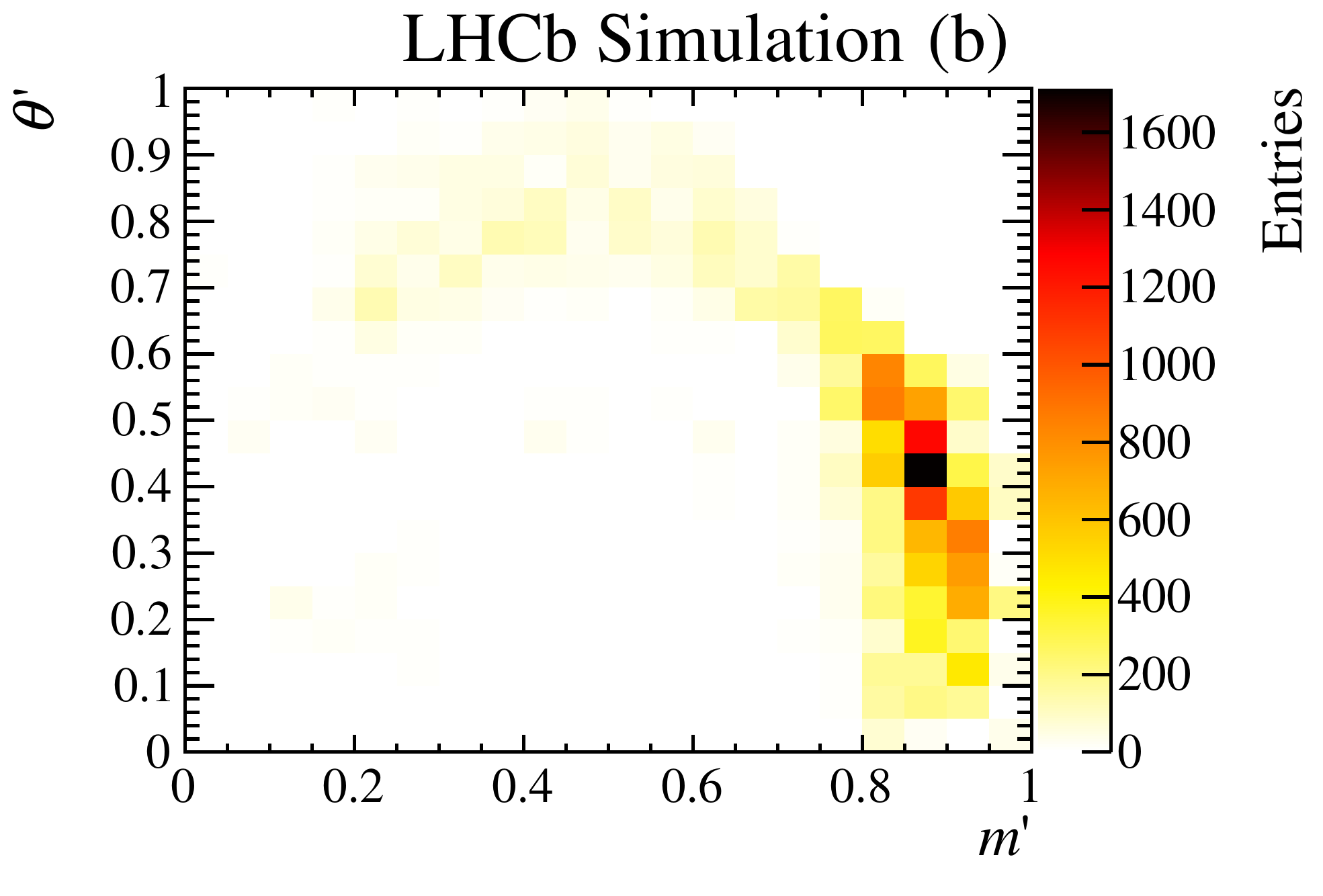}
 \includegraphics[scale=0.38]{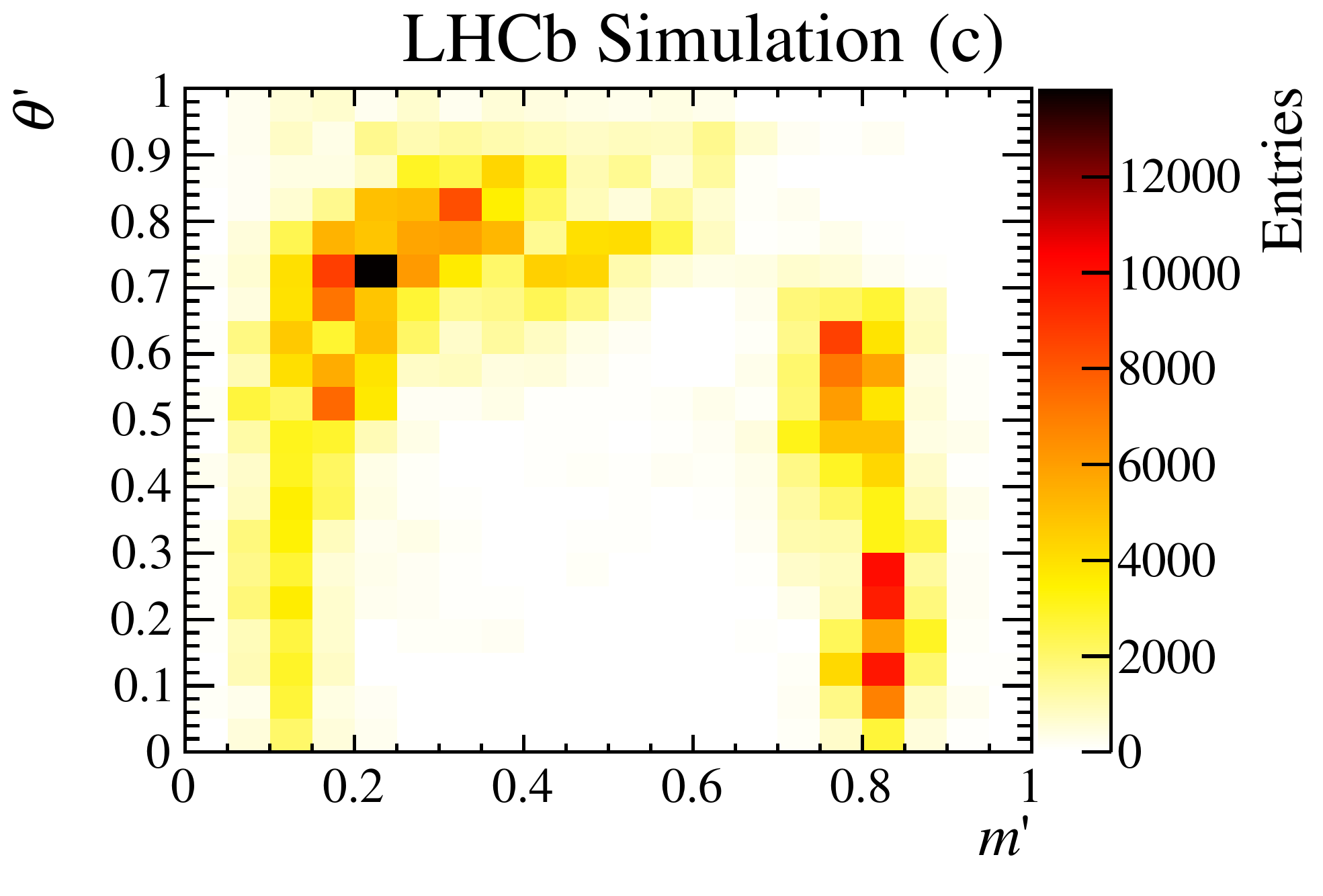}
 \caption{\small SDP distributions of the background contributions from (a) combinatorial, (b) $\Lbbar \to \DorDstarzb \antiproton\pip$ and (c) $\Bd\to\DorDstarzb \pip\pim$ backgrounds.}
 \label{fig:bkg}
\end{figure}

There are non-negligible background contributions in the signal region from combinatorial background and from $\Bd\to\DorDstarzb \pip\pim$ and $\Lbbar \to \DorDstarzb \antiproton\pip$ decays.
As shown in Table~\ref{tab:yields}, these sources correspond to $7.4\,\%$, $2.8\,\%$ and $2.3\,\%$ of the total number of candidates in the signal region, respectively, and therefore their Dalitz plot distributions need to be modelled.
Small contributions from other sources of background are neglected.
The shapes of all background sources in the SDP are described by histograms and are shown in Fig.~\ref{fig:bkg}.

The combinatorial background distribution is obtained from candidates in a high $\Bs$ mass sideband, in the range $5500$--$5900\mevcc$.
The result of the invariant mass fit described in Sec.~\ref{sec:MassFit} shows that this region contains only combinatorial background and a small amount of $\Bd\to\DorDstarzb \pip\pim$ decays.
The latter component is modelled using simulated decays as described below and subtracted from the sideband distribution.
A sample of $\Dzb\Kpm\pipm$ candidates is used to verify that the SDP distribution of combinatorial background does not depend significantly on the $\Bs$ candidate invariant mass, and therefore the sideband distribution can be considered a reliable description of the background in the signal region.

The SDP distributions of the $\Lbbar \to \DorDstarzb \antiproton\pip$ and $\Bd\to\DorDstarzb\pip\pim$ backgrounds are derived from simulated events.
In each shape, the components from the final states containing \Dzb and \Dstarzb mesons are combined and the simulated samples reweighted as described in Sec.~\ref{sec:MassFit}.
The dominant contribution in the signal region comes, for both shapes, from the
final state with a \Dzb, not a \Dstarzb, meson.
\subsection{Efficiency variation across the square Dalitz plot}
\label{sec:DalitzEfficiency}

Variation of the signal efficiency across the SDP is induced by the detector acceptance and by trigger, selection and PID requirements.
The variation of the efficiency is studied using simulated samples of signal events generated uniformly over the SDP, with several data driven corrections. Statistical fluctuations from limited sample size are smoothed out by fitting the efficiency functions to a two-dimensional cubic spline across the SDP.

Corrections are applied for known differences between data and simulation in
the tracking, trigger and PID efficiencies.
A tracking correction is obtained from $\jpsi\to\mumu$ decays for each of the four final state tracks as a function of $\eta$ and $p$.
The total correction is obtained from the product of the factors for each track.

The trigger efficiency correction is different for two mutually exclusive subsamples of the selected candidates.
The first includes candidates that are triggered at hardware level by clusters in the hadronic calorimeter created by one or more of the final state particles, and the second contains those triggered only by particles in the rest of the event.
For the first subsample, a correction is calculated from the probability of an energy deposit in the hadronic calorimeter to fire the trigger, evaluated from calibration data samples as a function of particle type, dipole magnet polarity, transverse energy and position in the calorimeter.
In the second subsample, a smaller correction is applied to account for the requirement that the signal decay products did not fire the hadronic calorimeter hardware trigger.
The efficiency is evaluated for each subsample as a function of SDP position, and these are combined into a single efficiency map according to their proportions in data.

The PID efficiency is evaluated using a calibration sample of $\Dzb\to\Kp\pim$ decays as described in Sec.~\ref{sec:Selection}.
Efficiencies for background-subtracted samples of kaons and pions are obtained as functions of their $p$, $\pt$ and of the number of tracks in the event.
The kinematic properties of the four final state signal particles are obtained from simulation while the distribution of the number of tracks in the event is taken from data.
Efficiencies for each of the final state particles are evaluated and their product gives the efficiency for the candidate accounting for possible correlations between the kinematics of the four tracks.

\begin{figure}[!tb]
 \centering
 \includegraphics[scale=0.38]{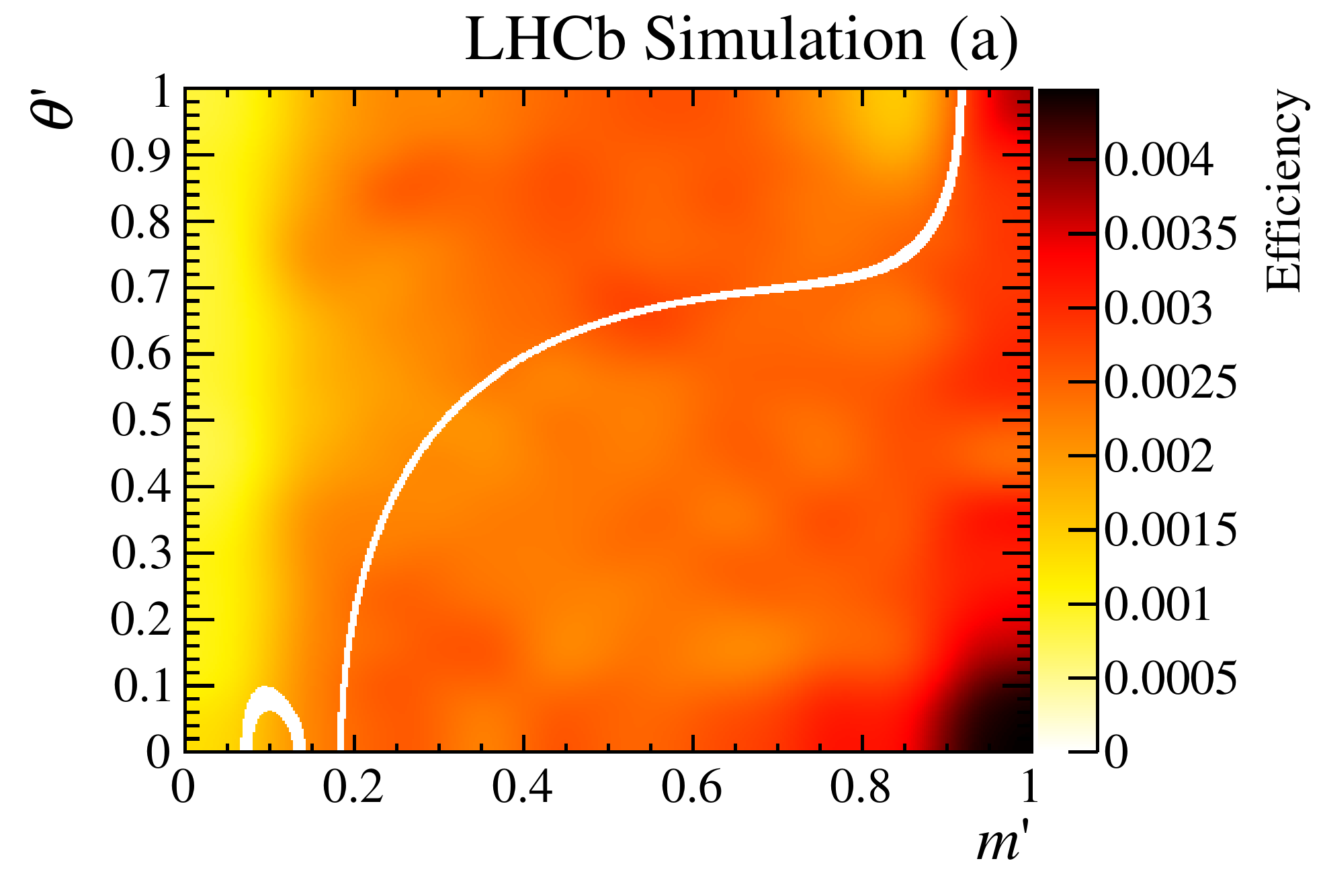}
 \includegraphics[scale=0.38]{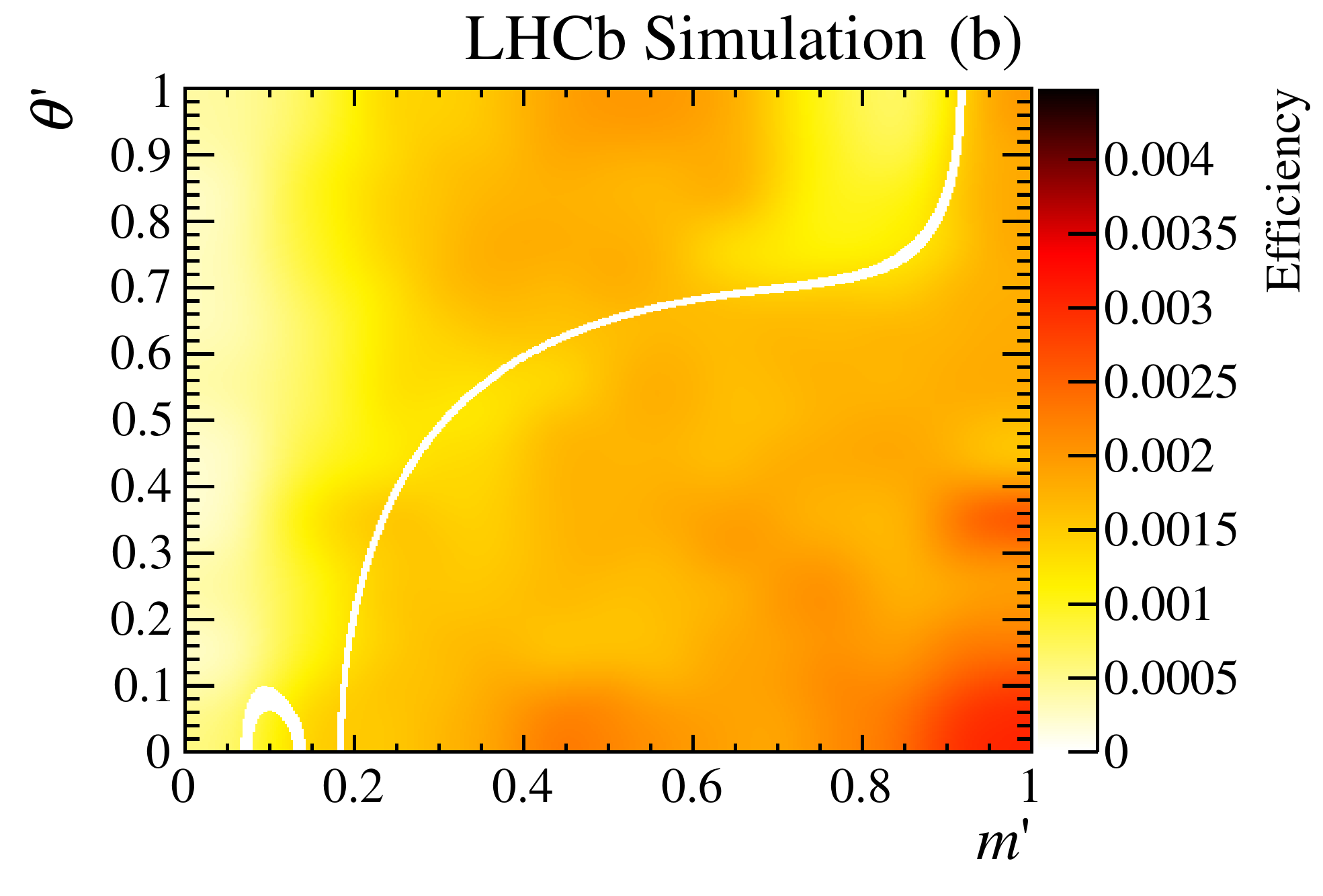}
 \caption{\small
   Signal efficiency across the SDP for (a) events triggered by signal decay products and (b) the rest of the event.
   The relative uncertainty at each point is typically $5\,\%$.
   The effect of the \Dz veto can be seen as a curved band running across the
   SDP, while the \Dstar veto appears in the bottom left corner of the SDP.
}
 \label{fig:eff}
\end{figure}

Contributions from the various sources are then combined into a single efficiency map across the SDP that is used as an input to the Dalitz plot fit and is shown in Fig.~\ref{fig:eff}.
The largest source of variation arises due to the reconstruction, which causes a rapid drop of the efficiency at the smallest values of \mpr, which corresponds to high $m(\Dzb\Km)$ and hence slow $\pip$ tracks.
The largest source of efficiency variation induced by the selection arises due to the PID requirements, which lead to a maximum efficiency variation of about $\pm 20\,\%$ across the SDP.

\subsection{\boldmath Amplitude model for $\Bs \to \Dzb \Km \pip$ decays}
\label{sec:DalitzBaseline}

The Dalitz plot fit is performed using the {\sc Laura++}~\cite{Laura++} package.
The likelihood function that is optimised is given by
\begin{equation}
  {\cal L} =
  \prod_i^{N_c}
  \Bigg[
  \sum_k N_k {\cal P}_k\left(m^2_i(\Dzb\Km),m^2_i(\Km\pip)\right)
  \Bigg] \,,
\end{equation}
where the indices $i$ and $k$ run over the $N_c$ selected candidates and the signal and background categories, respectively.
The signal and background yields $N_k$ are given in Table~\ref{tab:yields}.
The signal probability density function ${\cal P}_{\rm sig}$ is a modified
version of Eq.~(\ref{eq:SigDPLike}), where factors of $|{\cal A}\left(m^2(\Dzb\Km), m^2(\Km\pip)\right)|^2$ in both numerator and in the integral
in the denominator are multiplied by the efficiency function described in
Sec.~\ref{sec:DalitzEfficiency}.
The mass resolution is below $2 \mevcc$, much less than the width of the
narrowest structures on the Dalitz plot, and therefore has negligible effect
on the likelihood.
The background SDP distributions are discussed in Sec.~\ref{sec:DalitzBackground} and shown in Fig.~\ref{fig:bkg}.

The free parameters of the fit are the real and imaginary parts of the complex coefficients, $c_j$ in Eq.~(\ref{eqn:amp}), for each amplitude included in the fit model, except for the $D^{*}_{s2}(2573)^{-}$ component for which the real and imaginary parts of the amplitude are fixed to 1 and 0, respectively, as a reference.
Several parameters of the lineshapes are also determined from the fit, as described below.
Results for the complex amplitudes are also presented in terms of their magnitudes and phases, and in addition the fit fractions and interference fit fractions are determined.
Uncertainties on these derived quantities are determined using large samples of simulated pseudoexperiments to correctly account for correlations between the fit parameters.
This approach allows effects of non-trivial correlations between fit parameters to be appropriately treated.

It is possible for the minimisation procedure to find a local minimum of the negative logarithm of the likelihood (NLL) function.
Therefore to find the true global minimum the fit is repeated many times with randomised initial values of the complex amplitude.

\begin{table}[!tb]
 \centering
 \caption{\small
   Contributions to the fit model. Resonances labelled with subscript $v$ are virtual.
   Parameters and uncertainties are taken from Ref.~\cite{PDG2012} except where indicated otherwise.
   Details of these models are given in Sec.~\ref{sec:DalitzGeneralities}.
}
 \label{tab:dpmodel}
 \begin{tabular}{lcccc}
 \hline
 Resonance & Spin & Dalitz plot axis & Model & Parameters $(\!\mevcc)$ \\
 \hline \\ [-2.5ex]
 $\Kstarb(892)^{0}$ &1& $m^2(\Km\pip)$ & RBW & $m_0 = 895.81 \pm 0.19 $, $\Gamma_0 = 47.4 \pm 0.6 $ \\
 $\Kstarb(1410)^{0}$ &1& $m^2(\Km\pip)$ & RBW & $m_0 = 1414 \pm 15 $, $\Gamma_0 = 232 \pm 21 $ \\
 $\Kstarbsubz(1430)^{0}$ &0& $m^2(\Km\pip)$ & LASS & See text\\
 $\Kstarbsubt(1430)^{0}$ &2& $m^2(\Km\pip)$ & RBW & $m_0 = 1432.4 \pm 1.3 $, $\Gamma_0 = 109 \pm 5 $  \\
 $\Kstarb(1680)^{0}$ &1& $m^2(\Km\pip)$ & RBW & $m_0 = 1717 \pm 27 $, $\Gamma_0 = 322 \pm 110 $ \\
 $\Kstarbsubz(1950)^{0}$ &0& $m^2(\Km\pip)$ & RBW & $m_0 = 1945 \pm 22 $, $\Gamma_0 = 201 \pm 90 $\\
 $D^{*}_{s2}(2573)^{-}$ &2& $m^2(\Dzb\Km)$ & RBW & See text\\
 $D^{*}_{s1}(2700)^{-}$ &1& $m^2(\Dzb\Km)$ & RBW & $m_0 = 2709 \pm 4 $, $\Gamma_0 = 117 \pm 13 $  \\
 $D^{*}_{sJ}(2860)^{-}$ &1& $m^2(\Dzb\Km)$ & RBW & See text\\
 $D^{*}_{sJ}(2860)^{-}$ &3& $m^2(\Dzb\Km)$ & RBW & See text\\
 \hline \\ [-2.5ex]
 Nonresonant & & $m^2(\Dzb\Km)$ & EFF & See text\\
 \hline \\ [-2.5ex]
 $D^{*-}_{s\,v}$ & 1 & $m^2(\Dzb\Km)$ & RBW & $m_0 = 2112.3 \pm 0.5 $, $\Gamma_0 = 1.9$ \\
 $D^{*}_{s0\,v}(2317)^{-}$ & 0 & $m^2(\Dzb\Km)$ & RBW & $m_0 = 2317.8 \pm 0.6 $, $\Gamma_0 = 3.8$ \\
 $B^{*+}_{v}$ & 1 & $m^2(\Dzb\pip)$ & RBW & $m_0 = 5325.2 \pm 0.4 $, $\Gamma_0 = 0$ \\
 \hline
 \end{tabular}
\end{table}

The baseline amplitude model for $\Bs \to \Dzb \Km \pip$ decays is defined by
considering many possible resonant, virtual or nonresonant contributions and removing those that do not significantly affect the fit.
Resonances with unnatural spin-parity, that do not decay to two pseudoscalars,
are not considered.
The resulting signal fit model consists of the contributions shown in Table~\ref{tab:dpmodel}.
There are a total of fourteen components: six $\Km\pip$ resonances, four $\Dzb\Km$ resonances, three virtual resonances and a $\Dzb\Km$ nonresonant contribution.
The majority are modelled with the RBW lineshape, the exceptions being: (i) the $\Km\pip$ S-wave, including the $\Kstarbsubz(1430)^{0}$ resonance, which is modelled by the LASS lineshape with an additional contribution from the $\Kstarbsubz(1950)^{0}$ state; and (ii) the $\Dzb\Km$ nonresonant component, which is modelled with an EFF.

As discussed further in Sec.~\ref{sec:Results}, a highly significant improvement in
the likelihood is obtained when including two resonances, one spin-1 and
another spin-3, both with $m(\Dzb\Km) \approx 2.86 \gevcc$.
Previous studies of the $D_{sJ}^*(2860)^-$
state~\cite{Aubert:2009ah,LHCb-PAPER-2012-016}, have assumed a single
resonance in this region, and therefore values of the mass and width obtained
from those analyses cannot be used in the fit.
Instead, the parameters of these states are obtained from the data.
The sensitivity of the data to the parameters of the $D^{*}_{s2}(2573)^-$
resonance exceeds that of previous measurements~\cite{PDG2012}, and therefore
these parameters are also obtained from the fit.

The slope parameter, $\alpha$, of the EFF model for the $\Dzb\Km$ nonresonant
contribution, and the parameters of the LASS shape
are also determined from the data.
The values that are obtained are $\alpha = 0.412 \pm 0.024~(\!\gevcc)^{-2}$,
$m_0 = 1.552 \pm 0.010\gevcc$,  $\Gamma_0 = 0.195 \pm 0.012 \gevcc$,
$a = 4.9 \pm 0.6 \gevcc$ and $r = 0.0 \pm 0.2 \gevcc$,
where the uncertainties are statistical only.
The LASS model is considered as providing an effective description of the
$\Km\pip$ S-wave, and the parameters should not be compared to other measurements from different processes.
Alternative models for the $\Dzb\Km$ and $\Km\pip$ S-waves are used to evaluate associated systematic uncertainties, as discussed in Sec.~\ref{sec:Systematics}.

The results of the fit to the baseline Dalitz plot model are shown in
Table~\ref{tab:cfitfrac} for the fit fractions and complex coefficients,
and in Table~\ref{tab:mG-stat} for the masses and widths.
Results for the interference fit fractions are presented in App.~\ref{app:IFF-results}.
In Table~\ref{tab:cfitfrac}, and for all results for fit fractions, values are given both for the nonresonant and $\Kstarbsubz(1430)^{0}$ parts of the LASS function separately and for the two combined taking into account their interference.
The interference effects between the components of the $\Km\pip$ S-wave explain most of the excess of the total fit fraction from unity.
Other local minima of the NLL function are found to be separated from the global minimum by at least 10 units.

\begin{figure}[!tb]
\centering
 \includegraphics[scale=0.50]{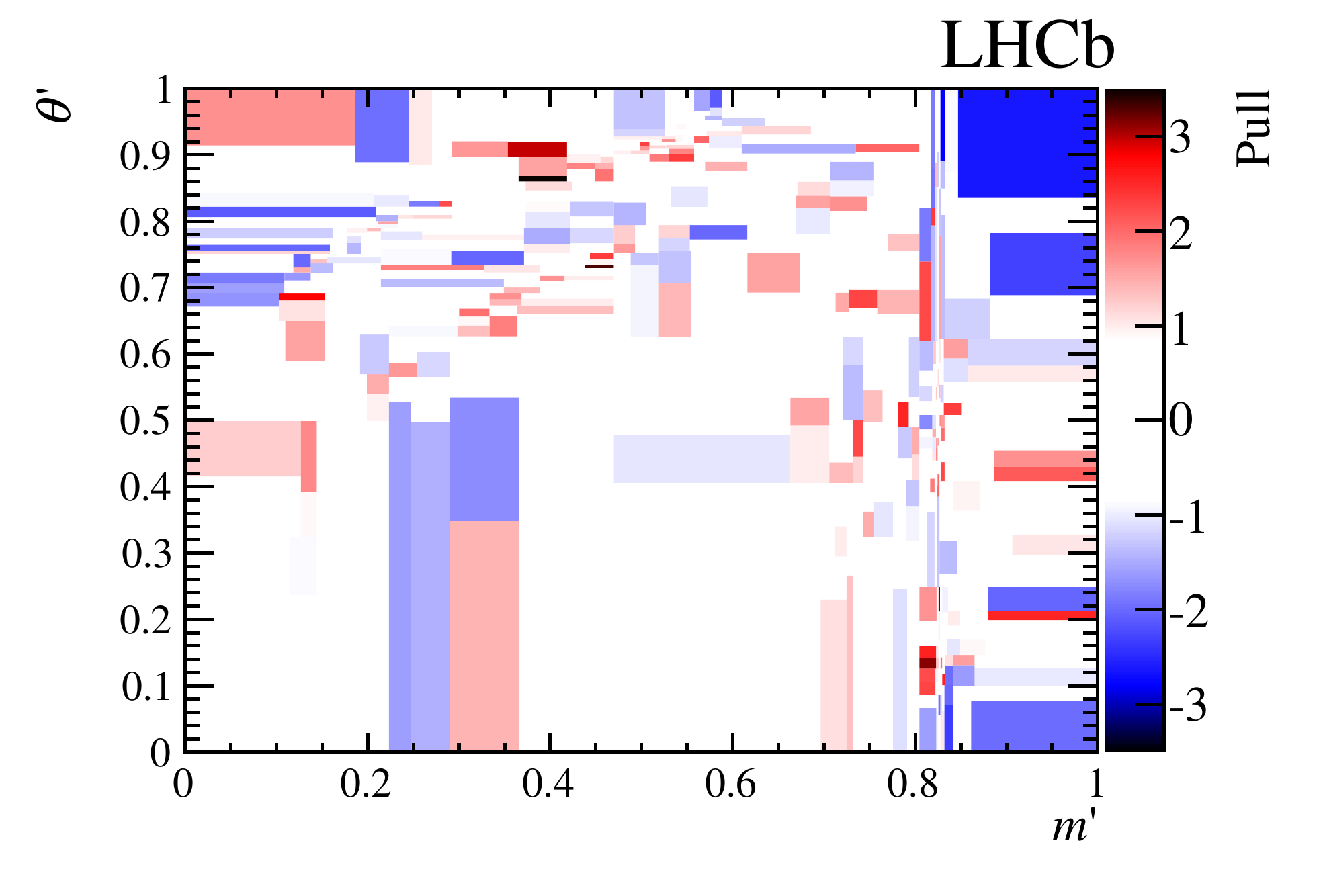}
\caption{\small
  Distribution of the pull between data and the fit result as a function of
  SDP position.
}
\label{fig:pulls}
\end{figure}

The fit quality is evaluated by determining a $\chisq$ value by comparing the
data and the fit model in $N_{\rm bins} = 576$ SDP bins that are defined
adaptively to ensure approximately equal population  with a minimum bin
content of 21 entries.
The effective number of degrees of freedom of the $\chisq$ is bounded by
$N_{\rm bins} - N_{\rm pars} - 1$ and $N_{\rm bins} - 1$, where $N_{\rm pars}$
is the number of parameters determined by the data.
The former choice gives a higher reduced $\chisq$ value of 1.21, where only
statistical uncertainties are included in the calculation.
The effects of systematic uncertainties on the $\chisq$ value are discussed at the end of Sec.~\ref{sec:Systematics}.
The distribution across the SDP of the pull, defined as the difference between the data and the fit model divided by the uncertainty, is shown in Fig.~\ref{fig:pulls}.
Other unbinned tests~\cite{Williams:2010vh} of the fit quality also show that
the fit provides a good, but not perfect, model of the data.

\begin{table}[!tb]
\centering
\caption{\small
  Fit fractions and complex coefficients determined from the Dalitz plot fit.
  Uncertainties are statistical only and are obtained as described in the text.
}
\label{tab:cfitfrac}
\resizebox{\textwidth}{!}{
\begin{tabular}{lccccc}
\hline
Resonance & Fit fraction (\%) & Real part & Imaginary part & Magnitude & Phase (radians)
\\
\hline \\ [-2.5ex]
$\Kstarb(892)^{0}$         & $          28.6 \pm 0.6$ & $          -0.75 \pm 0.08$ & $\phantom{-}0.74 \pm 0.08$ & $1.06 \pm 0.02$ & $\phantom{-}2.36 \pm 0.13$\\
$\Kstarb(1410)^{0}$        & $\phantom{1}1.7 \pm 0.5$ & $          -0.25 \pm 0.03$ & $          -0.04 \pm 0.05$ & $0.25 \pm 0.04$ & $          -2.96 \pm 0.21$\\
LASS nonresonant           & $          13.7 \pm 2.5$ & $          -0.43 \pm 0.09$ & $\phantom{-}0.59 \pm 0.06$ & $0.73 \pm 0.06$ & $\phantom{-}2.19 \pm 0.16$\\
$\Kstarbsubz(1430)^{0}$    & $          20.0 \pm 1.6$ & $          -0.49 \pm 0.10$ & $\phantom{-}0.73 \pm 0.07$ & $0.88 \pm 0.04$ & $\phantom{-}2.16 \pm 0.20$\\
\ \ \ LASS total          & $          21.4 \pm 1.4$ &                            &                            &                 &                           \\
$\Kstarbsubt(1430)^{0}$    & $\phantom{1}3.7 \pm 0.6$ & $\phantom{-}0.09 \pm 0.05$ & $          -0.37 \pm 0.03$ & $0.38 \pm 0.03$ & $          -1.34 \pm 0.10$\\
$\Kstarb(1680)^{0}$        & $\phantom{1}0.5 \pm 0.4$ & $          -0.08 \pm 0.04$ & $\phantom{-}0.12 \pm 0.04$ & $0.14 \pm 0.06$ & $\phantom{-}2.16 \pm 0.26$\\
$\Kstarbsubz(1950)^{0}$    & $\phantom{1}0.3 \pm 0.2$ & $\phantom{-}0.11 \pm 0.03$ & $          -0.01 \pm 0.04$ & $0.11 \pm 0.04$ & $          -0.09 \pm 0.41$\\
$D^{*}_{s2}(2573)^-$    & $          25.7 \pm 0.7$ & $\phantom{-}1.00$ & $\phantom{-}0.00$ & $1.00$ & $\phantom{-}0.00$\\
$D^{*}_{s1}(2700)^-$    & $\phantom{1}1.6 \pm 0.4$ & $          -0.22 \pm 0.04$ & $          -0.13 \pm 0.04$ & $0.25 \pm 0.04$ & $          -2.61 \pm 0.17$\\
$D^{*}_{s1}(2860)^-$    & $\phantom{1}5.0 \pm 1.2$ & $          -0.41 \pm 0.05$ & $\phantom{-}0.16 \pm 0.06$ & $0.44 \pm 0.05$ & $\phantom{-}2.78 \pm 0.20$\\
$D^{*}_{s3}(2860)^-$    & $\phantom{1}2.2 \pm 0.1$ & $\phantom{-}0.27 \pm 0.02$ & $          -0.12 \pm 0.03$ & $0.29 \pm 0.02$ & $          -0.42 \pm 0.07$\\
\hline
Nonresonant               & $          12.4 \pm 2.7$ & $\phantom{-}0.58 \pm 0.07$ & $          -0.39 \pm 0.06$ & $0.70 \pm 0.08$ & $          -0.59 \pm 0.10$\\
\hline
$D^{*-}_{s\,v}$           & $\phantom{1}4.7 \pm 1.4$ & $\phantom{-}0.36 \pm 0.04$ & $\phantom{-}0.23 \pm 0.05$ & $0.43 \pm 0.05$ & $\phantom{-}0.57 \pm 0.12$\\
$D^{*}_{s0\,v}(2317)^-$ & $\phantom{1}2.3 \pm 1.1$ & $\phantom{-}0.18 \pm 0.08$ & $\phantom{-}0.24 \pm 0.04$ & $0.30 \pm 0.06$ & $\phantom{-}0.91 \pm 0.21$\\
$B^{*+}_{v}$              & $\phantom{1}1.9 \pm 1.2$ & $          -0.09 \pm 0.10$ & $          -0.26 \pm 0.05$ & $0.27 \pm 0.09$ & $          -1.90 \pm 0.40$\\
\hline
Total fit fraction        &124.3\\
\hline
\end{tabular}
}
\end{table}

\begin{table}[!tb]
\centering
\caption{\small
  Resonance parameters of the $D^{*}_{s2}(2573)^-$, $D^{*}_{s1}(2860)^-$
  and $D^{*}_{s3}(2860)^-$ states from the Dalitz plot fit (statistical uncertainties only).
}
\label{tab:mG-stat}
\begin{tabular}{lcc}
\hline
Resonance & Mass $(\!\mevcc)$ & Width $(\!\mevcc)$ \\
\hline
$D^{*}_{s2}(2573)^-$
& $2568.39\pm0.29$ & $16.9\pm0.5$ \\
$D^{*}_{s1}(2860)^-$
& $\phantom{.00}2859\pm12\phantom{.0}$ & $159\pm23$ \\
$D^{*}_{s3}(2860)^-$
& $\phani2860.5\pm2.6\phani$ & $53\pm7$ \\
\hline
\end{tabular}
\end{table}

\begin{figure}[!tb]
\centering
 \includegraphics[scale=0.35]{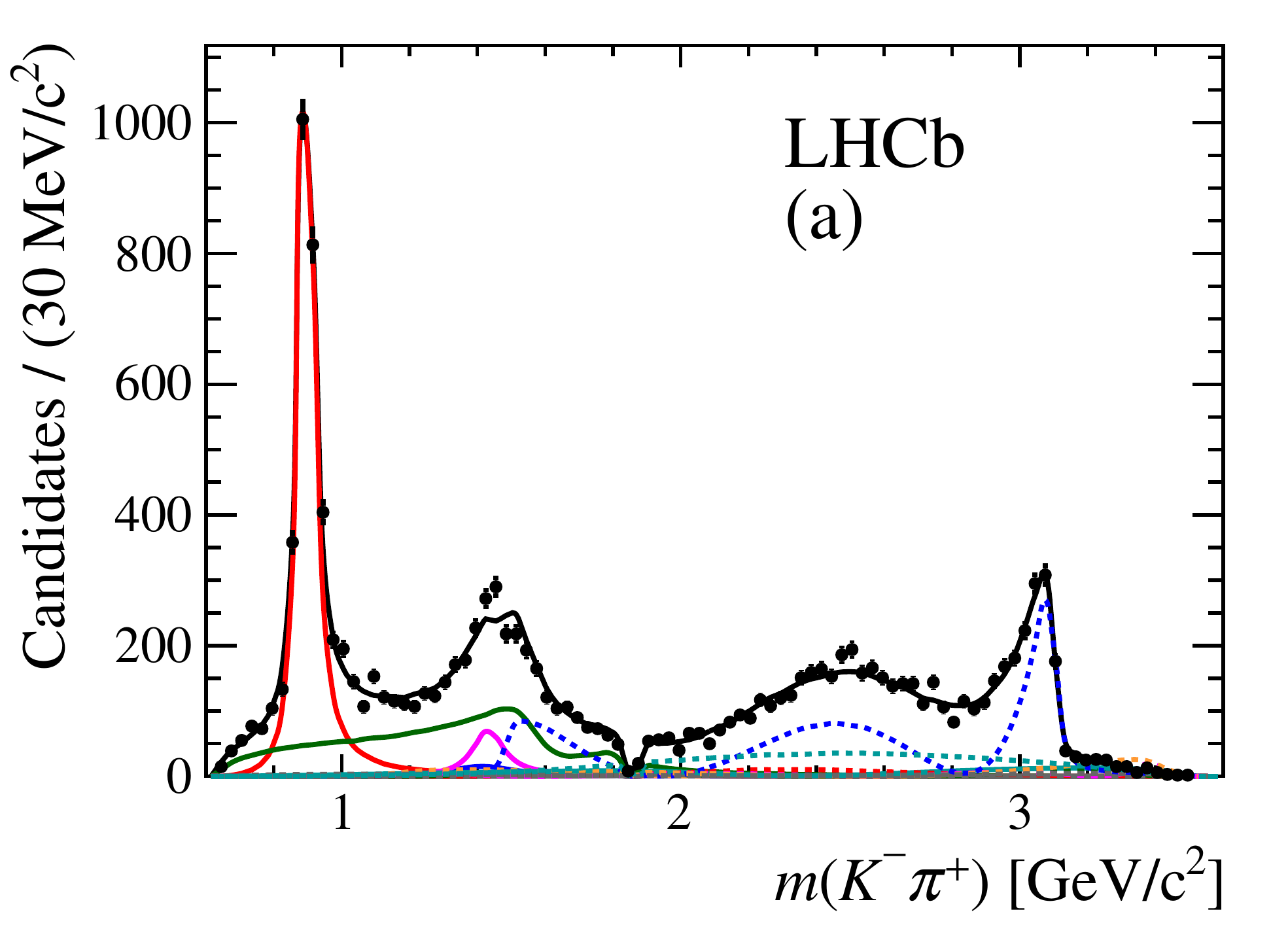}
 \includegraphics[scale=0.35]{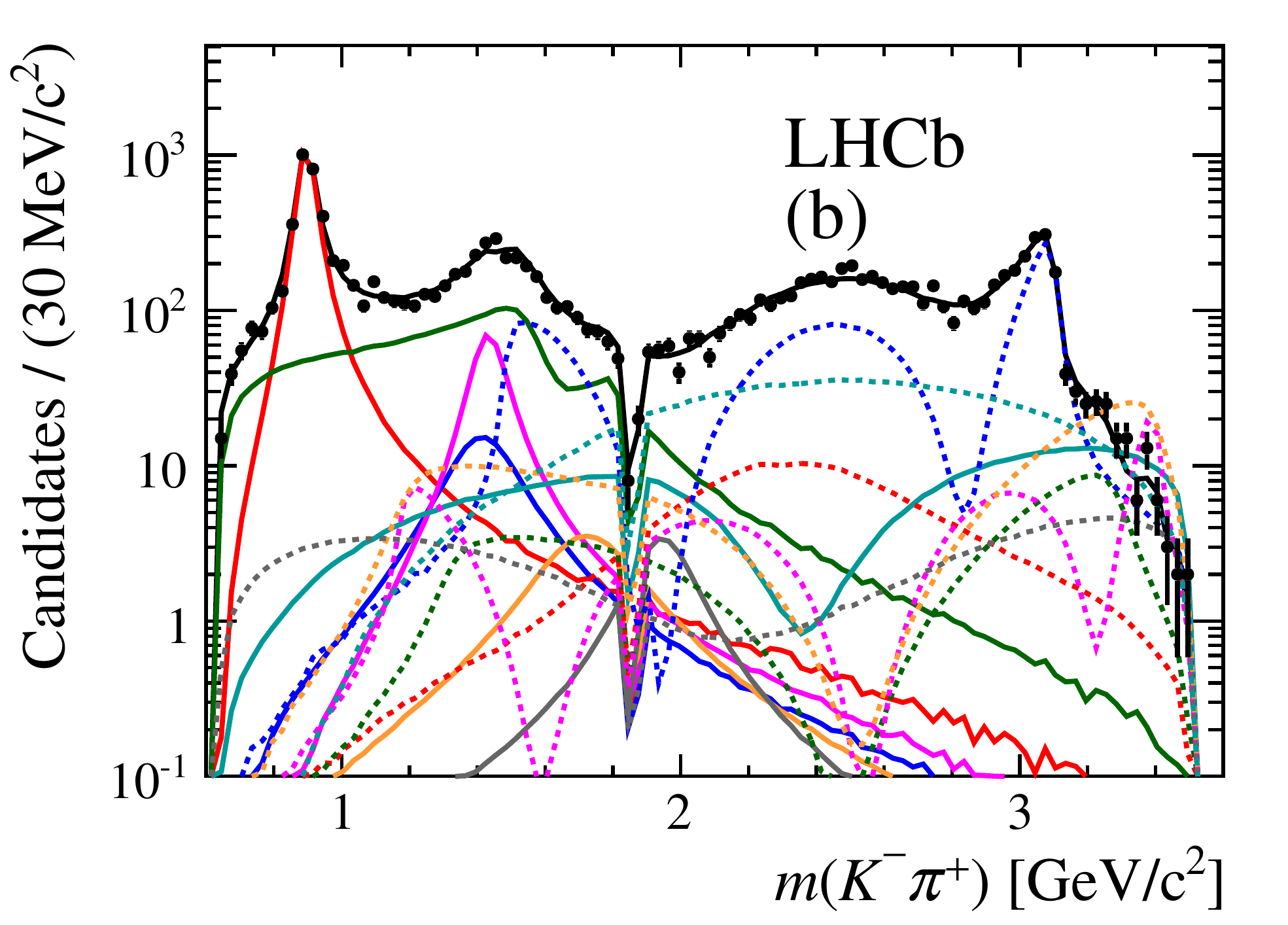}
 \includegraphics[scale=0.35]{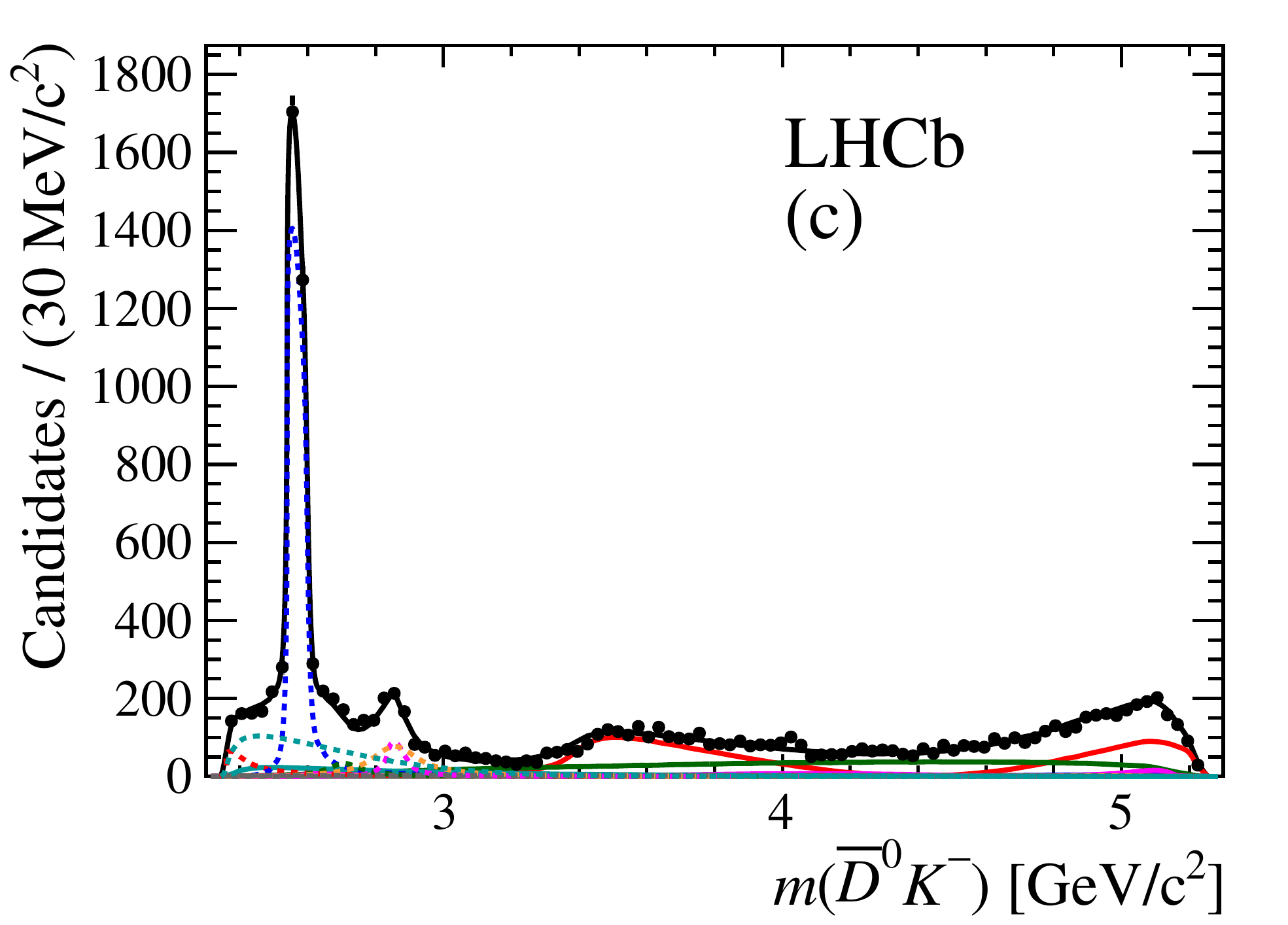}
 \includegraphics[scale=0.35]{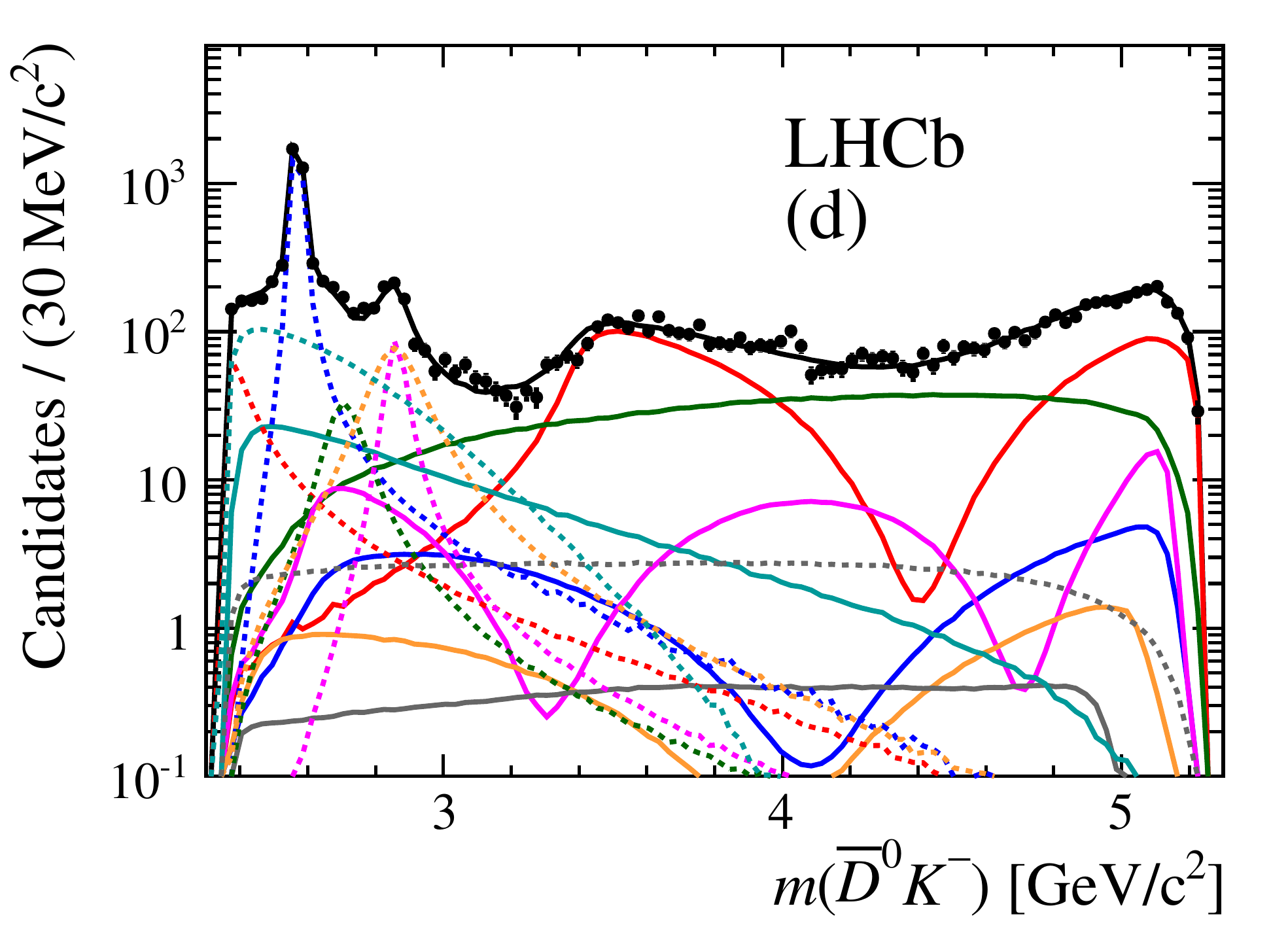}
 \includegraphics[scale=0.35]{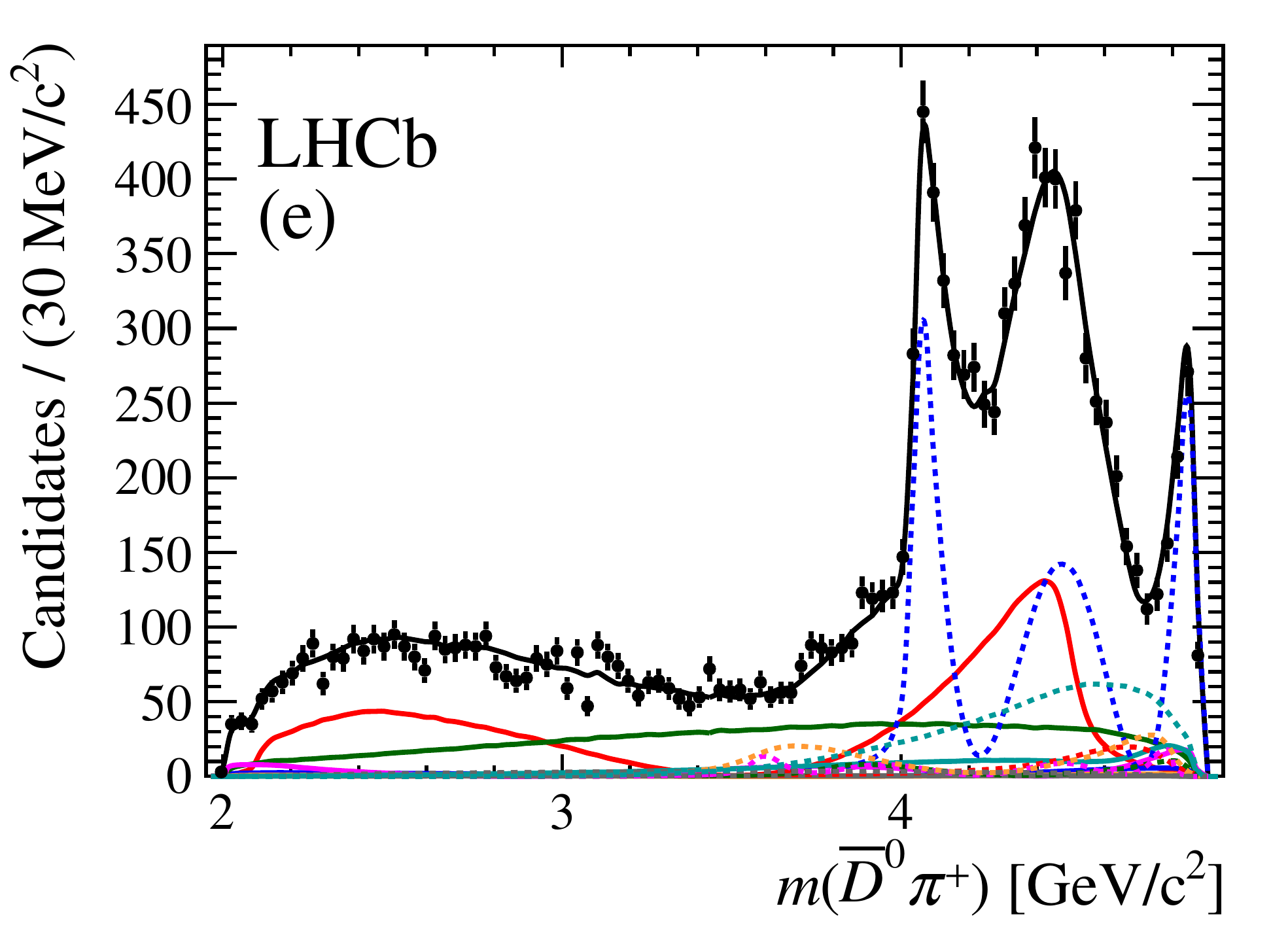}
 \includegraphics[scale=0.35]{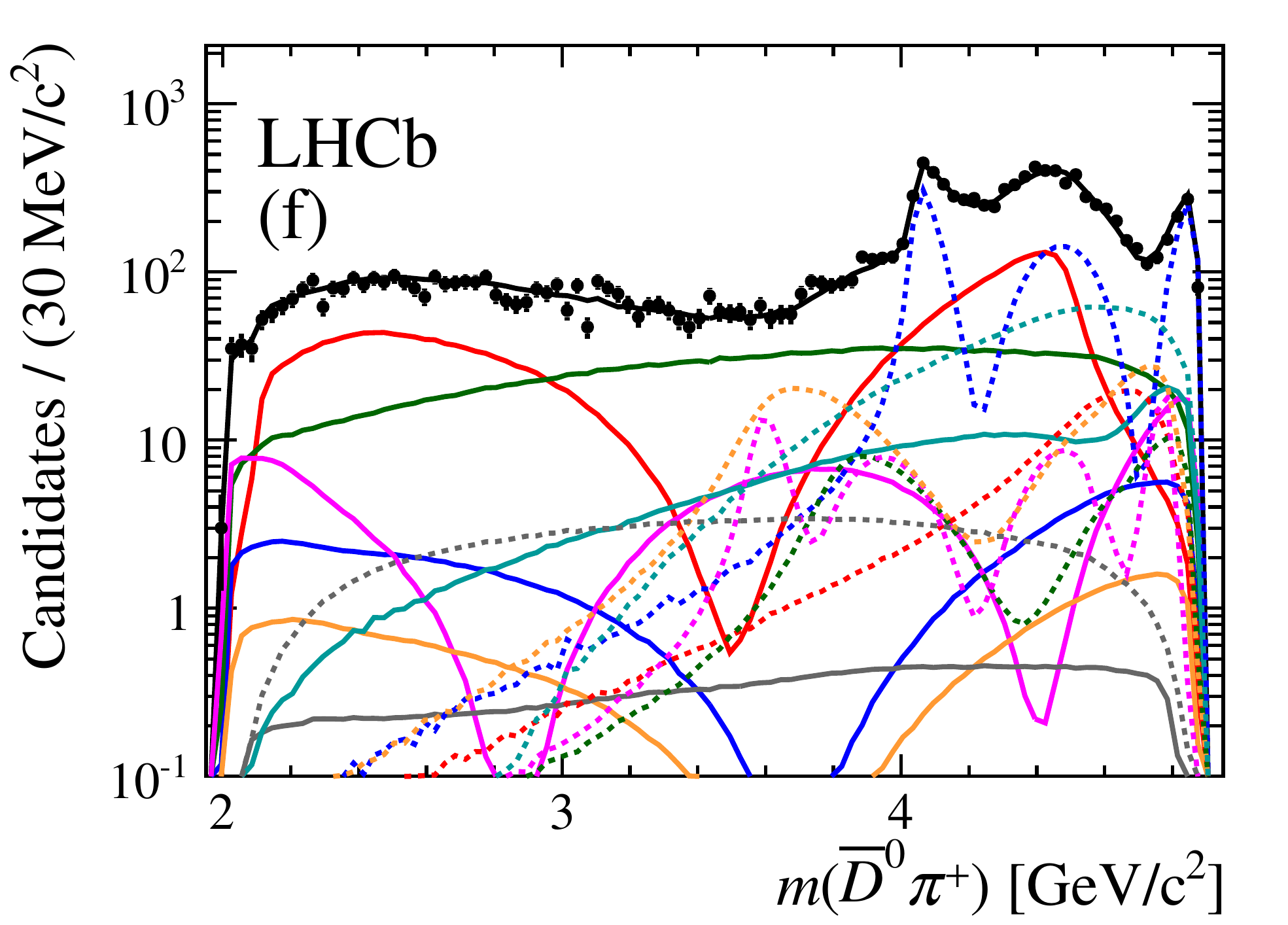}
 \includegraphics[scale=0.50]{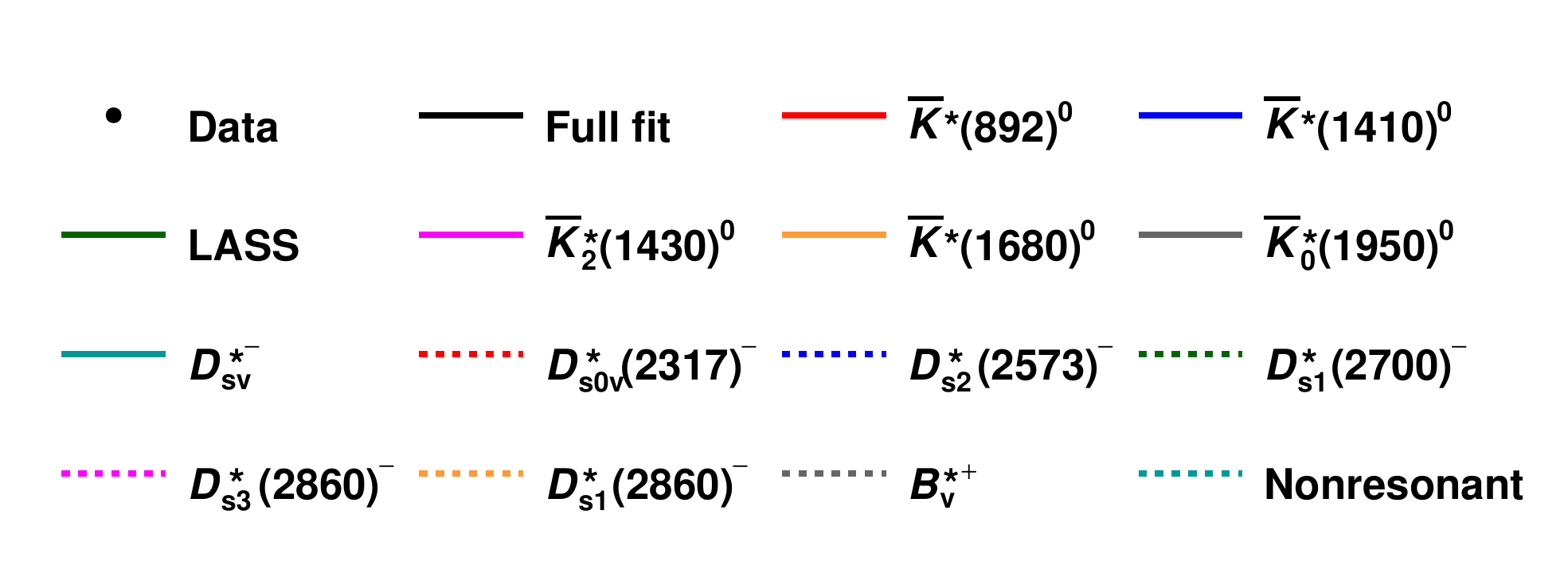}
\caption{\small
  Projections of the data and the Dalitz plot fit result onto
  (a) $m(\Km\pip)$, (c) $m(\Dzb\Km)$ and (e) $m(\Dzb\pip)$, with the same projections
  shown with a logarithmic $y$-axis scale in (b), (d) and (f), respectively.
  The components are as described in the legend (small background components are not shown).
}
\label{fig:dpproj}
\end{figure}

\begin{figure}[!tb]
\centering
 \includegraphics[scale=0.38]{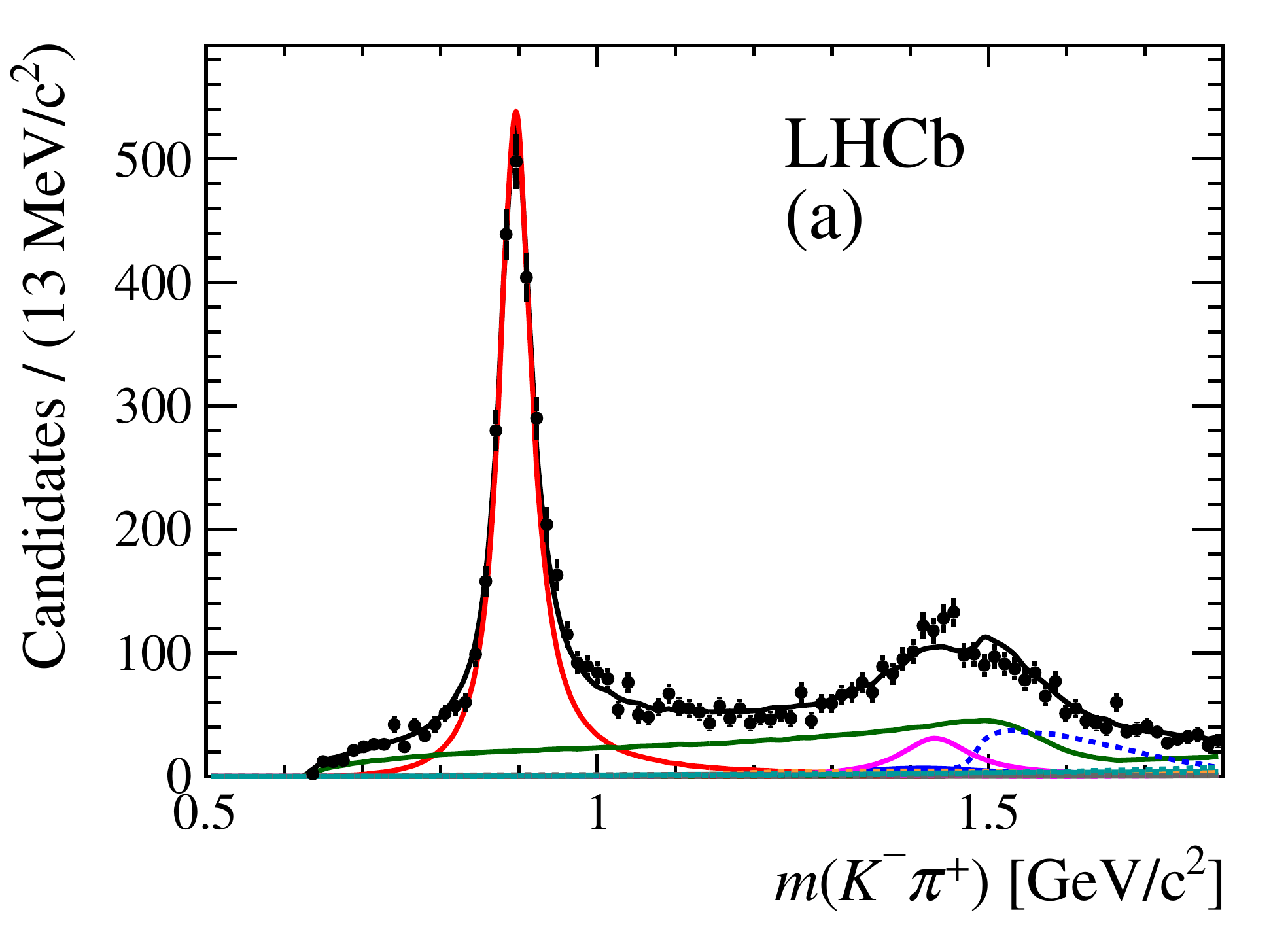}
 \includegraphics[scale=0.38]{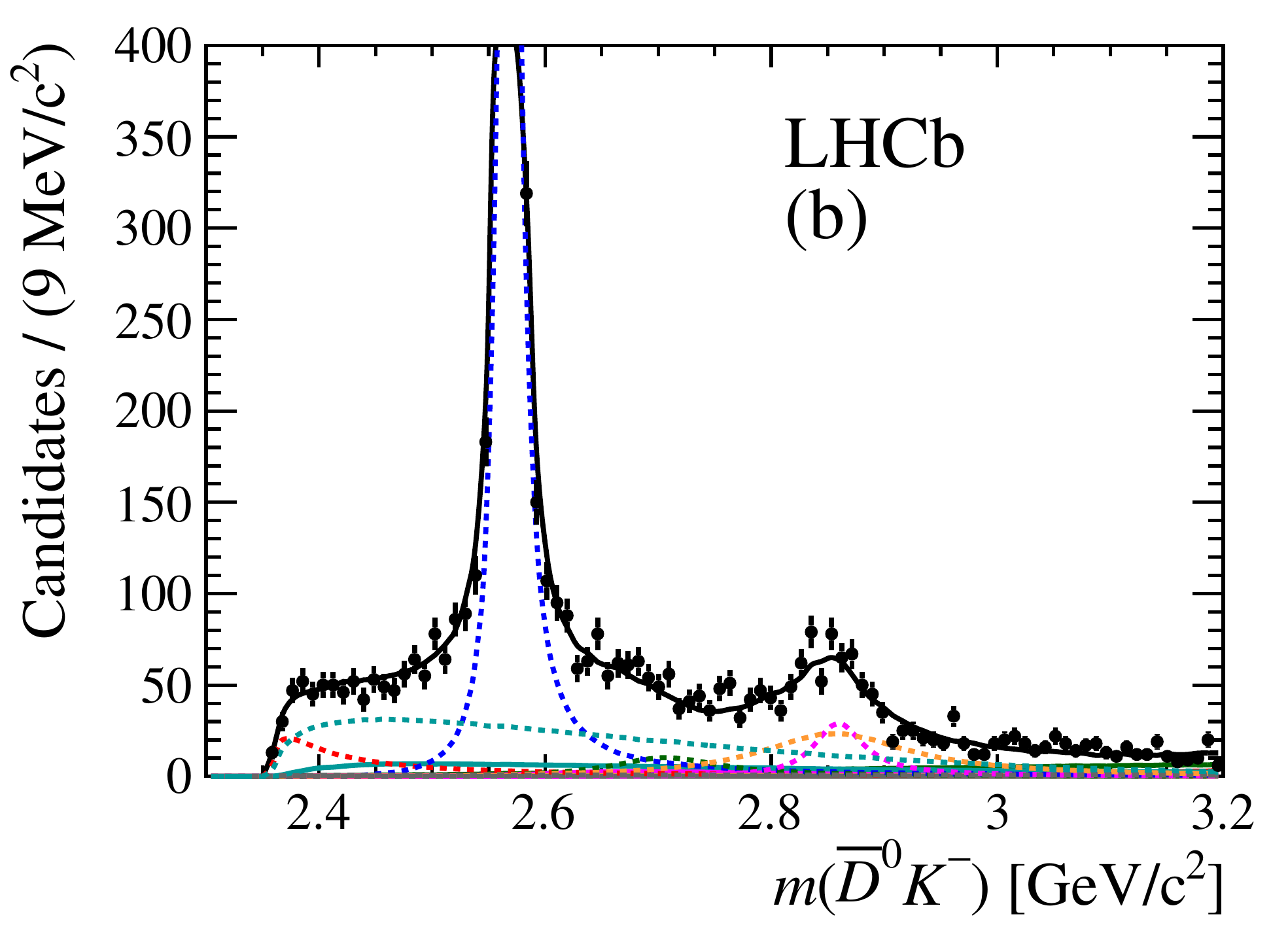}
 \includegraphics[scale=0.38]{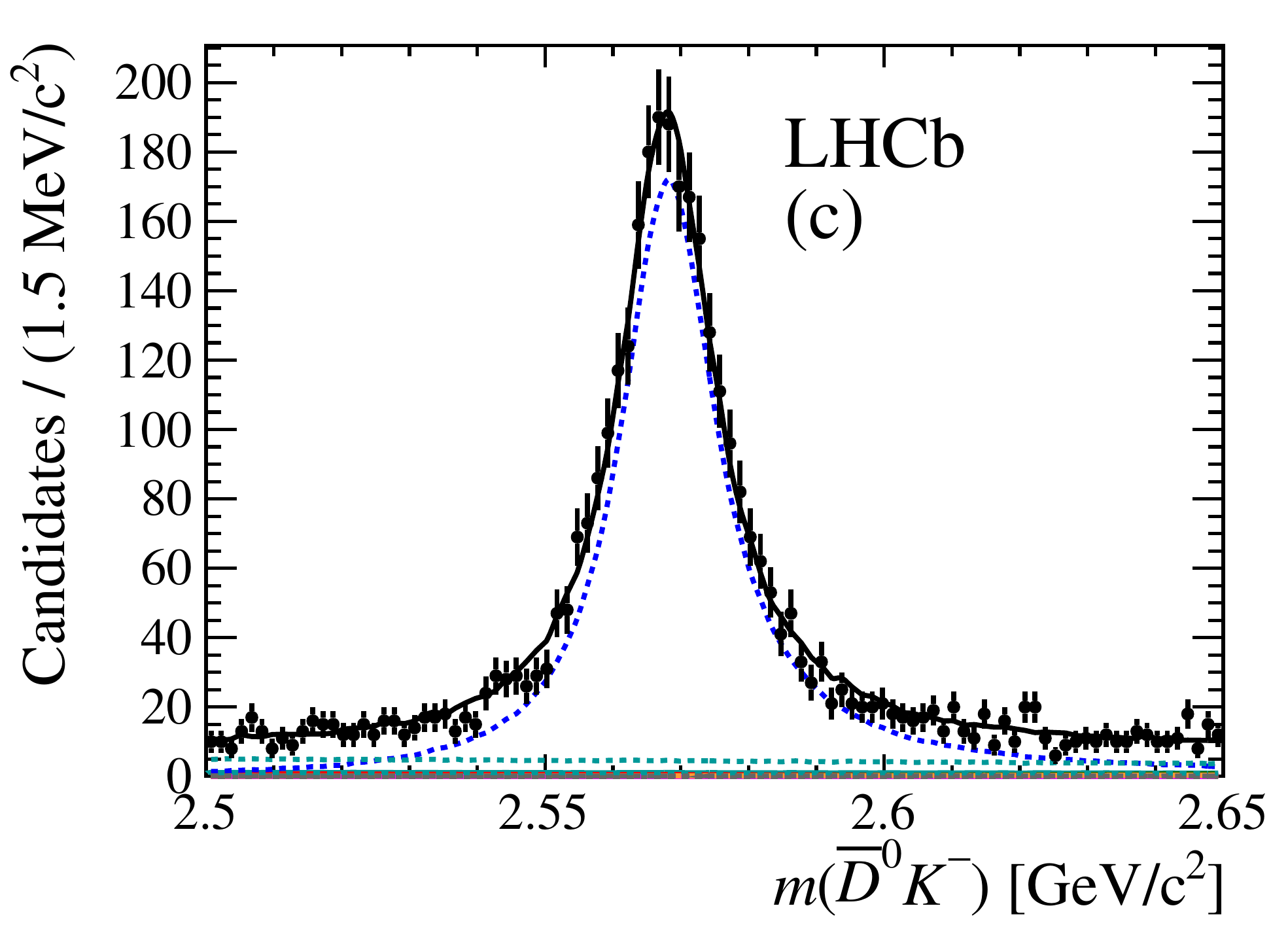}
 \includegraphics[scale=0.38]{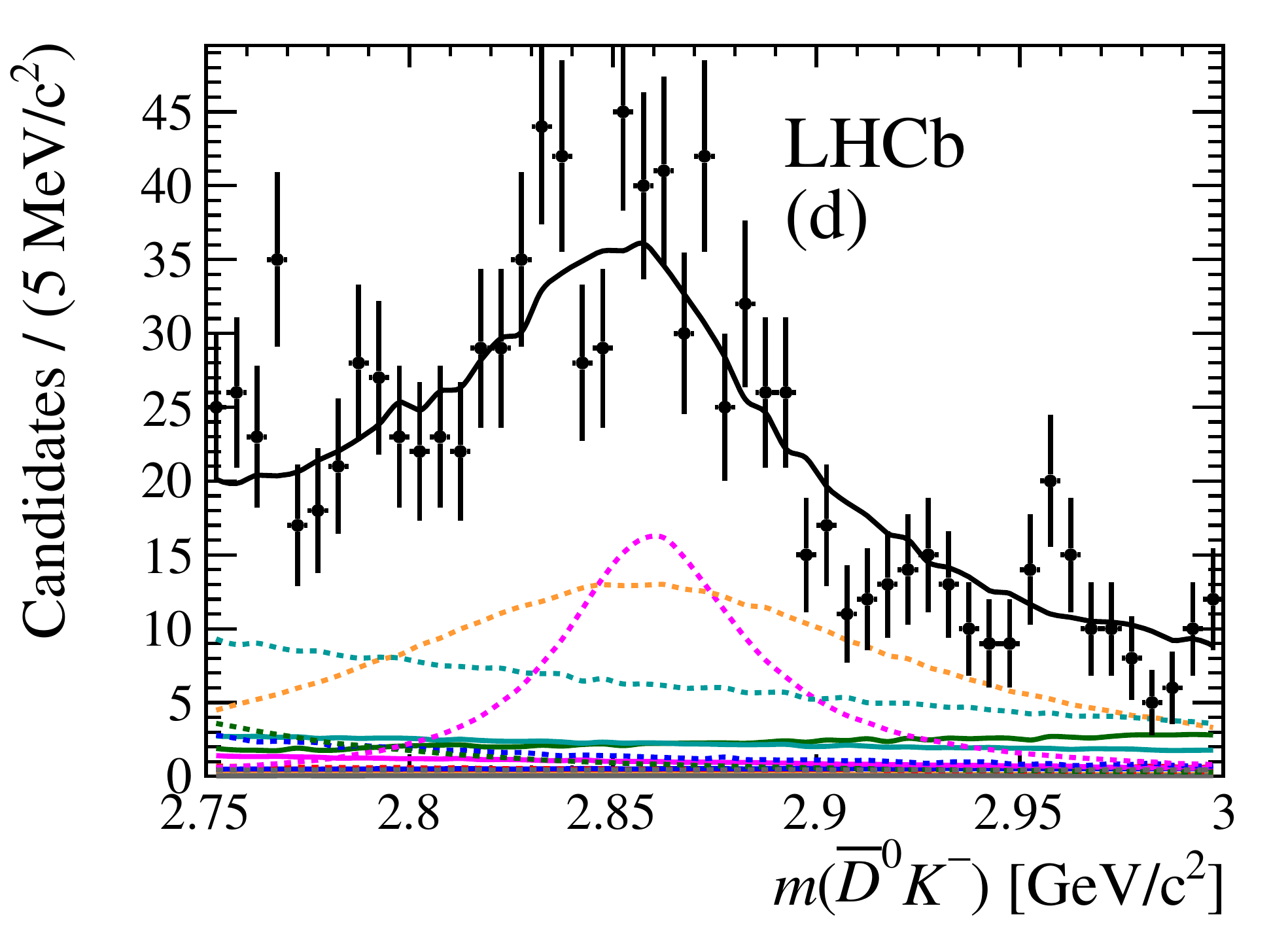}
\caption{\small
  Projections of the data and the Dalitz plot fit result onto
  (a) $m(\Km\pip)$ in the range $0.5$--$1.8\gevcc$, (b) $m(\Dzb\Km)$ between
  $2.2\gevcc$ and $3.2\gevcc$, (c) $m(\Dzb\Km)$ around the $D^{*}_{s2}(2573)^-$
  resonance and (d) the $D^{*}_{sJ}(2860)^-$ region.
  Discrepancies between the data and the model are discussed at the end of Sec.~\ref{sec:Systematics}.
  The components are as described in the legend for Fig~\ref{fig:dpproj}.
}
\label{fig:zooms}
\end{figure}

Projections of the data and the baseline fit result onto $m(\Km\pip)$, $m(\Dzb\Km)$ and $m(\Dzb\pip)$ are shown in Fig.~\ref{fig:dpproj}.
The dip visible in $m(\Km\pip)$ is due to the $\Dz$ veto described in Sec.~\ref{sec:Selection}.
Zooms around the main resonant contributions are shown in Fig.~\ref{fig:zooms}.
Good, but not perfect, agreement between the data and the fit is seen.

\begin{figure}
\centering
 \includegraphics[scale=0.38]{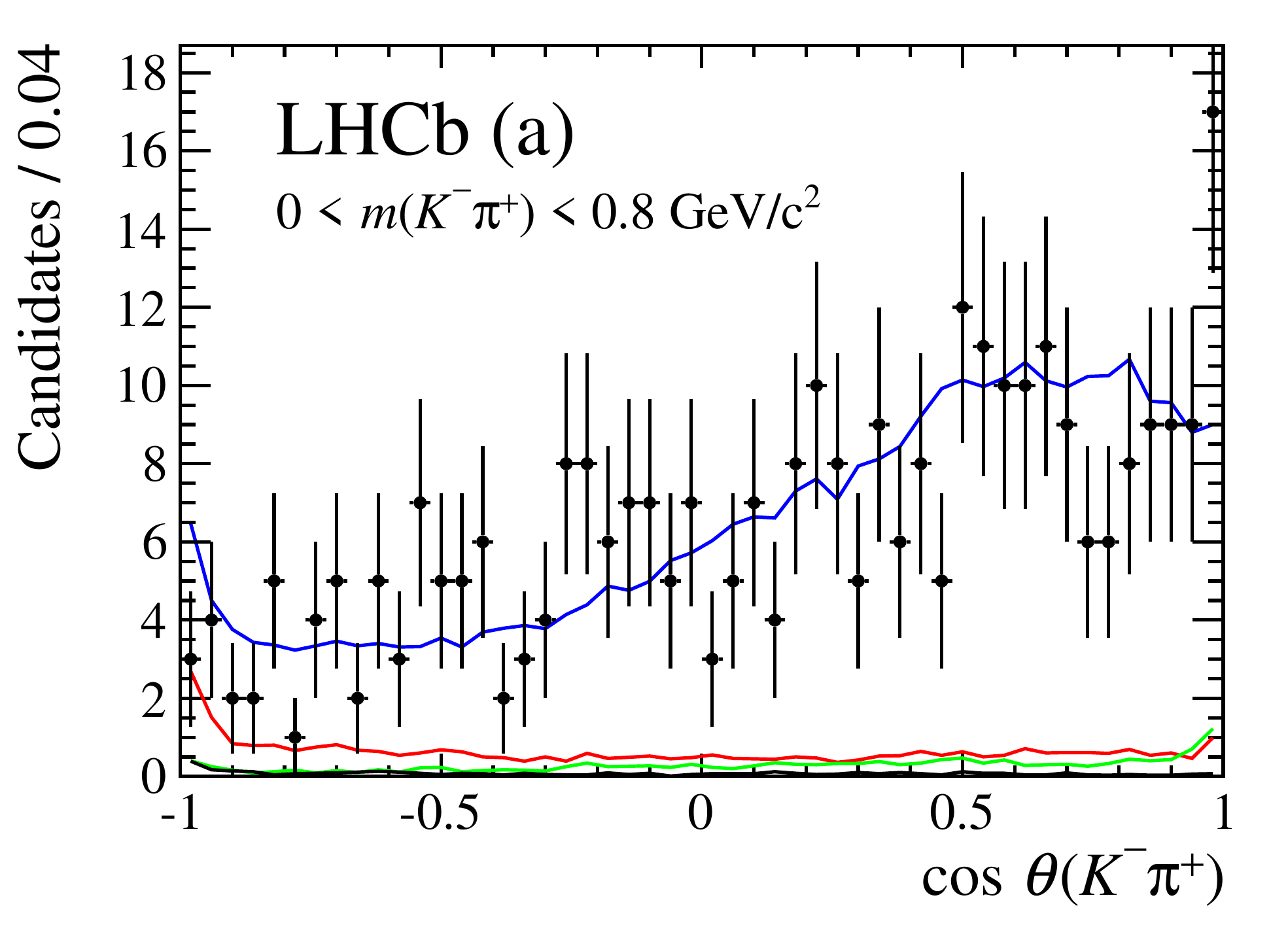}
 \includegraphics[scale=0.38]{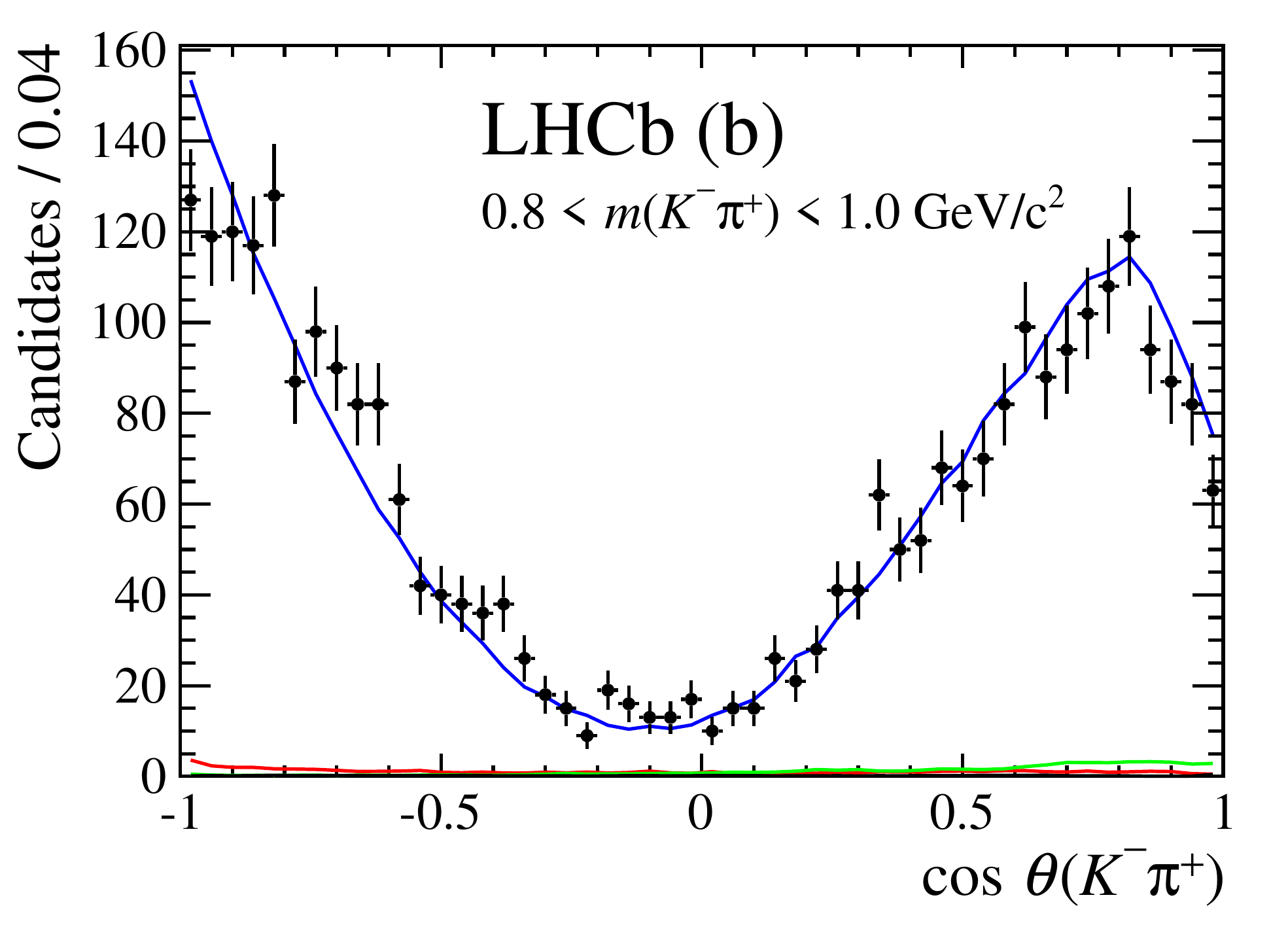}
 \includegraphics[scale=0.38]{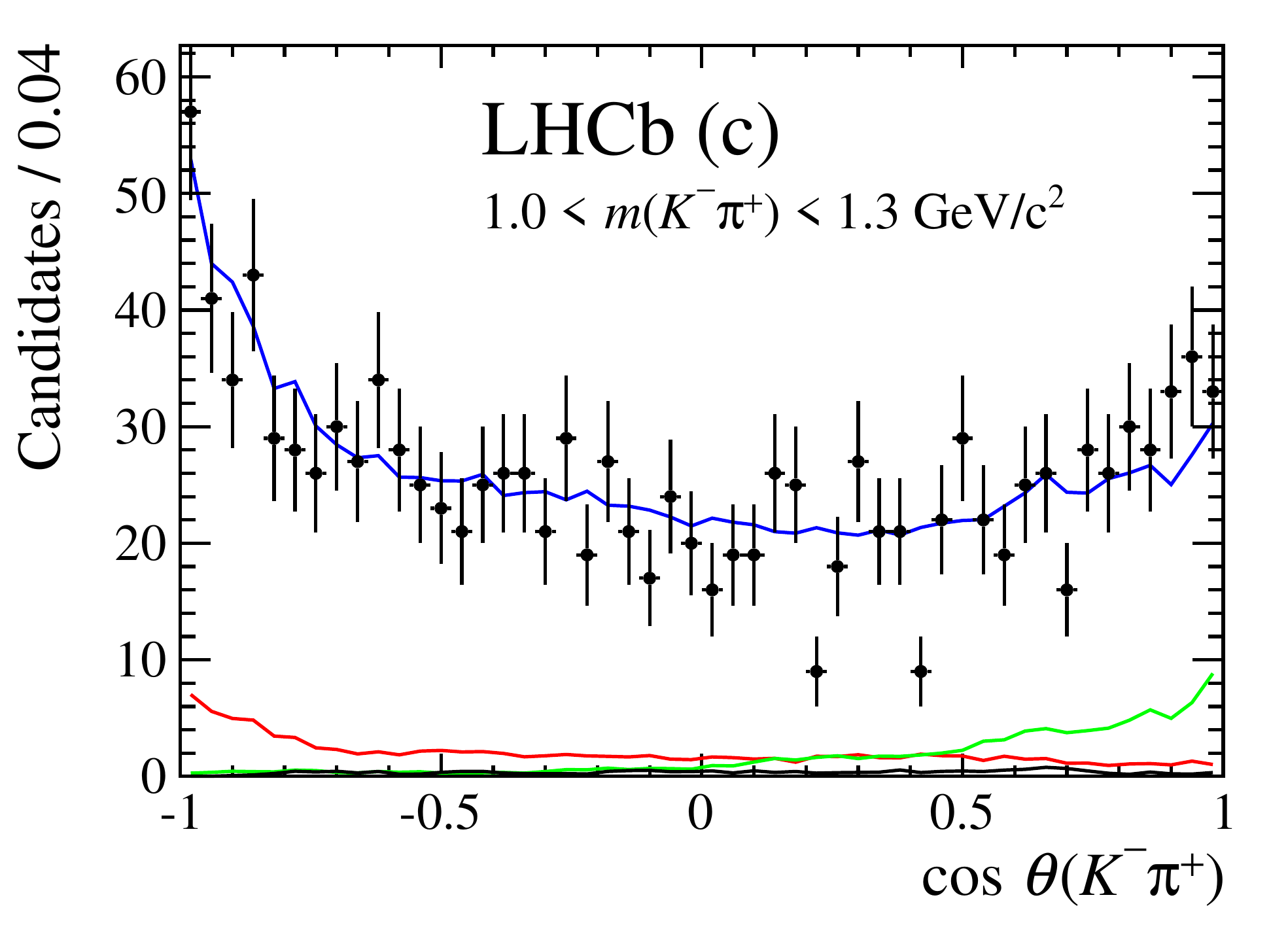}
 \includegraphics[scale=0.38]{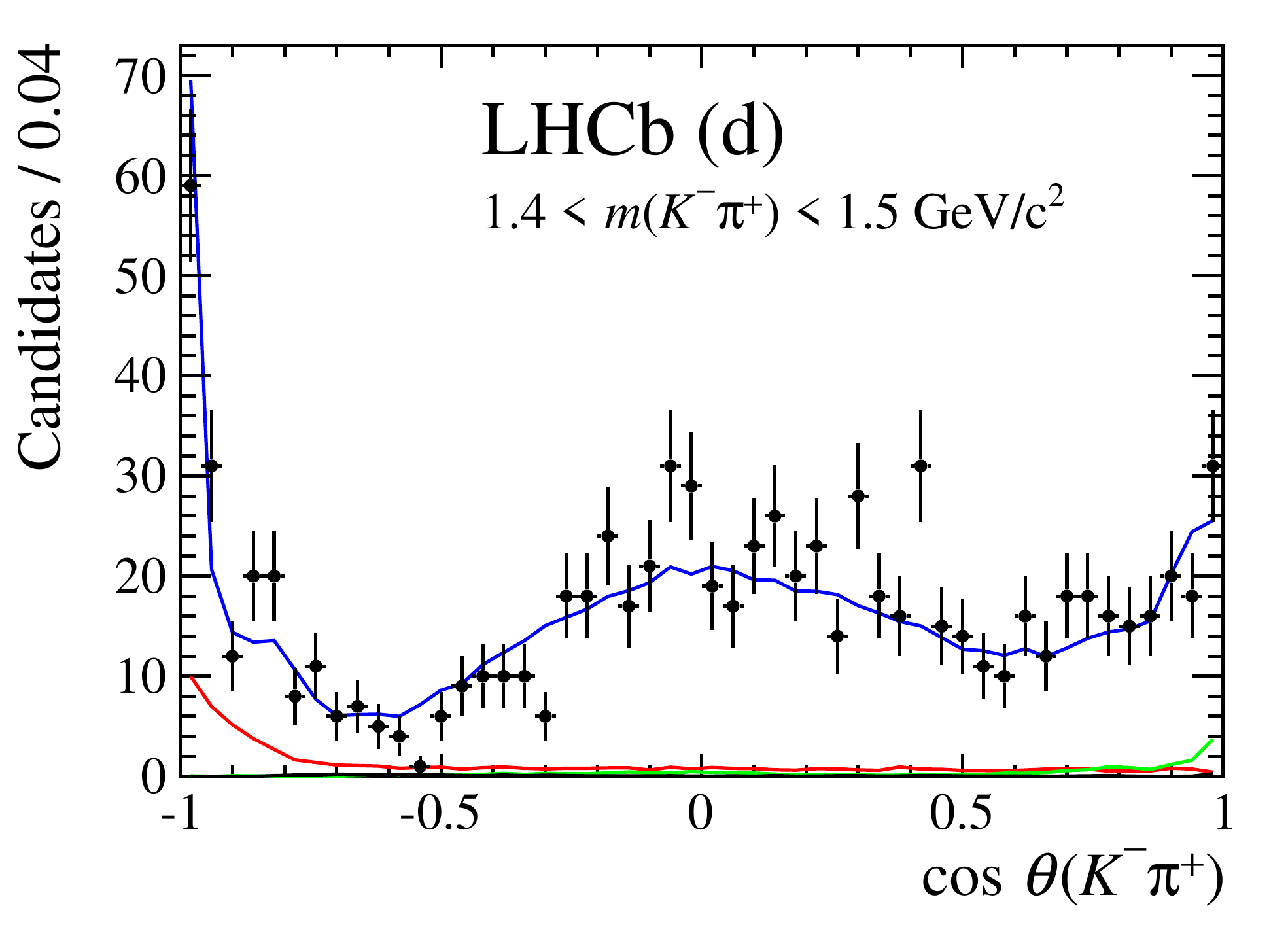}
\caption{\small
  Projections of the data and the Dalitz plot fit result onto the cosine of
  the helicity angle of the $\Km\pip$ system, $\cos\theta(\Km\pip)$, for
  $m(\Km\pip)$ slices of (a) $0$--$0.8\gevcc$, (b) $0.8$--$1.0\gevcc$, (c)
  $1.0$--$1.3\gevcc$ and (d) $1.4$--$1.5\gevcc$.
  The data are shown as black points, the total fit result as a solid blue
  curve, and the small contributions from
  $\Bd\to \DorDstarzb \pip\pim$, $\Lbbar\to \DorDstarzb \antiproton\pip$ and combinatorial
  background shown as green, black and red curves, respectively.
}
\label{fig:coshelkpi}
\end{figure}

\begin{figure}
\centering
 \includegraphics[scale=0.38]{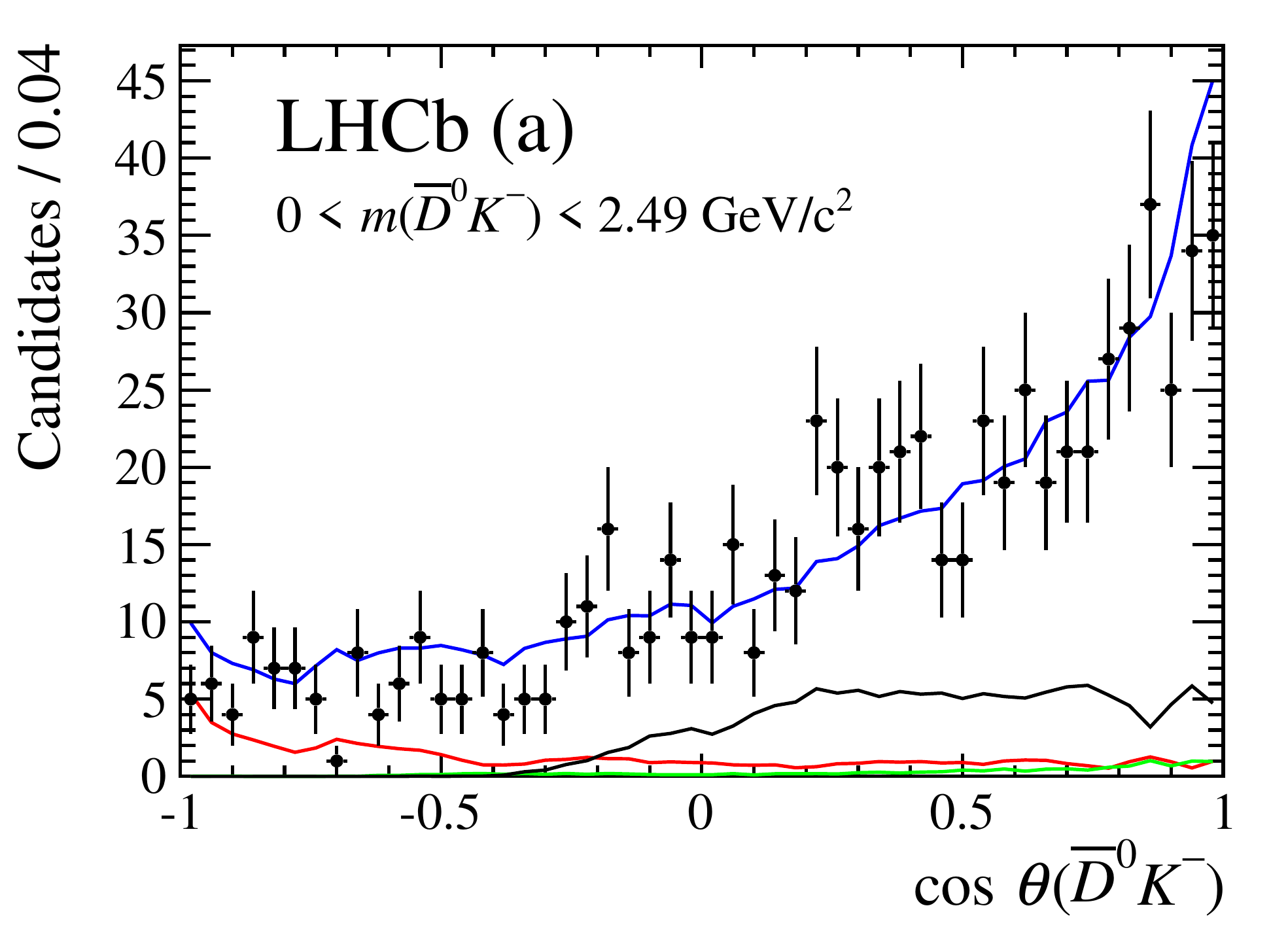}
 \includegraphics[scale=0.38]{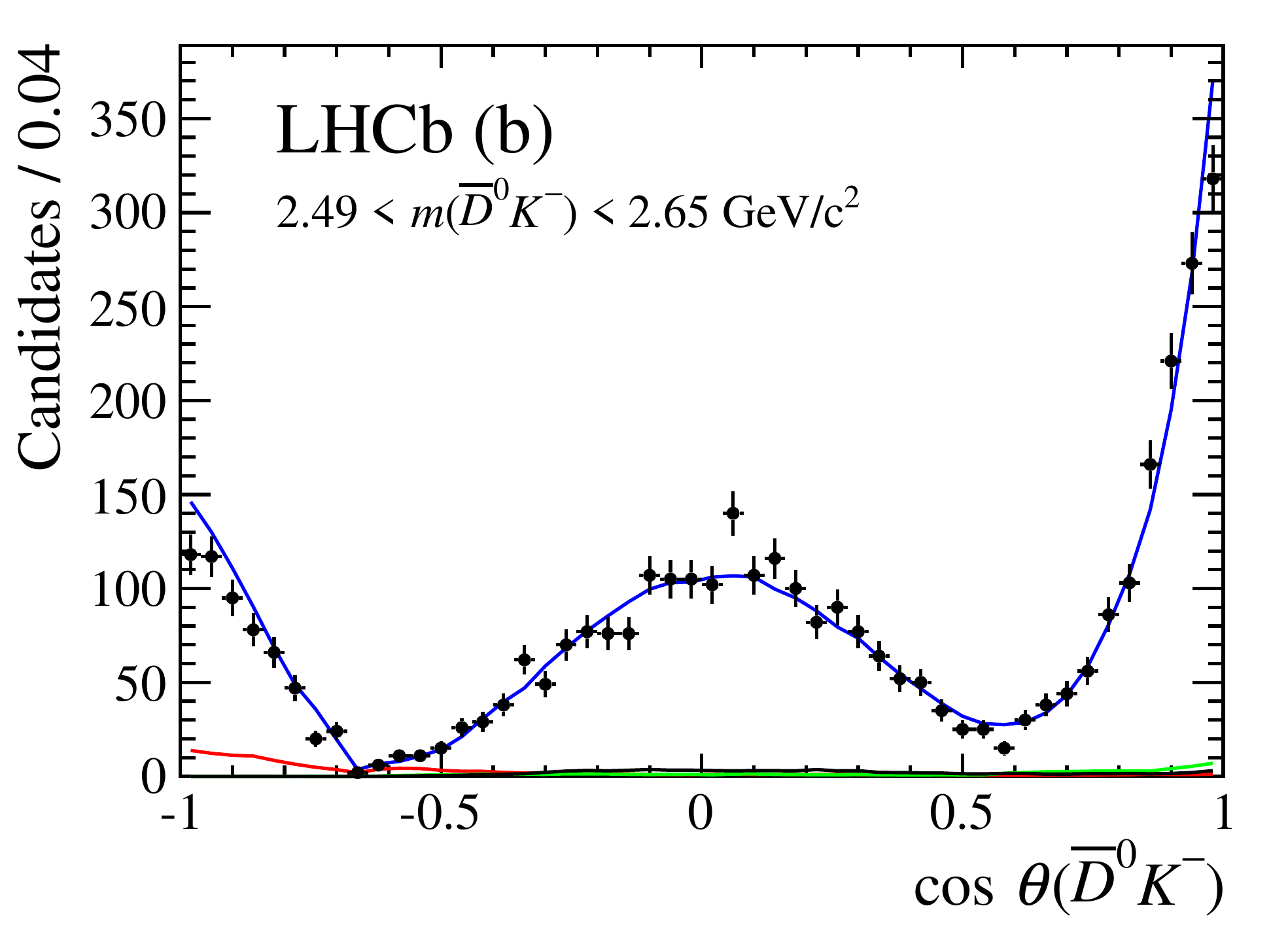}
 \includegraphics[scale=0.38]{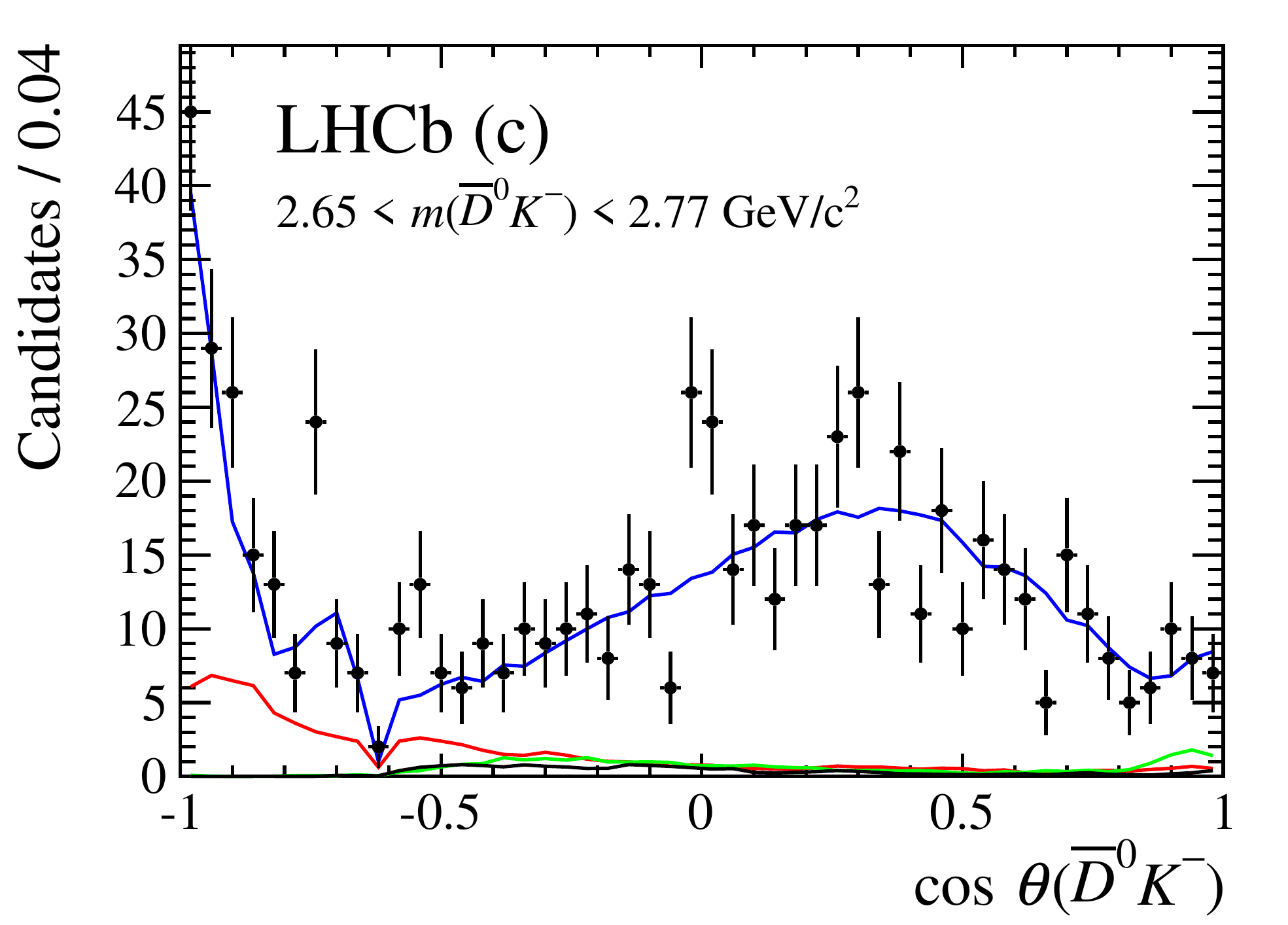}
 \includegraphics[scale=0.38]{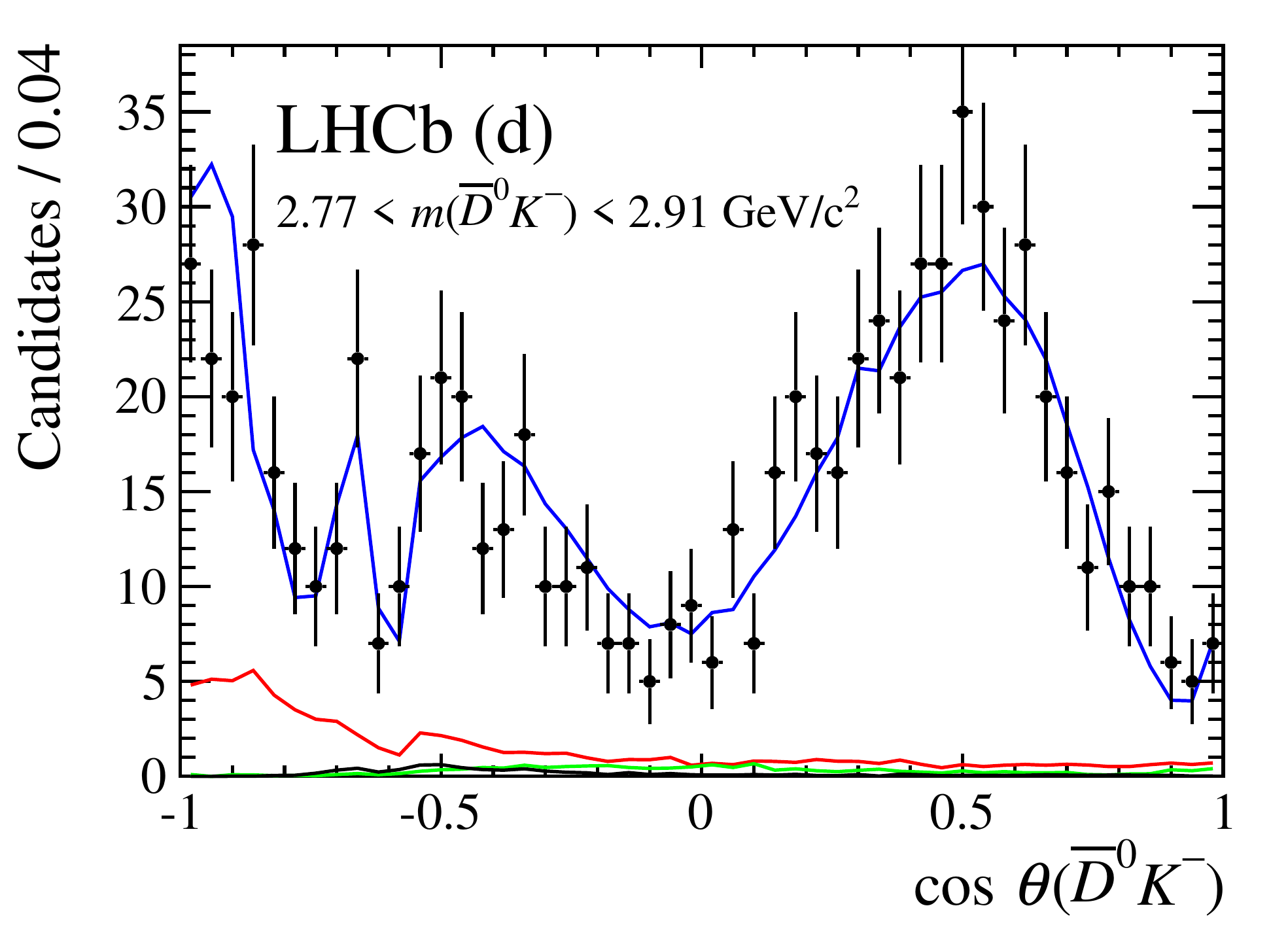}
\caption{\small
  Projections of the data and the Dalitz plot fit result onto the cosine of
  the helicity angle of the $\Dzb\Km$ system, $\cos\theta(\Dzb\Km)$, for $m(\Dzb\Km)$
  slices of (a) $0$--$2.49\gevcc$, (b) $2.49$--$2.65\gevcc$, (c)
  $2.65$--$2.77\gevcc$ and (d) $2.77$--$2.91\gevcc$.
  The data are shown as black points, the total fit result as a solid blue
  curve, and the small contributions from
  $\Bd\to \DorDstarzb \pip\pim$, $\Lbbar\to \DorDstarzb \antiproton\pip$ and combinatorial
  background shown as green, black and red curves, respectively.
}
\label{fig:cosheldk}
\end{figure}

Further comparisons of regions of the data with the fit result are given in
Figs.~\ref{fig:coshelkpi} and~\ref{fig:cosheldk}.
These show projections of the cosine of the helicity angle of the $\Km\pip$
and $\Dzb\Km$ systems, respectively, and show that the spin content of the fit
model matches well that of the data.
In particular, Fig.~\ref{fig:cosheldk}(d) shows that the region around the
$D^{*}_{sJ}(2860)^-$ states is well modelled by a combination of spin-1
and spin-3 states.
This is confirmed by the $\chisq$ value of 56 that is found by comparing the data and the fit model in only the 70 SDP bins, defined with the adaptive binning scheme discussed above, that overlap or are contained in this region of phase-space ($0.71 < \mpr < 0.77$).
The distinctive angular distribution of the spin-3 state enables the comparatively precise determination of its properties (Table~\ref{tab:mG-stat}).

To test whether any other combination of resonances can provide a comparably
good description of the data, the fit is repeated with different hypotheses.
The results are shown in Table~\ref{tab:DsJDNLL}.
The values of $\sqrt{2\Delta{\rm NLL}}$ are given as a crude indication of the
significance but are not otherwise used in the analysis --- numerical values of
the significance are instead obtained from pseudoexperiments as described in Sec.~\ref{sec:Results}.
Some of the results in Table~\ref{tab:DsJDNLL} are labelled with * to indicate that the fit prefers to position one of the resonances in a different mass region from the discussed peak region.
For spin-0 this is subthreshold, for spin-2 it is either very near to the $D^{*}_{s2}(2573)^-$ mass or at higher mass.

\begin{figure}[!tb]
\centering
\includegraphics[scale=0.50]{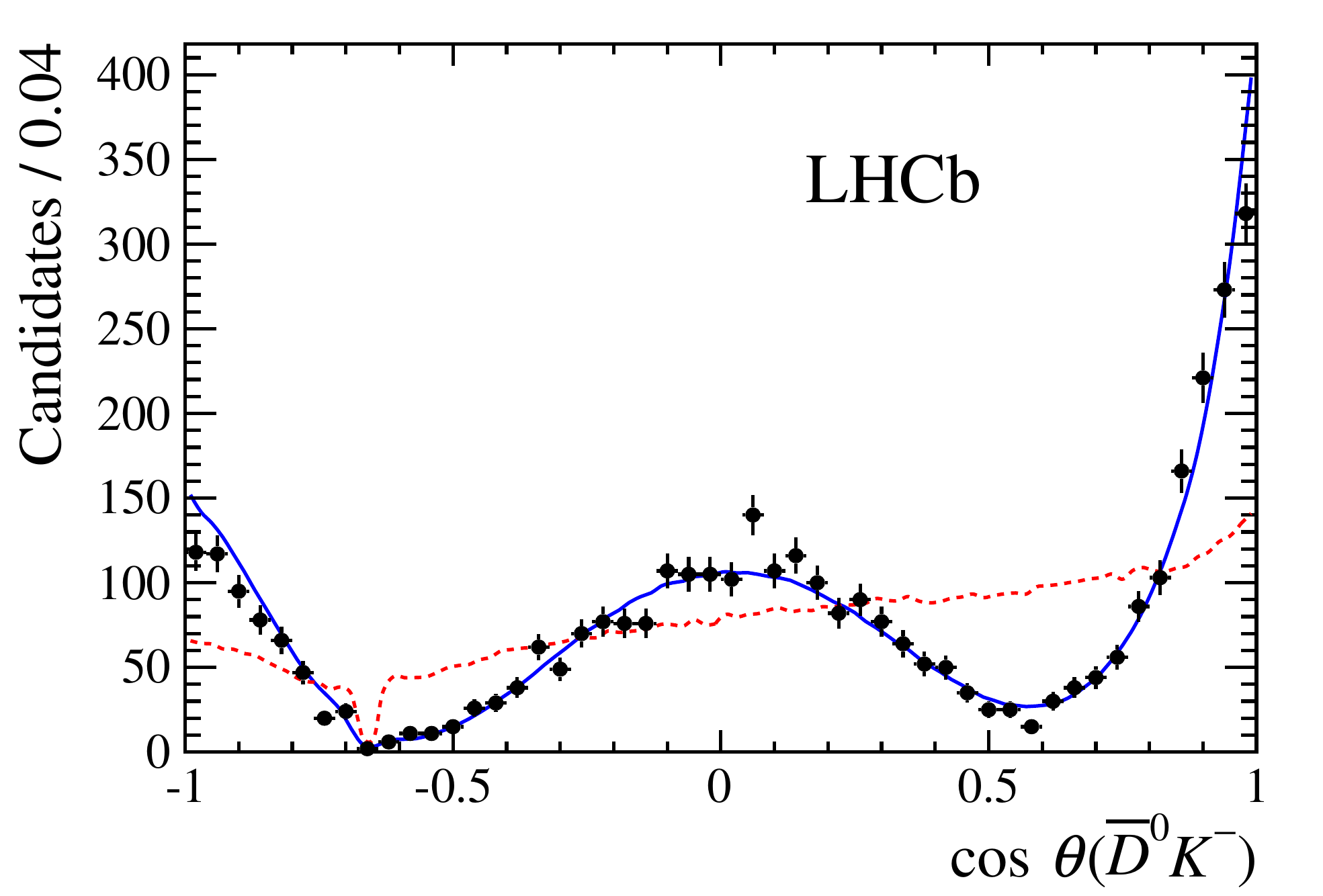}
\caption{\small
  Projections of the data and Dalitz plot fit results with alternative models
  onto the cosine of the helicity angle of the $\Dzb\Km$ system,
  $\cos\theta(\Dzb\Km)$, for $2.49 < m(\Dzb\Km) < 2.65\gevcc$.
  The data are shown as black points, the result of the baseline fit with a spin-2 resonance is given as a solid blue curve, and the result of
  the fit from the best model with a spin-0 resonance is shown as a dashed red
  line.
}
\label{fig:ds2alt}
\end{figure}

The spin of the $D^{*}_{s2}(2573)^-$ state has not previously been determined experimentally~\cite{PDG2012}.
As seen in Fig.~\ref{fig:cosheldk}(b), the helicity angle distribution in this region follows closely the expectation for a spin-2 state.
No alternative spin hypothesis can give a reasonable description of the data --- the closest is a fit assuming spin-0, which gives a value of $\sqrt{2\Delta{\rm NLL}}$ above 40.
The helicity angle distributions for the best fits with spin-2 and spin-0 hypotheses are compared to the data in Fig.~\ref{fig:ds2alt}.

\begin{table}
\centering
\caption{\small
  Changes in NLL from fits with different
  hypotheses for the state(s) at $m(\Dzb\Km)=2860\mevcc$.
  Units of $\mevcc$ are implied for the masses and widths.
  When two pairs of mass and width values are given, the first corresponds to
  the lower spin state.  Values marked * are discussed further in the text.
  There are two entries for spin-2 because two solutions were found.
}
\label{tab:DsJDNLL}
\begin{tabular}{lcccccc}
\hline \\ [-2.5ex]
Spin hypothesis & $\Delta{\rm NLL}$ & $\sqrt{2\Delta{\rm NLL}}$ &
\multicolumn{4}{c}{Masses and widths} \\
\hline
1+3     & 0      &   ---   & \multicolumn{4}{c}{See Table~\ref{tab:mG-stat}} \\
\hline
0       & 141.0  &   16.8  & 2862  &  57   &        &       \\
0+1     & 113.2  &   15.0  & \phani2446* &  250  &  2855  &  96   \\
0+2     & 155.1  &   17.6  & 2870  &  61   &  \phani2569* &  17  \\
0+3     & 105.1  &   14.5  & \phani2415* &  188  &  2860  &  52   \\
1       & 156.8  &   17.7  & 2866  &  92   &        &       \\
1+2     & 138.6  &   16.6  & 2851  &  99   &  \phani3134* &  174 \\
2       & 287.9  &   24.0  & \phani3243* &  81   &        &       \\
2       & 365.5  &   27.0  & \phani2569* &  17   &        &       \\
2+3     & 131.2  &   16.2  & 2878  &  12   &  2860  &  56   \\
3       & 136.5  &   16.5  & 2860  &  57   &        &       \\
\hline
\end{tabular}
\end{table}

Another approach to assess the agreement between the data and the fit result is to compare their angular moments, obtained by weighting the events in each $m(\Dzb\Km)$ ($m(\Km\pip)$) bin by the Legendre polynomial of order $L$ in
$\cos\theta(\Dzb\Km)$ ($\cos\theta(\Km\pip)$), where $\theta(\Dzb\Km)$
($\theta(\Km\pip)$) is the angle between the $\pip$ and the $\Dzb$ meson
(the $\Dzb$ and the $\Km$ meson) in the $\Dzb\Km$ ($\Km\pip$) rest frame.
This approach is very powerful in the case that resonances are only present in
one invariant mass combination, since then structures are seen in moments up to
$2\times J_{\rm max}$, where $J_{\rm max}$ is the highest spin of the
contributing resonances.
When resonances in other invariant mass combinations cause reflections, higher
moments are introduced in a way that is hard to interpret.

\begin{figure}[!tb]
\centering
 \includegraphics[scale=0.36]{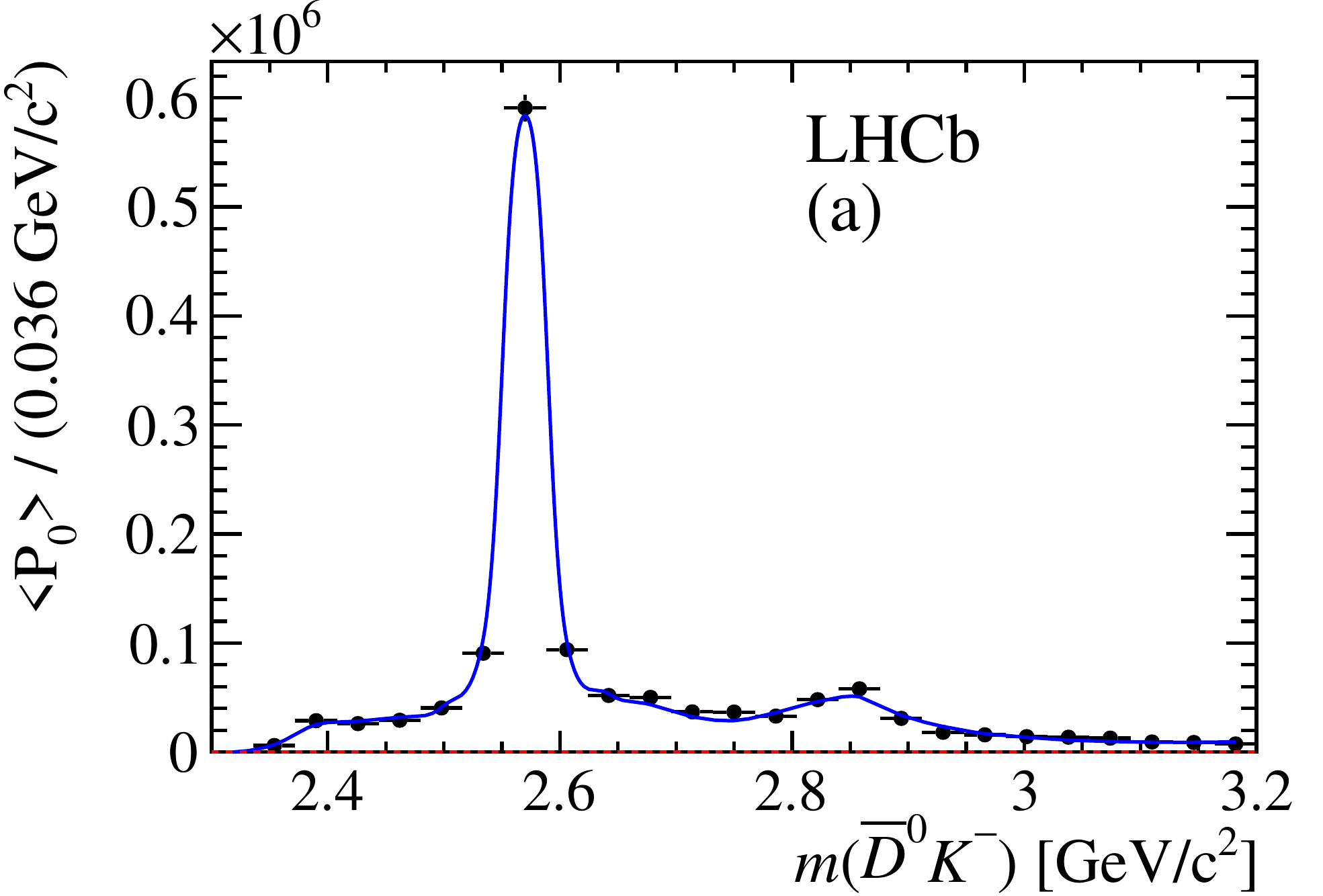}
 \includegraphics[scale=0.36]{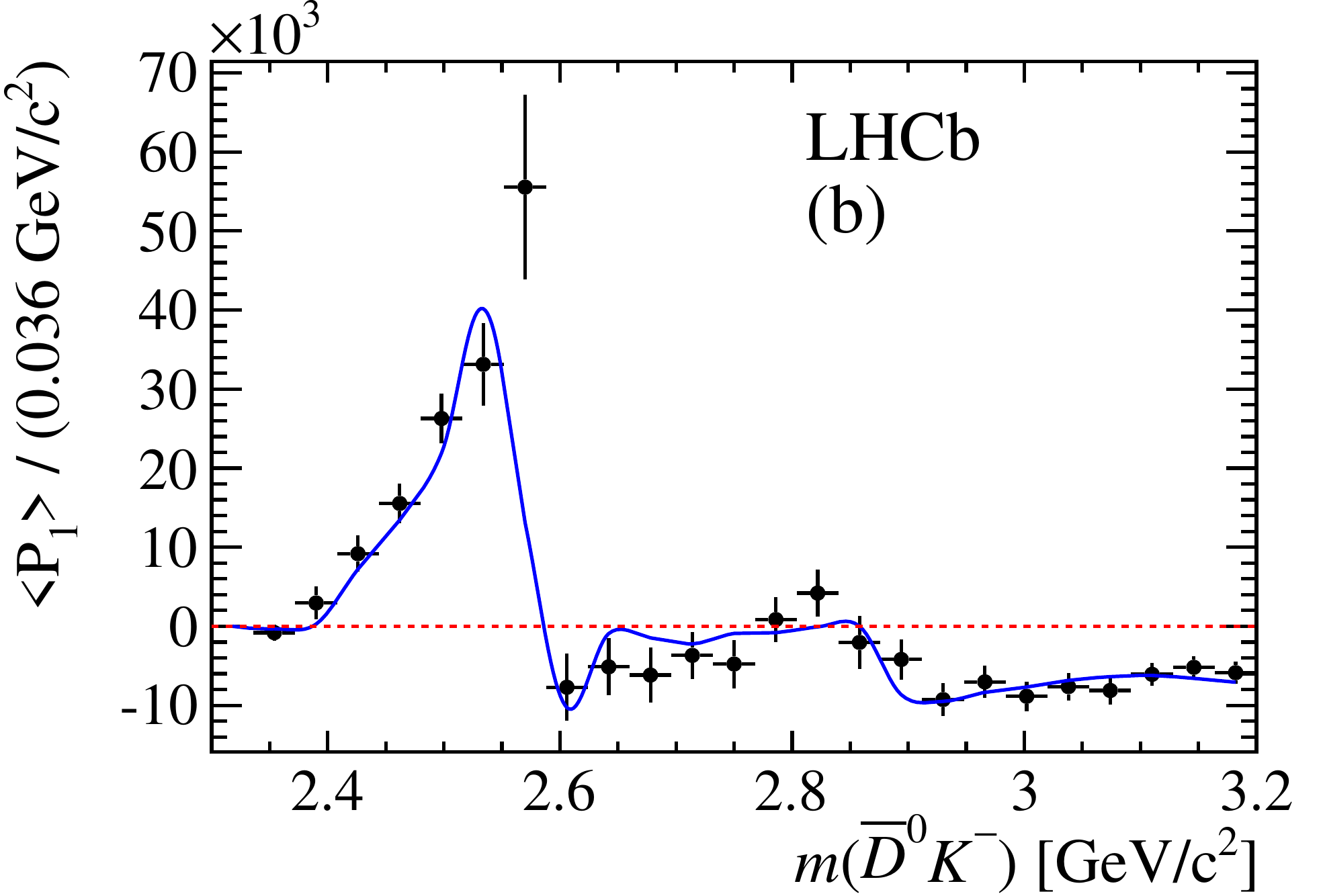}
 \includegraphics[scale=0.36]{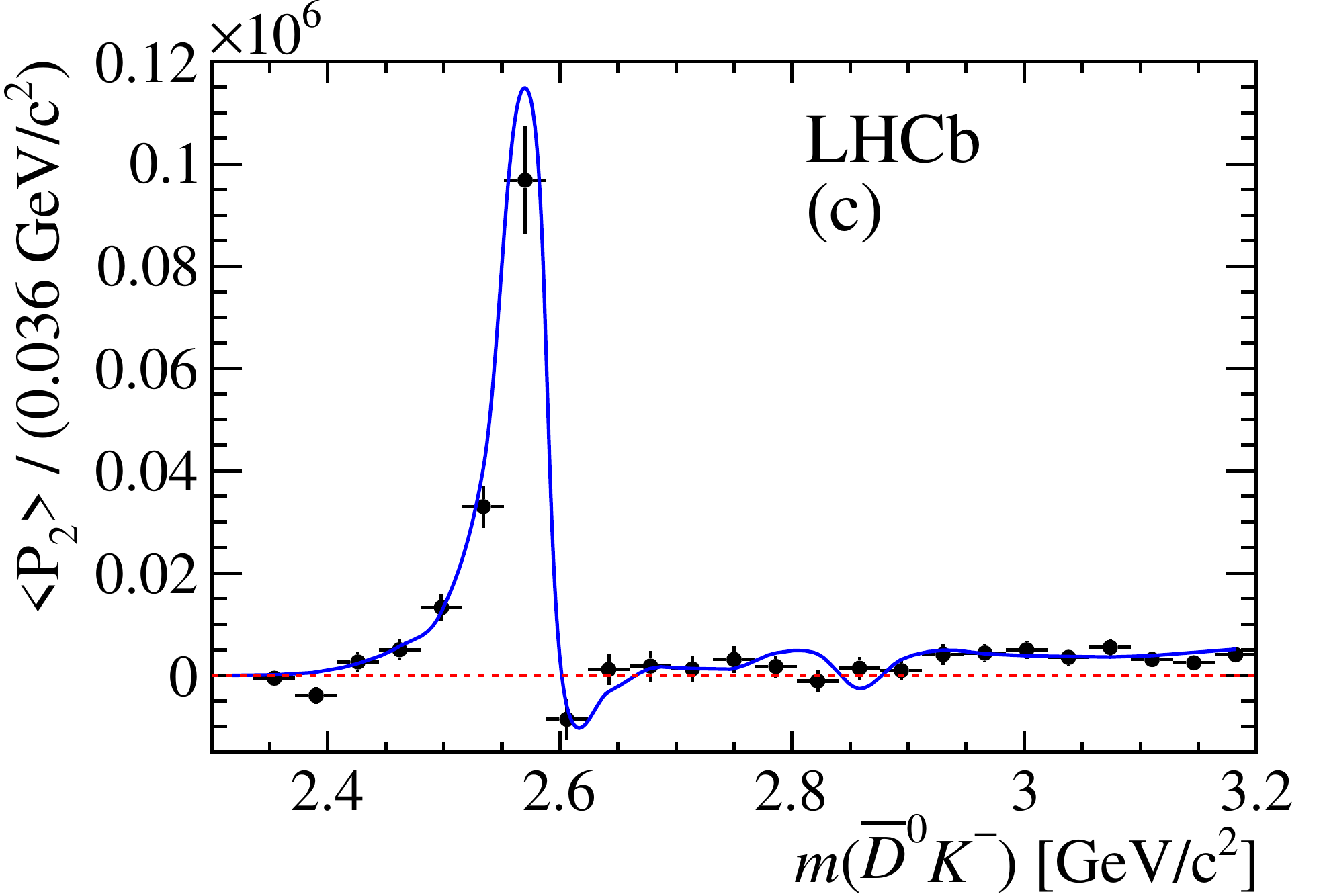}
 \includegraphics[scale=0.36]{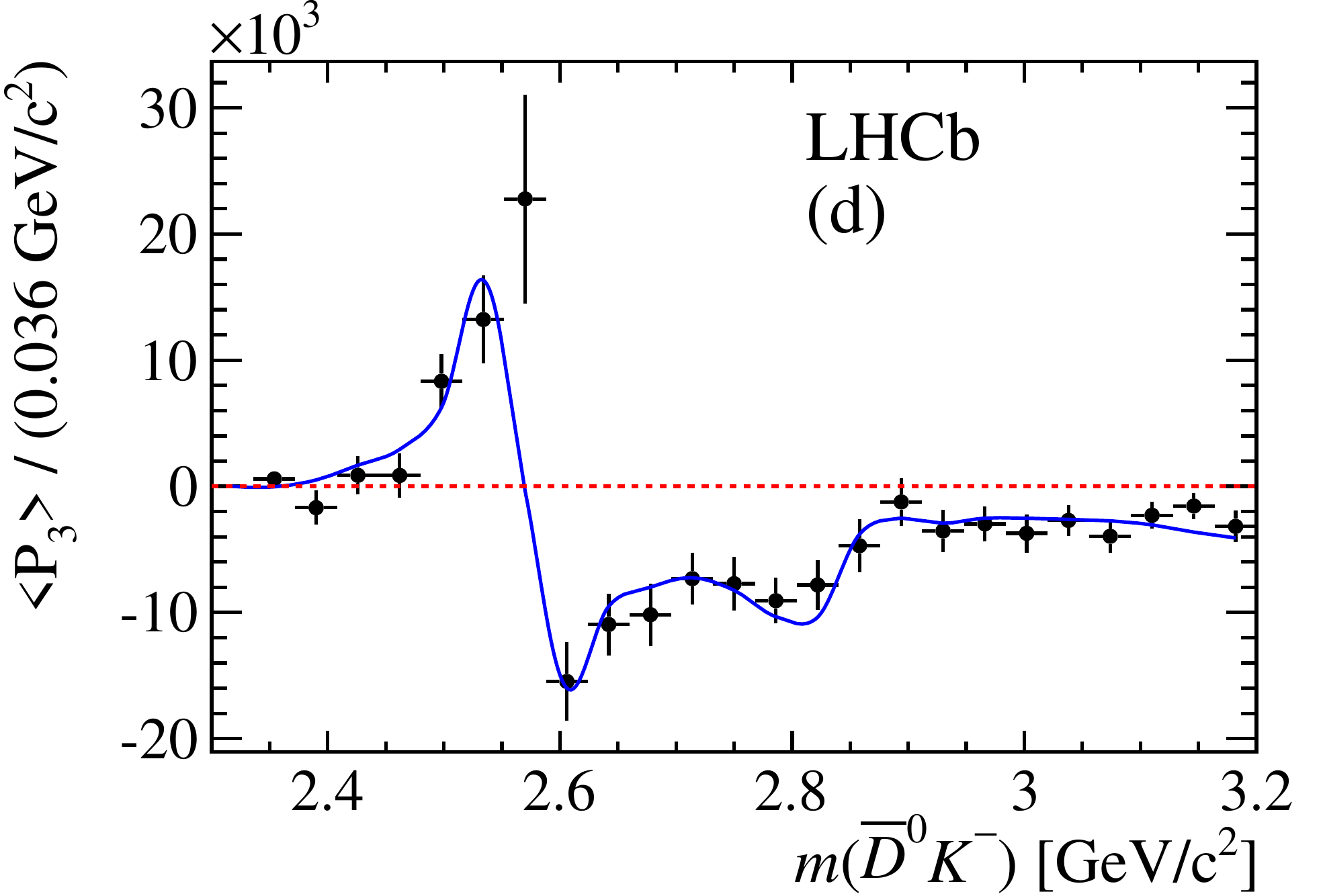}
 \includegraphics[scale=0.36]{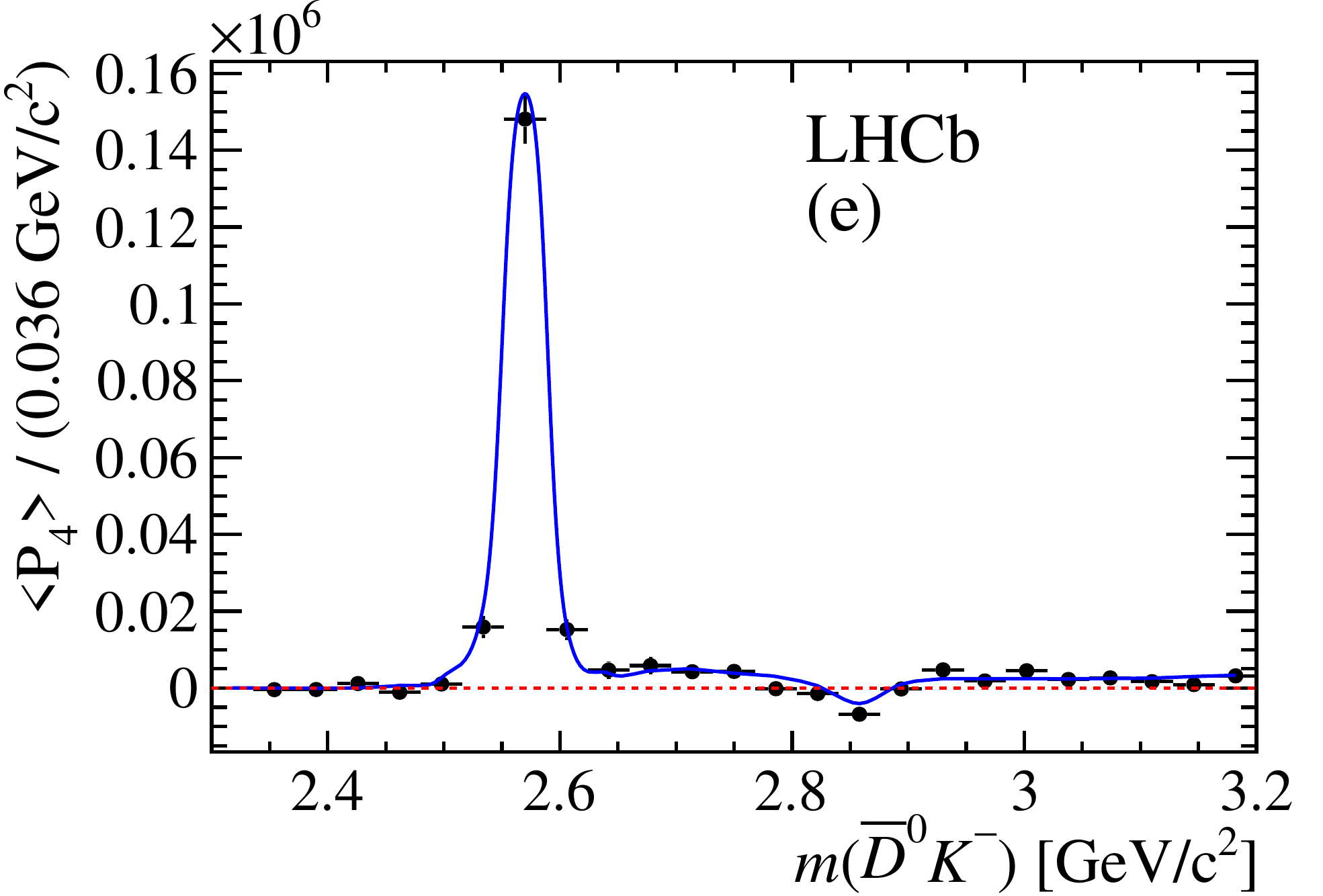}
 \includegraphics[scale=0.36]{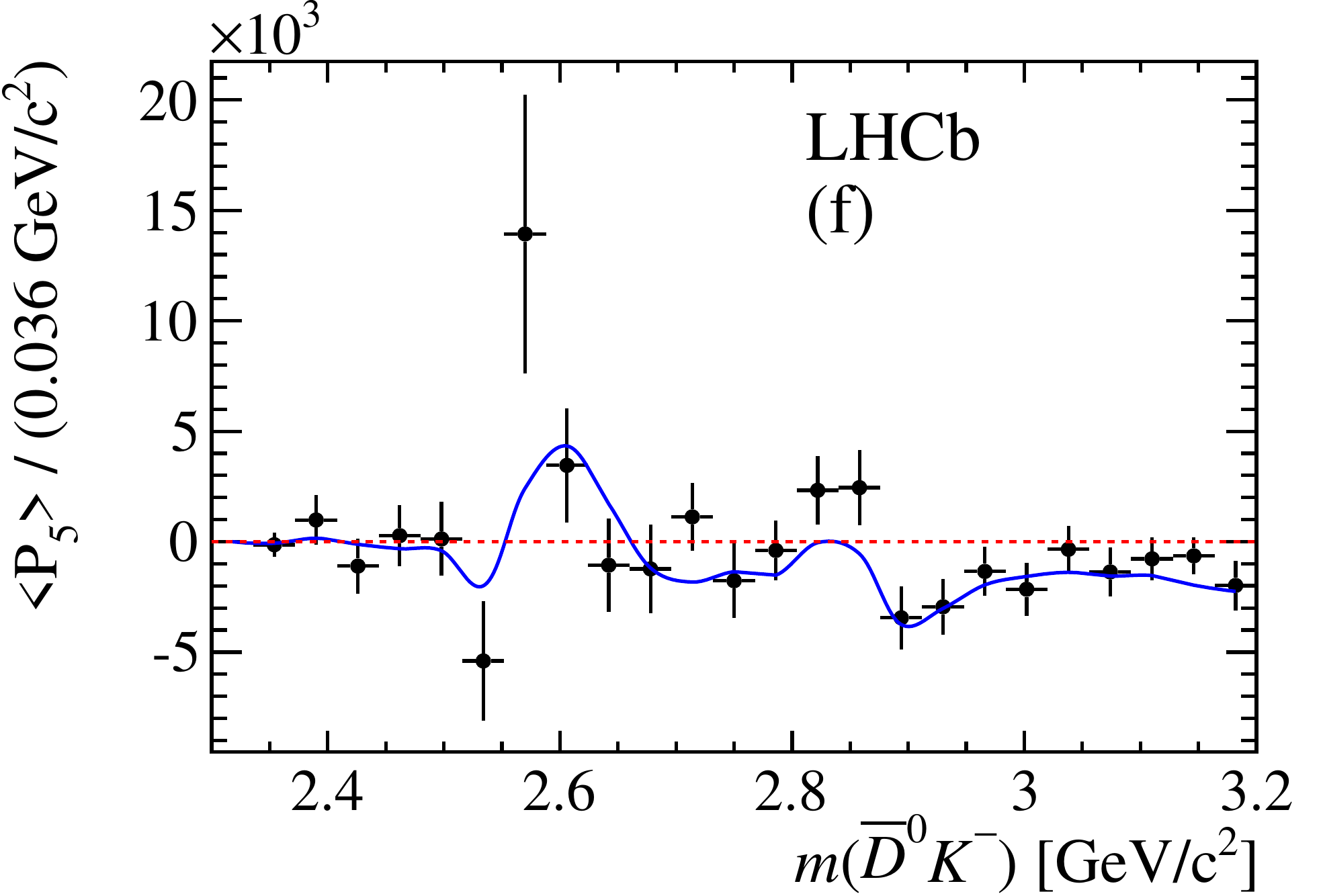}
 \includegraphics[scale=0.36]{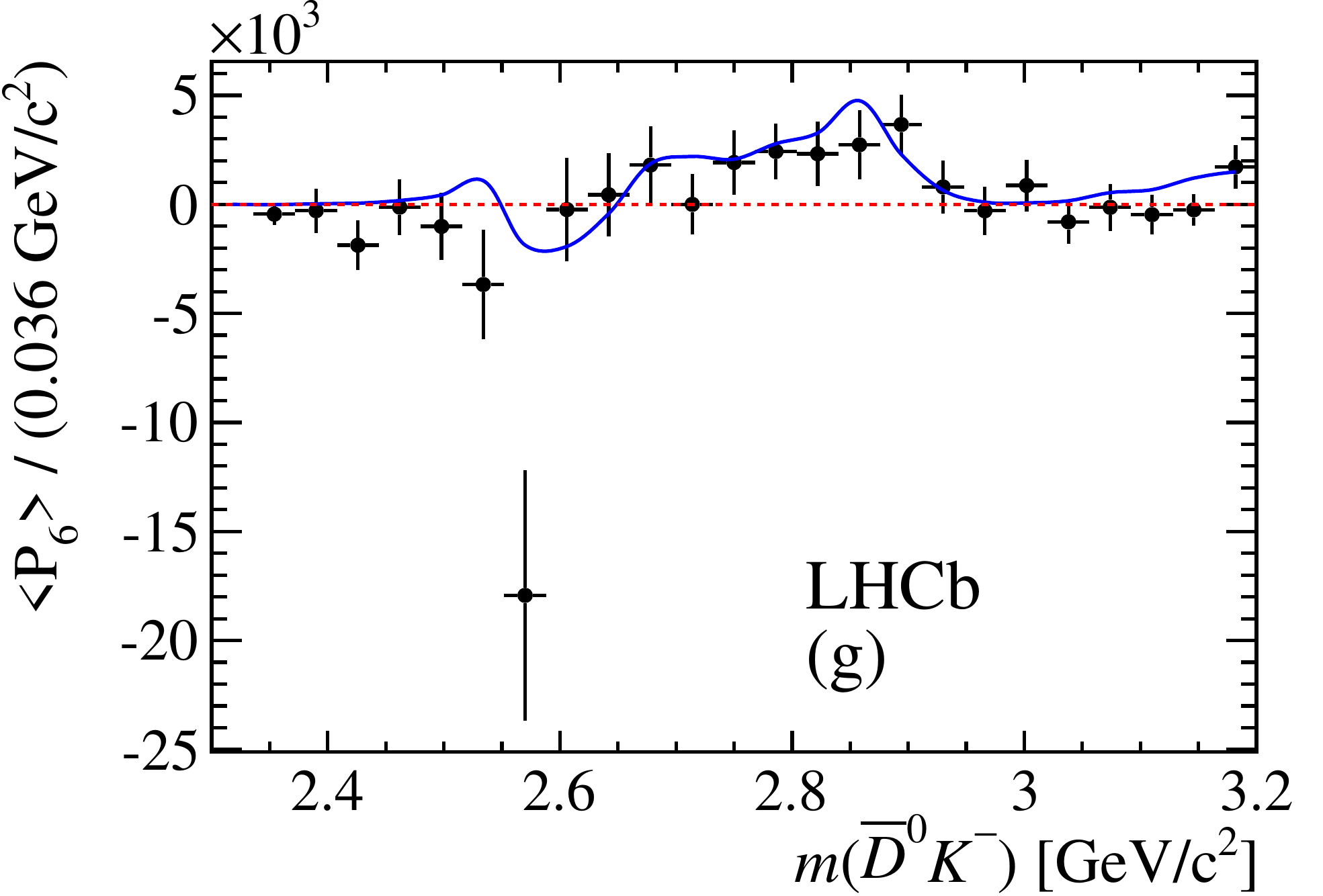}
 \includegraphics[scale=0.36]{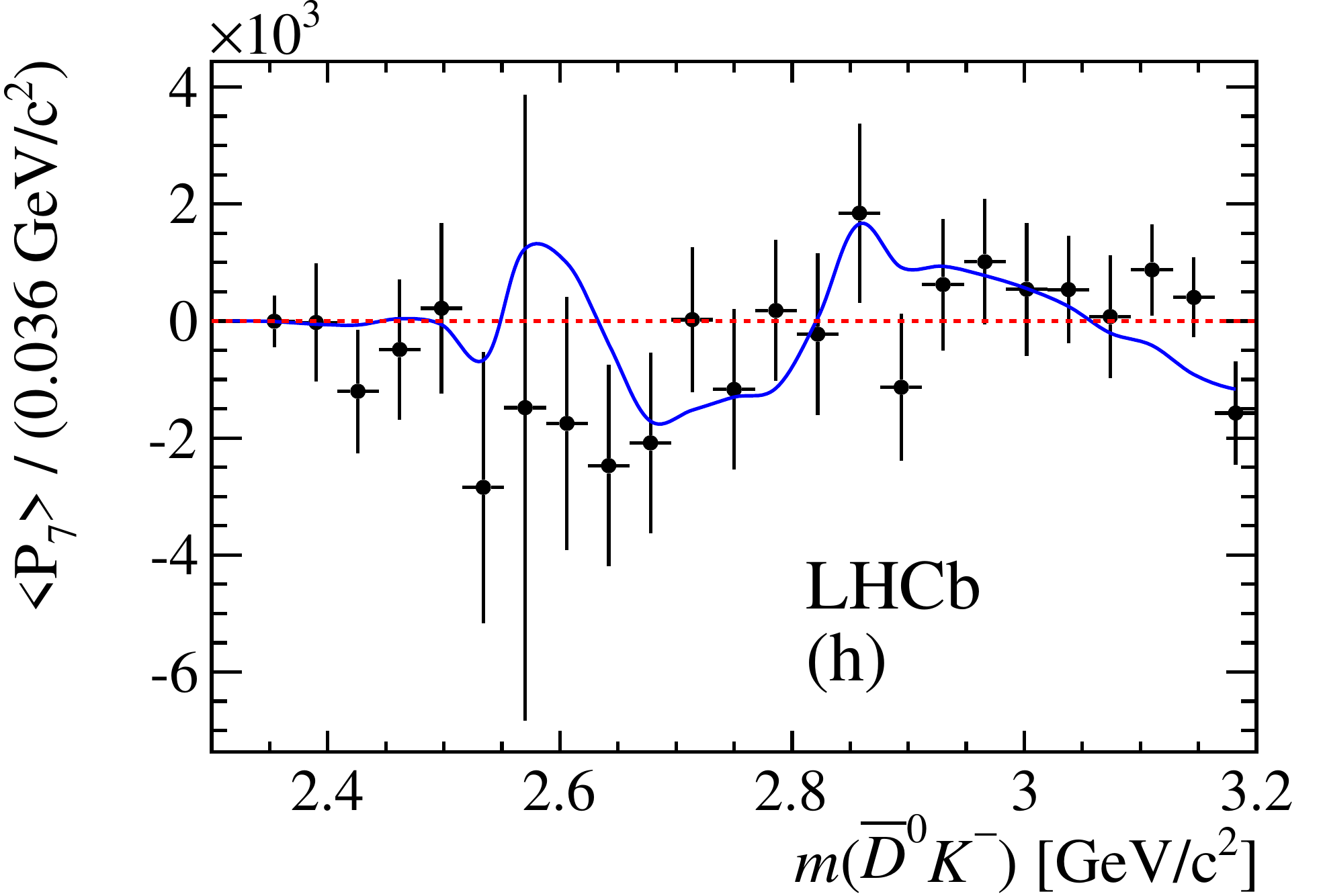}
\caption{\small
  Legendre moments up to order 7 calculated as a function of $m(\Dzb\Km)$ for
  data (black data points) and the fit result (solid blue curve).}
\label{fig:momentDKzoom}
\end{figure}

\begin{figure}[!tb]
\centering
 \includegraphics[scale=0.36]{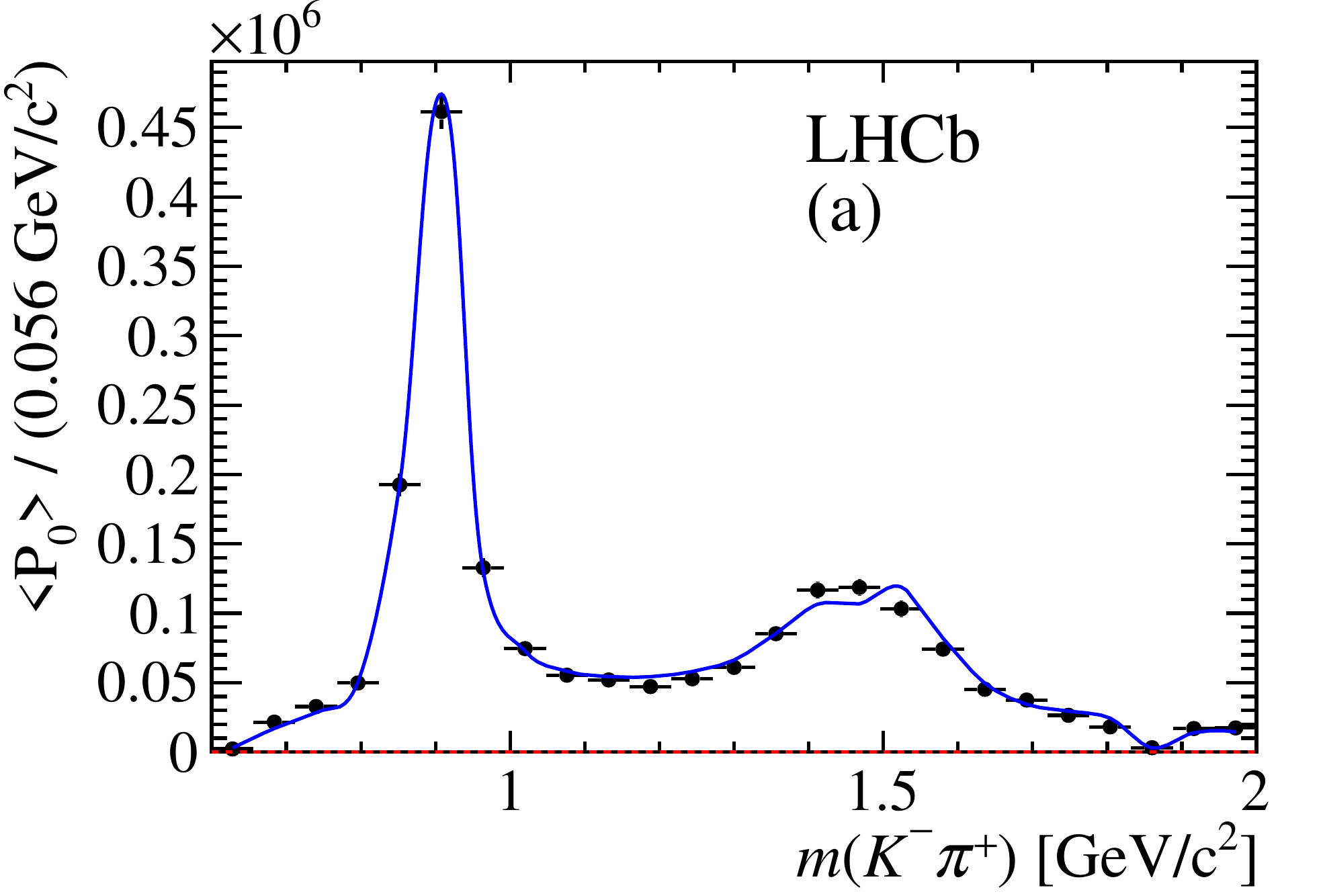}
 \includegraphics[scale=0.36]{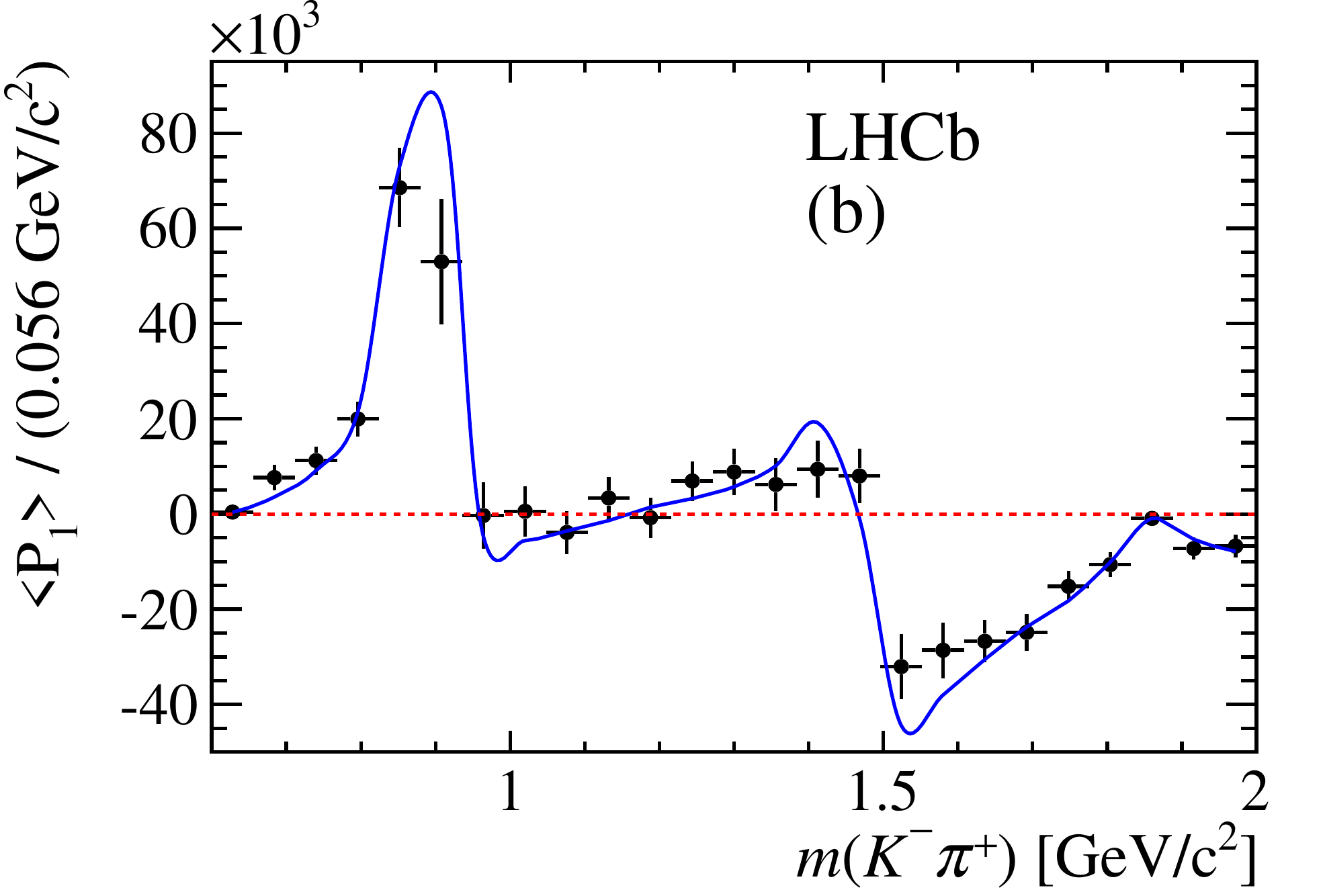}
 \includegraphics[scale=0.36]{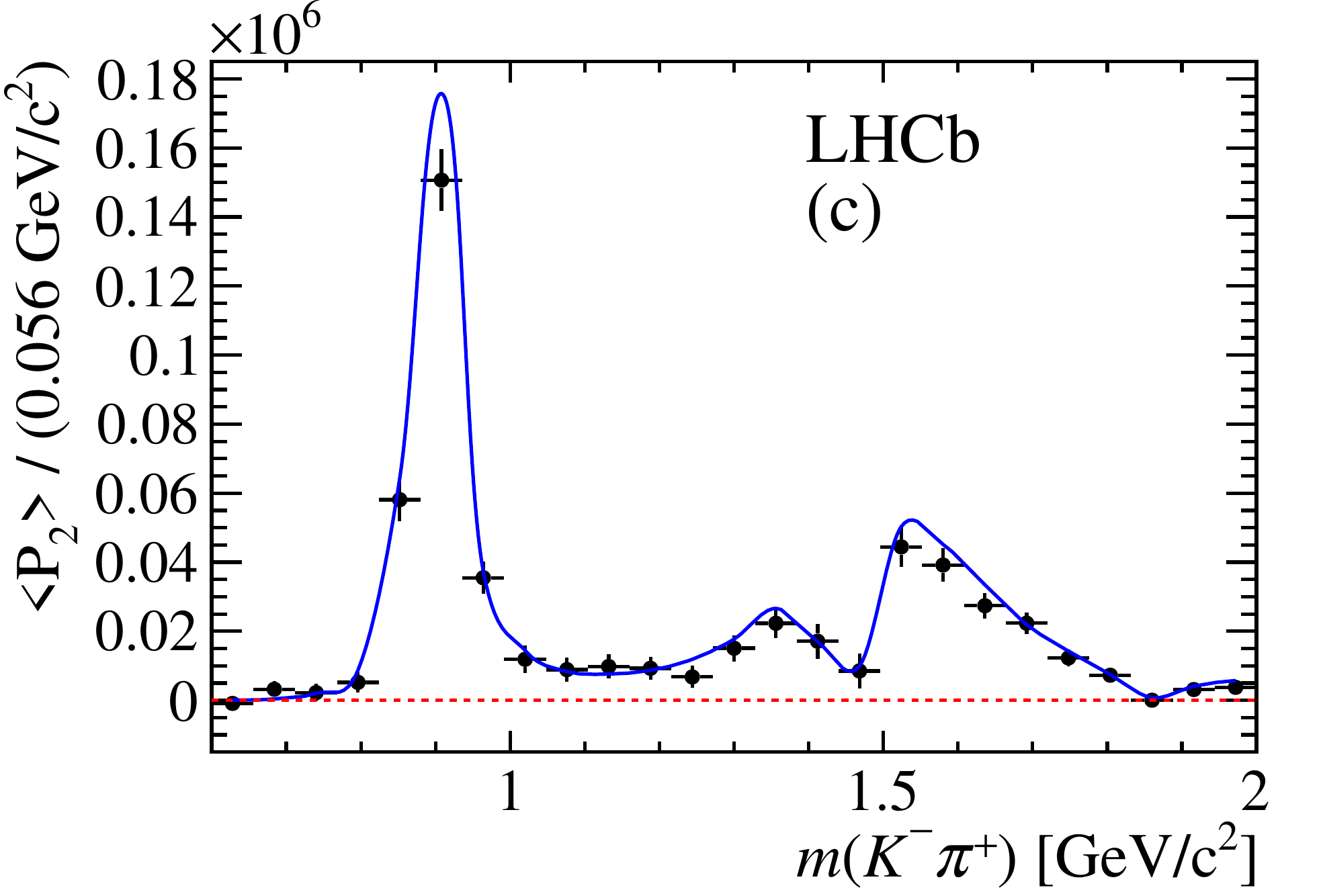}
 \includegraphics[scale=0.36]{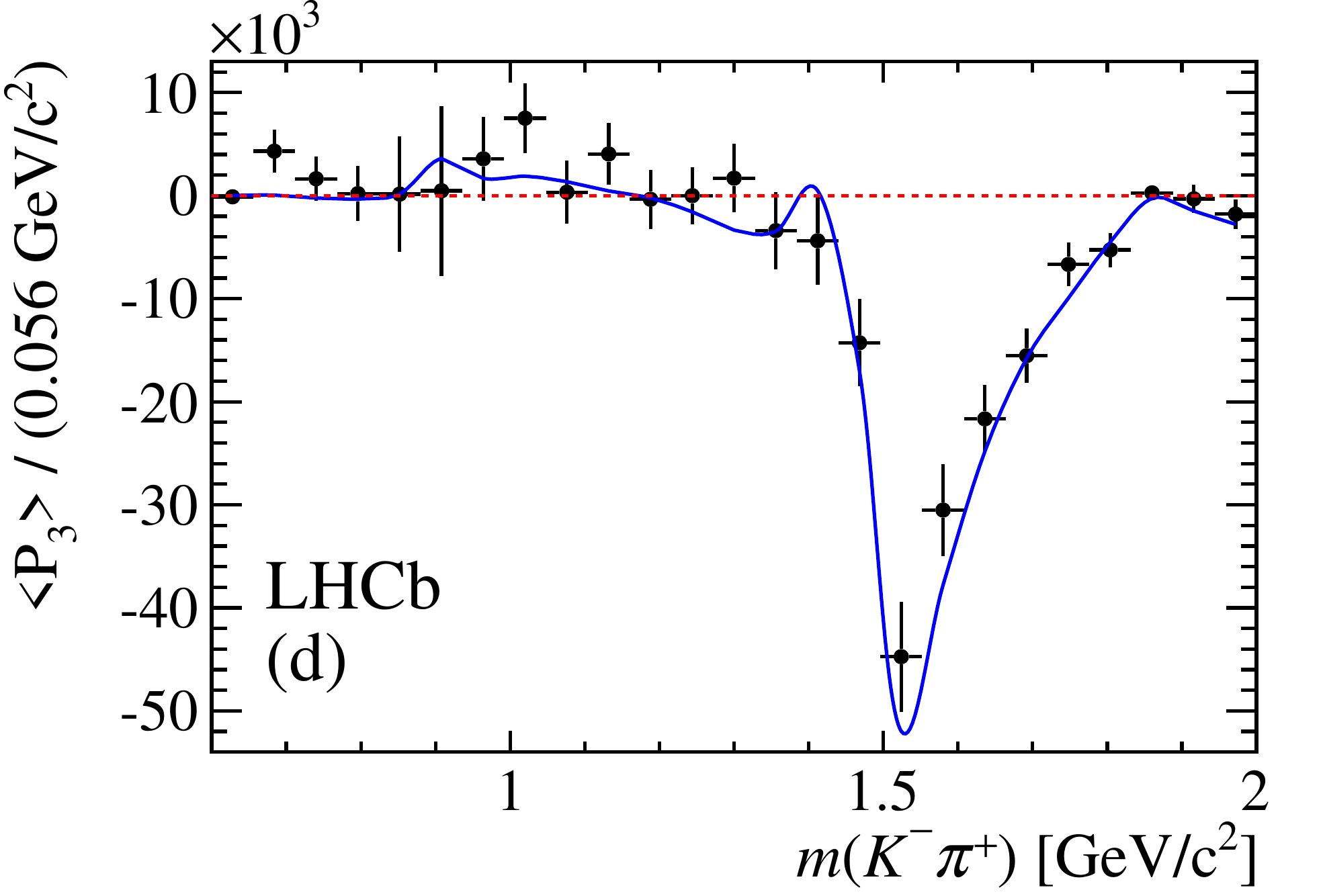}
 \includegraphics[scale=0.36]{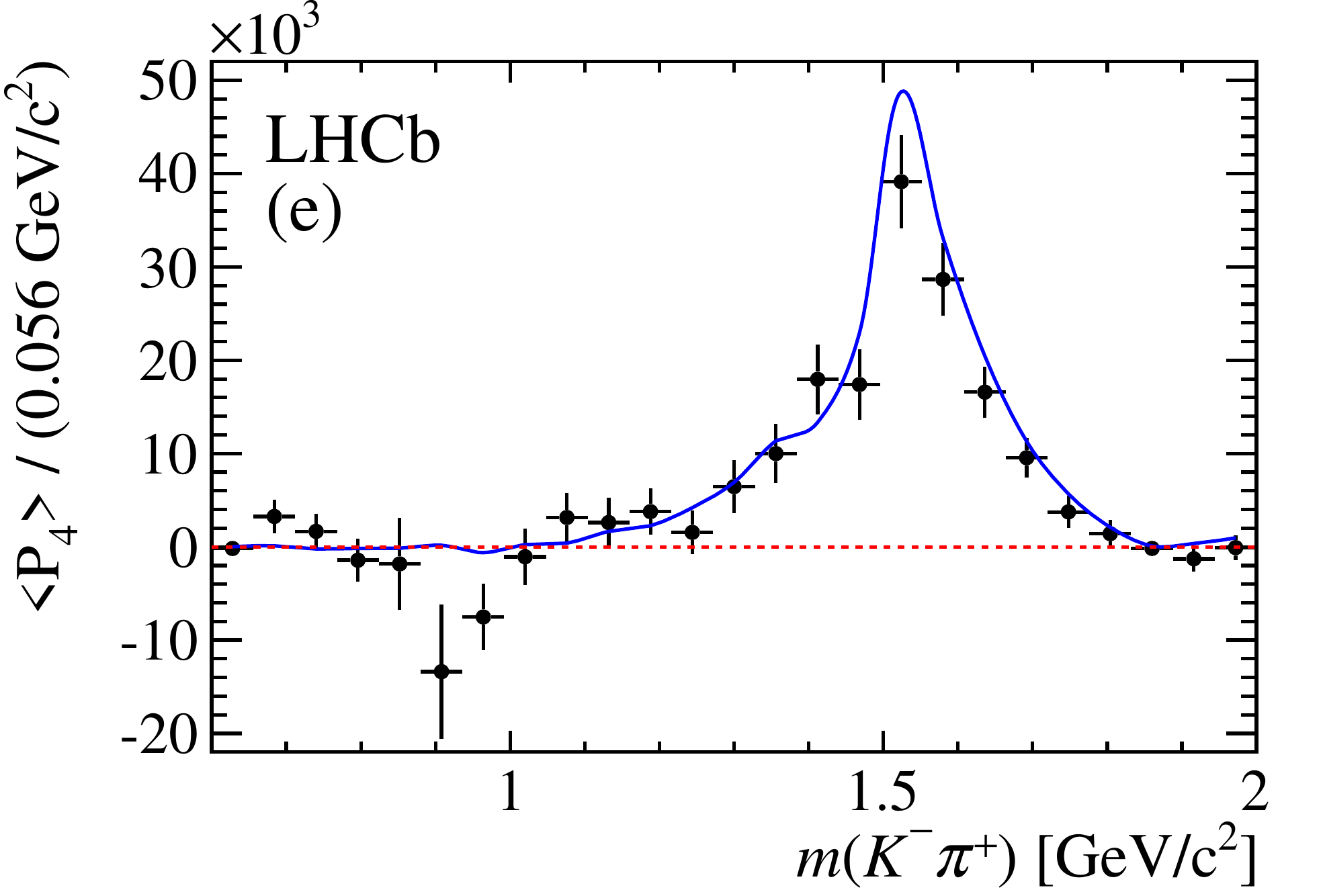}
 \includegraphics[scale=0.36]{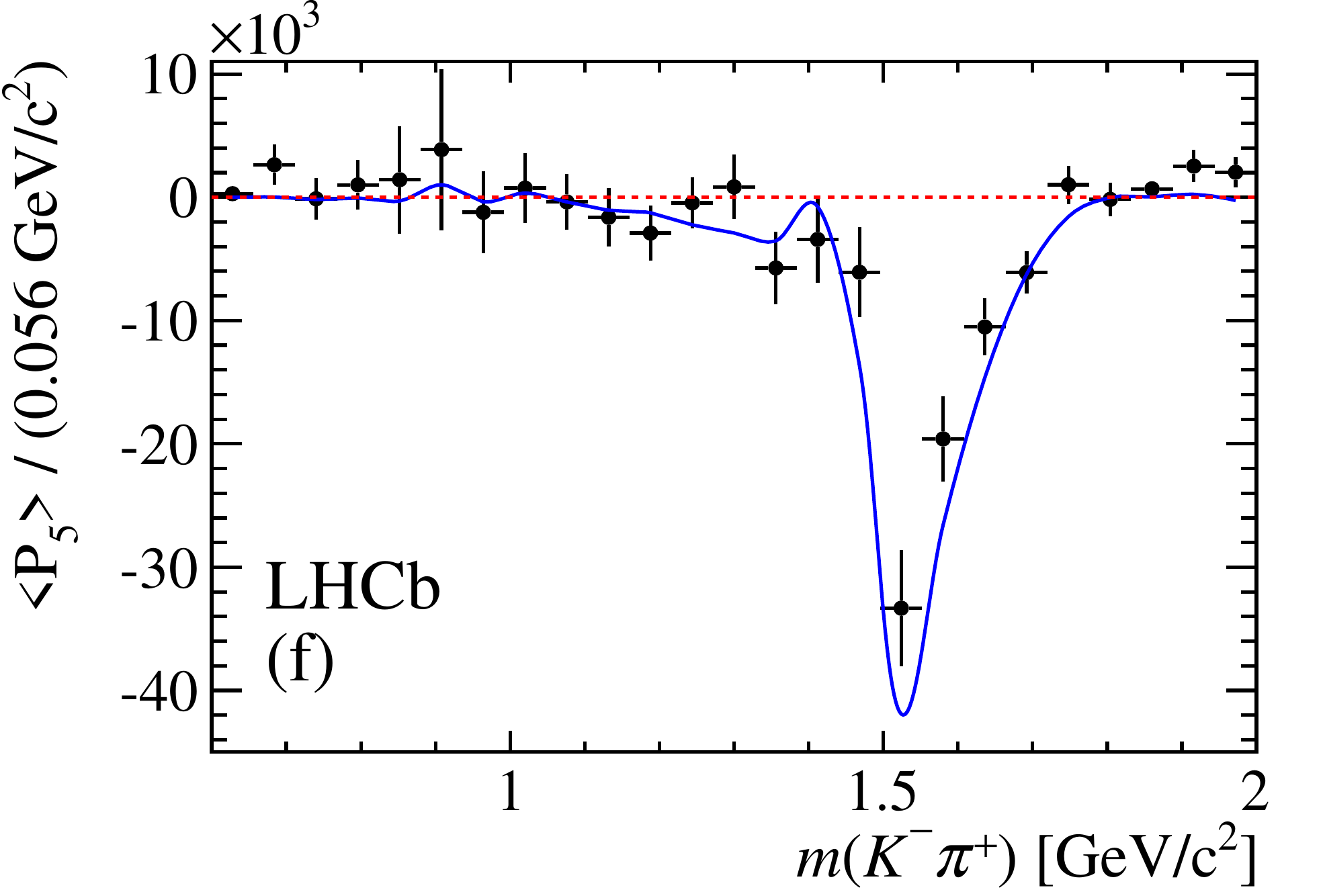}
 \includegraphics[scale=0.36]{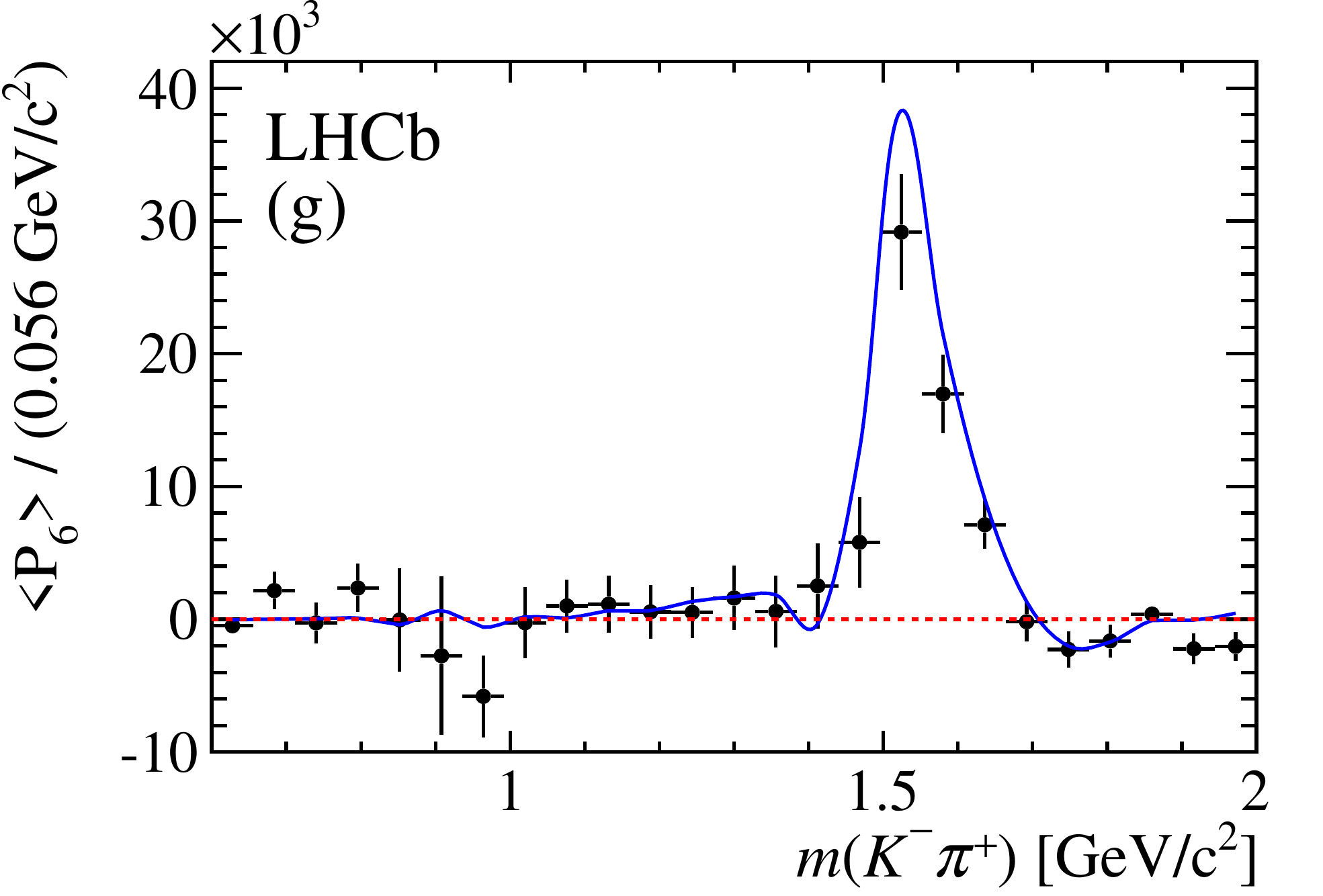}
 \includegraphics[scale=0.36]{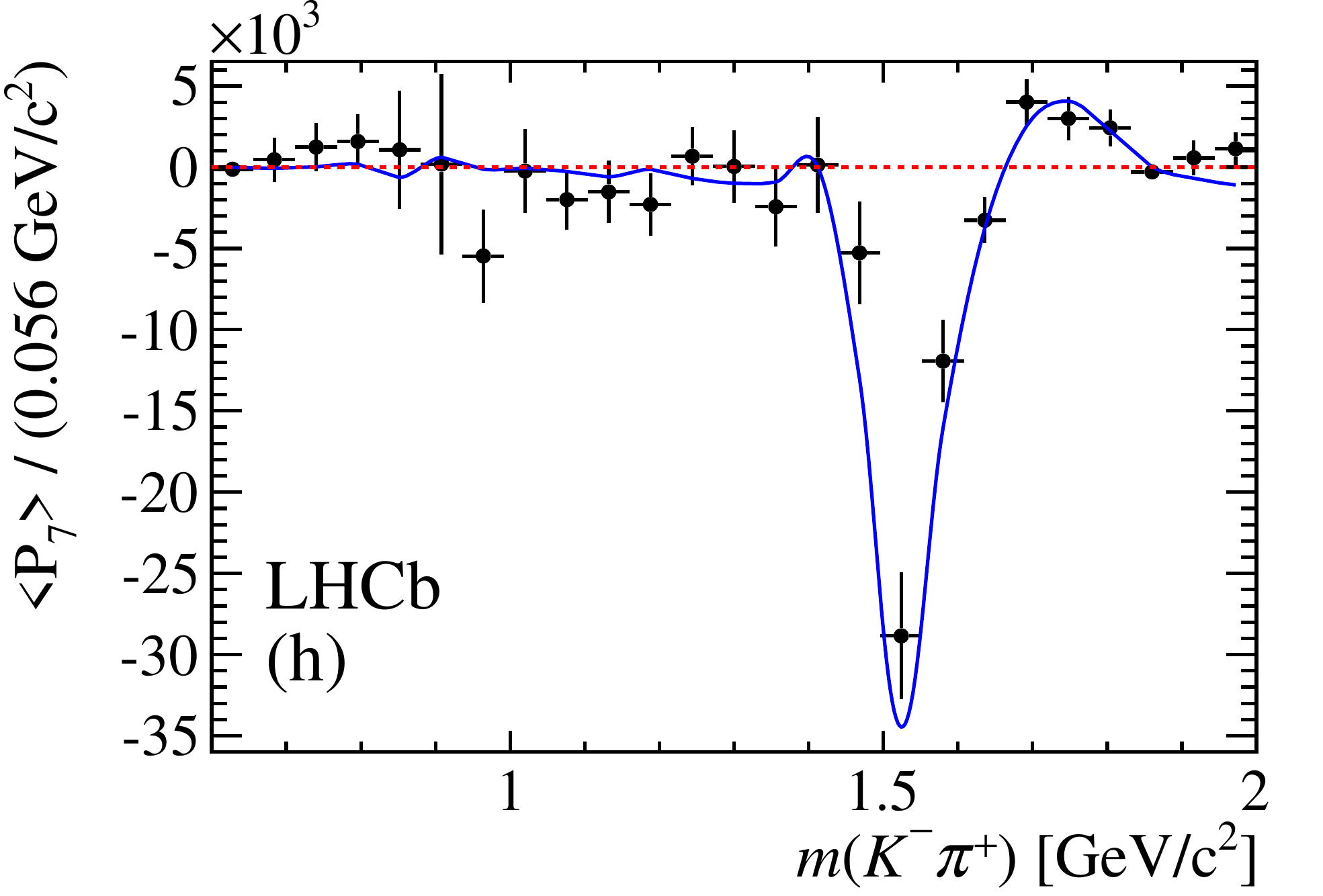}
\caption{\small
  Legendre moments up to order 7 calculated as a function of $m(\Km\pip)$ for
  data (black data points) and the fit result (solid blue curve).}
\label{fig:momentKpizoom}
\end{figure}

The angular moments of the data and the fit model in $m(\Dzb\Km)$ and $m(\Km\pip)$ are compared in Figs.~\ref{fig:momentDKzoom} and~\ref{fig:momentKpizoom}, respectively.
Significant structures in the $\Kstarb(892)^0$ peak region are observed in moments up to order 2, as expected for a spin-1 resonance in the absence of reflections.
The moments in the regions of other resonances are affected by reflections, as can be seen in the Dalitz plot (Fig.~\ref{fig:signalevents}).
Nonetheless, the large structures in the $D_s^*(2573)^-$ peak region in moments up to order 4 unambiguously determine that its spin is 2.
At higher masses, interpretation of the moments becomes more difficult.
Nonetheless, the reasonable agreement between data and the fit model provides confidence that the two-dimensional structures in the data are well described.

\section{Systematic uncertainties}
\label{sec:Systematics}

The considered sources of systematic uncertainty are divided into two main categories: experimental and model uncertainties.
The experimental systematic uncertainties arise from imperfect knowledge of:
the relative amount of signal and background in the selected events;
the distributions of each of the background components across the SDP;
the variation of the efficiency across the SDP;
the possible bias induced by the fit procedure;
the momentum calibration;
the fixed masses of the $\Bs$ and $\Dzb$ mesons used to define the boundaries of the Dalitz plot.
Model uncertainties occur due to:
fixed parameters in the Dalitz plot model;
the decision to include or exclude marginal components in the baseline fit model;
the choice of models for the $\Km\pip$ S-wave and the $\Dzb\Km$ S- and P-waves.
The systematic uncertainties from each source are combined in quadrature.

The yields of signal and background components in the signal region are given by the result of the fit to the $\Bs$ candidate invariant mass.
Both statistical and systematic uncertainties on these values are considered, where the latter are evaluated as in Ref.~\cite{LHCb-PAPER-2013-022}.
The signal and background yields are varied appropriately and the effects on the results of the Dalitz plot fit are assigned as uncertainties.

The uncertainty due to the imperfect knowledge of the background distributions across the SDP is estimated by varying the histograms used to model the shape within their statistical uncertainties.
In addition, the relative contributions from decays with $\Dzb$ and $\Dstarzb$ mesons in the $\Lbbar \to \DorDstarzb\antiproton\pi$ and $\Bz \to \DorDstarzb\pip\pim$ backgrounds are varied.
The effect on the results of not reweighting the SDP distribution of the $\Dstarzb$ component in these backgrounds is also included as a source of systematic uncertainty.
Other systematic uncertainties due to uncertainties on the weights applied to obtain the background distributions are negligible.

The uncertainty arising due to the imperfect knowledge of the efficiency variation across the SDP is determined by varying the content of the histogram from which the spline function used in the fit is obtained.
Since sources of systematic bias may affect the bins of this histogram in a
correlated way, only the central bin in each cell of $3\times3$ bins is
varied, and interpolation is used to obtain the values of the adjacent bins.
The effects on the results of the Dalitz plot fit are assigned as uncertainties.
In addition, the effect of binning the $\Dz \to \Km\pip$ control sample used to obtain the PID efficiencies is evaluated by varying the binning scheme.

An ensemble of pseudoexperiments is used to search for intrinsic bias in the fit procedure.
The differences between the inputs and the mean values obtained from the ensemble are all found to be small.
Systematic uncertainties are assigned as the sum in quadrature of the difference between the input and output values with the uncertainty on the mean from the fit to the ensemble of pseudoexperiments.

The uncertainty due to the momentum calibration is estimated by varying the calibration factor within its uncertainty~\cite{LHCb-PAPER-2012-048,LHCB-PAPER-2013-011}.
The differences with respect to the default results are assigned as the corresponding systematic uncertainties.

The masses of the $\Bs$ and $\Dzb$ mesons are fixed to their known values~\cite{PDG2012} when the Dalitz plot coordinates are calculated.
The analysis is repeated after varying the $\Bs$ and $\Dzb$ meson masses up
and down by one standard deviation independently, and the changes in the
fitted values are taken as the corresponding uncertainty.

The uncertainties due to fixed model parameters are evaluated by repeating the
fit after varying these parameters within their uncertainties.
The parameters that are modified are the masses and widths given in
Table~\ref{tab:dpmodel} and the Blatt--Weisskopf radius parameter, which is varied between $3$ and $5 \gev^{-1}$.
As a cross-check, different Blatt--Weisskopf radius parameters are used for the
$\Km\pip$ and $\Dzb\Km$ resonances, and the likelihood is minimised with
respect to these parameters with results
$r_{\rm BW}(\Km\pip) = \left( 3.6 \,^{+1.1}_{-0.7} \right)\gev^{-1}$ and
$r_{\rm BW}(\Dzb\Km) = \left( 4.1 \,^{+0.8}_{-0.5} \right)\gev^{-1}$
where the uncertainties are statistical only.
This confirms that the nominal value of $4.0 \gev^{-1}$ for both sets of
resonances is reasonable, and that the range of values for the systematic
variation is conservative.

The least significant components in the fit are the $\Kstarb(1680)$, $\Kstarbsubz(1950)$, $D^{*}_{s0\,v}(2317)^-$ and $B^{*+}_{v}$ terms.
The effects on the other parameters when each of these marginal components is removed individually from the model are assigned as uncertainties.
The effect of introducing the $\Kstarb_3(1780)^0$ and $\Kstarb_4(2045)^0$ resonances into the model is also considered.
The results of these fits are used to set upper limits on the corresponding
branching fractions (see Sec.~\ref{sec:Results}) as well as to determine contributions to the model uncertainty.

The models used to describe the $\Km\pip$ S-wave and the $\Dzb\Km$ S- and P-waves are known to be approximate forms, and therefore additional uncertainties are assigned due to the changes in the fitted values of the other parameters when these are replaced with alternative models.
The LASS shape is replaced with a Flatt\'e shape~\cite{Flatte:1976xu} for the $\Kstarbsubz(1430)$ and a resonant term with a modified mass-dependent width for the $\kappa$ (or $\Kstarbsubz(800)$) resonance at low $m(\Km\pip)$~\cite{bugg}.
The alternative model for the $\Km\pip$ S-wave given in Ref.~\cite{lesniak} is also used to fit the data, with the larger variation from the two alternative models assigned as systematic uncertainty.
A K-matrix implementation of the $\Km\pip$ S-wave~\cite{kmatrix} is also
attempted but does not provide stable fit results.
As an alternative model for the $\Dzb\Km$ S-wave, the exponential form factor is
replaced with a power-law dependence.
To estimate the dependence of the results on the modelling of the $\Dzb\Km$ P-wave, the two broad spin-1 $\Dzb\Km$ resonances ($D^{*}_{s1}(2700)^-$ and
$D^{*}_{s1}(2860)^-$) are described with a modified version of the
Gounaris--Sakurai lineshape~\cite{Gounaris:1968mw} instead of relativistic Breit--Wigner functions.
In addition, the dependence of the results on the choice of description of the effective pole mass for virtual components [Eq.~(\ref{eqn:effmass})] is evaluated by using a constant width instead of Eq.~(\ref{eq:GammaEqn}).

Summaries of the experimental systematic uncertainties on the fit fractions and complex amplitudes are given in Table~\ref{tab:expt-syst}.
A breakdown is given in Table~\ref{tab:expt-summary-FF} for the fit fractions, and in Table~\ref{tab:expt-summary-mG} for the masses and widths.
Similarly, summaries of the model uncertainties on the fit fractions and complex amplitudes are given in Table~\ref{tab:model-syst}, with breakdowns for the fit fractions and masses and widths in Tables~\ref{tab:model-summary-FF} and~\ref{tab:model-summary-mG}, respectively.
The largest sources of experimental systematic uncertainties on the fit fractions are, in general, those due to the efficiency variation across the SDP, the signal and background fractions and the description of the background SDP distributions.
The largest sources of model uncertainties on these parameters are, in general, from the description of the $\Km\pip$ S-wave and from removing the $\Kstarb(1680)^{0}$ and $B^{*+}_{v}$ components from the model.
These are also the largest sources of uncertainty on the mass and width measurements.
The magnitudes of the complex amplitudes are more robust against systematic uncertainties than the relative phases.

\begin{table}[!tb]
\centering
\caption{\small
  Experimental systematic uncertainties on the fit fractions and complex amplitudes.}
\label{tab:expt-syst}
\resizebox{\textwidth}{!}{
\begin{tabular}{lccccc}
\hline
Resonance & Fit fraction (\%) & Real part & Imaginary part & Magnitude & Phase (radians)
\\
\hline \\ [-2.5ex]
$\Kstarb(892)^{0}$         & 0.74 & 0.16 & 0.13 & 0.03 & 0.20  \\
$\Kstarb(1410)^{0}$        & 0.16 & 0.02 & 0.12 & 0.02 & 0.50  \\
LASS nonresonant           & 1.52 & 0.16 & 0.06 & 0.05 & 0.26  \\
$\Kstarbsubz(1430)^{0}$    & 0.72 & 0.22 & 0.07 & 0.03 & 0.25  \\
\ \ \ LASS total          & 0.95 &  ---  &  ---  &  ---  &  ---   \\
$\Kstarbsubt(1430)^{0}$    & 0.39 & 0.08 & 0.02 & 0.02 & 0.20  \\
$\Kstarb(1680)^{0}$        & 0.26 & 0.06 & 0.02 & 0.04 & 0.32  \\
$\Kstarbsubz(1950)^{0}$    & 0.13 & 0.03 & 0.04 & 0.03 & 0.32  \\
$D^{*}_{s2}(2573)^-$    & 0.78 &  ---  &  ---  &  ---  &  ---   \\
$D^{*}_{s1}(2700)^-$    & 0.44 & 0.02 & 0.06 & 0.03 & 0.18  \\
$D^{*}_{s1}(2860)^-$    & 0.65 & 0.05 & 0.05 & 0.03 & 0.12  \\
$D^{*}_{s3}(2860)^-$    & 0.28 & 0.03 & 0.02 & 0.02 & 0.10  \\
\hline
Nonresonant               & 4.30 & 0.25 & 0.04 & 0.15 & 0.36  \\
\hline
$D^{*-}_{s\,v}$           & 1.09 & 0.04 & 0.05 & 0.05 & 0.08  \\
$D^{*}_{s0\,v}(2317)^-$ & 1.94 & 0.22 & 0.05 & 0.16 & 0.72  \\
$B^{*+}_{v}$              & 1.07 & 0.08 & 0.11 & 0.06 & 0.34  \\
\hline
\end{tabular}
}
\end{table}

\begin{table}[!tb]
\centering
\caption{\small
  Breakdown of experimental systematic uncertainties on the fit fractions
  (\%).
  The columns give the contributions from the different sources described in
  the text.
}
\label{tab:expt-summary-FF}
\resizebox{\textwidth}{!}{
\begin{tabular}{lcccccc@{\hspace{5mm}}c}
\hline \\ [-2.5ex]
Resonance & S/B frac. & Eff. & Bkgd. SDP & Fit bias & $p$ scale & $\Dzb$,$\Bs$ mass & Total\\
\hline \\ [-2.5ex]
$\Kstarb(892)^{0}$         & 0.24 & 0.61 & 0.09 & 0.13 & 0.10 & 0.29 & 0.74  \\
$\Kstarb(1410)^{0}$        & 0.06 & 0.11 & 0.06 & 0.07 & 0.00 & 0.04 & 0.16  \\
LASS nonresonant           & 0.37 & 0.68 & 0.72 & 0.93 & 0.15 & 0.55 & 1.52  \\
$\Kstarbsubz(1430)^{0}$    & 0.50 & 0.33 & 0.18 & 0.21 & 0.15 & 0.24 & 0.72  \\
\ \ \ LASS total          &  0.49 & 0.54 & 0.43 & 0.36 & 0.05 & 0.24 & 0.95  \\
$\Kstarbsubt(1430)^{0}$    & 0.22 & 0.18 & 0.13 & 0.22 & 0.01 & 0.09 & 0.39  \\
$\Kstarb(1680)^{0}$        & 0.18 & 0.10 & 0.05 & 0.05 & 0.00 & 0.14 & 0.26  \\
$\Kstarbsubz(1950)^{0}$    & 0.06 & 0.03 & 0.03 & 0.03 & 0.03 & 0.10 & 0.13  \\
$D^{*}_{s2}(2573)^-$    &  0.50 & 0.53 & 0.08 & 0.20 & 0.16 & 0.13 & 0.78  \\
$D^{*}_{s1}(2700)^-$    &  0.41 & 0.07 & 0.14 & 0.05 & 0.04 & 0.02 & 0.44  \\
$D^{*}_{s1}(2860)^-$    &  0.42 & 0.25 & 0.36 & 0.19 & 0.00 & 0.10 & 0.65  \\
$D^{*}_{s3}(2860)^-$    &  0.03 & 0.07 & 0.05 & 0.15 & 0.02 & 0.21 & 0.28  \\
\hline
Nonresonant               & 3.53 & 1.06 & 1.13 & 1.05 & 0.45 & 1.51 & 4.30  \\
\hline
$D^{*-}_{s\,v}$           & 0.63 & 0.48 & 0.44 & 0.24 & 0.08 & 0.55 & 1.09  \\
$D^{*}_{s0\,v}(2317)^-$ & 1.79 & 0.37 & 0.46 & 0.28 & 0.10 & 0.37 & 1.94  \\
$B^{*+}_{v}$              & 0.54 & 0.54 & 0.68 & 0.19 & 0.00 & 0.27 & 1.07  \\
\hline
\end{tabular}
}
\end{table}

\begin{table}[!tb]
\centering
\caption{\small
  Breakdown of experimental systematic uncertainties on the masses and widths.
  Units of \mevcc are implied.
  The columns give the contributions from the different sources described in the text.
}
\label{tab:expt-summary-mG}
\resizebox{\textwidth}{!}{
\begin{tabular}{l@{\hspace{5mm}}cccccc@{\hspace{5mm}}c}
\hline
Resonance & \multicolumn{7}{c}{Mass} \\
 & S/B frac. & Eff. & Bkgd. SDP & Fit bias & $p$ scale & $\Dzb$,$\Bs$ mass & Total \\
\hline
$D^{*}_{s2}(2573)^-$ &  0.10 & 0.04 & 0.02 & 0.05 & 0.02 & 0.14 &   0.19 \\
$D^{*}_{s1}(2860)^-$ &  2.69 & 0.78 & 1.12 & 3.55 & 0.54 & 2.79 &   5.5\phani \\
$D^{*}_{s3}(2860)^-$ &  1.20 & 0.83 & 0.39 & 0.41 & 0.03 & 1.83 &   2.5\phani \\
\hline
\hline
Resonance & \multicolumn{7}{c}{Width} \\
 & S/B frac. & Eff. & Bkgd. SDP & Fit bias & $p$ scale & $\Dzb$,$\Bs$ mass & Total \\
\hline
$D^{*}_{s2}(2573)^-$ & \phani0.18 & 0.03 & 0.04 & 0.32 & 0.02 & 0.09 & \phani0.4 \\
$D^{*}_{s1}(2860)^-$ &      22.43 & 6.73 & 6.26 & 4.21 & 1.85 & 4.01 &  27.2 \\
$D^{*}_{s3}(2860)^-$ & \phani2.45 & 1.22 & 0.78 & 1.21 & 0.96 & 0.93 & \phani3.6 \\
\hline
\end{tabular}
}
\end{table}

\begin{table}[!tb]
\centering
\caption{\small
  Model uncertainties on the fit fractions and complex amplitudes.}
\label{tab:model-syst}
\resizebox{\textwidth}{!}{
\begin{tabular}{lccccc}
\hline
Resonance & Fit fraction (\%) & Real part & Imaginary part & Magnitude & Phase (radians)
\\
\hline \\ [-2.5ex]
$\Kstarb(892)^{0}$         & 0.88 & 0.72 & 0.33 & 0.03 & 0.76  \\
$\Kstarb(1410)^{0}$        & 1.37 & 0.15 & 0.22 & 0.14 & 1.09  \\
LASS nonresonant           & 4.09 & 0.14 & 0.18 & 0.11 & 0.26  \\
$\Kstarbsubz(1430)^{0}$    & 3.32 & 0.14 & 0.08 & 0.07 & 0.16  \\
\ \ \ LASS total          & 4.69 &  ---  &  ---  &  ---  &  ---   \\
$\Kstarbsubt(1430)^{0}$    & 1.06 & 0.26 & 0.03 & 0.05 & 0.65  \\
$\Kstarb(1680)^{0}$        & 0.80 & 0.14 & 0.20 & 0.11 & 2.66  \\
$\Kstarbsubz(1950)^{0}$    & 2.42 & 0.21 & 0.23 & 0.22 & 1.71  \\
$D^{*}_{s2}(2573)^-$    & 1.05 &  ---  &  ---  &  ---  &  ---   \\
$D^{*}_{s1}(2700)^-$    & 0.54 & 0.06 & 0.13 & 0.04 & 0.53  \\
$D^{*}_{s1}(2860)^-$    & 3.28 & 0.24 & 0.09 & 0.17 & 0.52  \\
$D^{*}_{s3}(2860)^-$    & 0.42 & 0.05 & 0.04 & 0.03 & 0.18  \\
\hline
Nonresonant               & 7.64 & 0.28 & 0.28 & 0.19 & 0.48  \\
\hline
$D^{*-}_{s\,v}$           & 4.02 & 0.18 & 0.17 & 0.16 & 0.43  \\
$D^{*}_{s0\,v}(2317)^-$ & 2.30 & 0.18 & 0.09 & 0.13 & 0.43  \\
$B^{*+}_{v}$              & 1.83 & 0.25 & 0.31 & 0.13 & 1.53  \\
\hline
\end{tabular}
}
\end{table}

\begin{table}[!tb]
\centering
\caption{\small
  Breakdown of model uncertainties on the fit fractions (\%).  The columns
  give the contributions from the different sources described in the text.
}
\label{tab:model-summary-FF}
\begin{tabular}{lccc@{\hspace{5mm}}c}
\hline
Resonance & Fixed      & Marginal   & Alternative & Total\\
          & parameters & components & models      & \\
\hline \\ [-2.5ex]
$\Kstarb(892)^{0}$         & 0.63 & 0.43 & 0.43 &0.88 \\
$\Kstarb(1410)^{0}$        & 0.37 & 0.47 & 1.23 &1.37 \\
LASS nonresonant           & 0.85 & 3.78 & 1.32 &4.09 \\
$\Kstarbsubz(1430)^{0}$    & 0.90 & 3.19 & 0.26 &3.32 \\
\ \ \ LASS total          & 0.73 & 2.62 & 3.82 &4.69 \\
$\Kstarbsubt(1430)^{0}$    & 0.21 & 0.21 & 1.01 &1.06 \\
$\Kstarb(1680)^{0}$        & 0.63 & 0.26 & 0.42 &0.80 \\
$\Kstarbsubz(1950)^{0}$    & 0.14 & 0.22 & 2.40 &2.42 \\
$D^{*}_{s2}(2573)^-$    & 0.50 & 0.26 & 0.88 &1.05 \\
$D^{*}_{s1}(2700)^-$    & 0.26 & 0.31 & 0.36 &0.54 \\
$D^{*}_{s1}(2860)^-$    & 0.57 & 1.80 & 2.67 &3.28 \\
$D^{*}_{s3}(2860)^-$    & 0.12 & 0.29 & 0.28 &0.42 \\
\hline
Nonresonant               & 0.72 & 5.55 & 5.20 &7.64 \\
\hline
$D^{*-}_{s\,v}$           & 1.35 & 2.04 & 3.19 &4.02 \\
$D^{*}_{s0\,v}(2317)^-$ & 0.55 & 1.38 & 1.76 &2.30 \\
$B^{*+}_{v}$               & 0.40 & 1.53 & 0.91 &1.83 \\
\hline
\end{tabular}
\end{table}

\begin{table}[!tb]
\centering
\caption{\small
  Breakdown of model uncertainties on the masses and widths.
  Units of \mevcc are implied.
  The columns give the contributions from the different sources
  described in the text.
}
\label{tab:model-summary-mG}
\begin{tabular}{l@{\hspace{5mm}}ccc@{\hspace{5mm}}c}
\hline
Mass      & Fixed      & Marginal   & Alternative & Total\\
Resonance & parameters & components & models      & \\
\hline
$D^{*}_{s2}(2573)^-$ &  0.03 & 0.10 & \phani0.15 & \phani0.18\\
$D^{*}_{s1}(2860)^-$ &  4.14 & 3.79 & 22.65 & 23.3\phani\\
$D^{*}_{s3}(2860)^-$ &  0.89 & 1.45 & \phani5.73 & \phani6.0\phani\\
\hline
\hline
Width     & Fixed      & Marginal   & Alternative & Total\\
Resonance & parameters & components & models      & \\
\hline
$D^{*}_{s2}(2573)^-$ & \phani0.16 & \phani0.18 & \phani0.37 & \phani0.4\\
$D^{*}_{s1}(2860)^-$ &      19.55 &      42.85 &      54.21 & 71.8\\
$D^{*}_{s3}(2860)^-$ & \phani0.81 & \phani3.27 & \phani5.52 & \phani6.5\\
\hline
\end{tabular}
\end{table}

The reduced \chisq value of 1.21 obtained by comparing the data and the default fit model in SDP bins, discussed in Sec.~\ref{sec:DalitzBaseline}, corresponds to a tiny $p$-value, given the large number of degrees of freedom.
Such a situation is not uncommon for high statistics Dalitz plot analyses, see \eg\ Refs.~\cite{Abe:2003zm,Aubert:2009wg}.
Moreover, the \chisq is evaluated accounting only for statistical uncertainties.
Some disagreement between the data and the fit model is visible in the helicity angle projections in the regions of the peaks with the largest statistics, namely the $\Kstarb(892)^0$ (Fig.~\ref{fig:coshelkpi}(b)) and the $D_{s2}^*(2573)^-$ (Fig.~\ref{fig:cosheldk}(b)) resonances.
The latter is also visible in Fig.~\ref{fig:zooms}(a) as the reflection from one lobe of the $D_{s2}^*(2573)^-$ structure overlaps with $\Kstarzb$ resonances in the $m(\Km\pip) \approx 1430 \mevcc$ region.
These regions correspond to bins with large pulls in Fig.~\ref{fig:pulls}.
The small peak in Fig.~\ref{fig:zooms}(d) at $m(\Dzb\Km) \approx 2.96 \gevcc$ is not statistically significant.

As seen in this section, both experimental systematic and model uncertainties are comparable in size to the statistical uncertainties on the parameters associated with those resonances, suggesting that these uncertainties may significantly affect the \chisq value.
In addition, certain aspects of the modelling, such as the description of the $\Km\pip$ and $\Dzb\Km$ S-waves, are known to be approximations.
The default model gives the best agreement with the data among the alternatives considered.
Nonetheless, the change in reduced \chisq value when alternative models are used, which is typically in the range 0.05--0.10, gives an estimate of how much the approximations used may affect the goodness-of-fit.
Therefore, the description of the data is considered to be acceptable.

A number of cross checks are performed to test the stability of the results.
The dataset is divided based on: the year of data-taking; the polarity of the magnet; the flavour ($\Bs$ or $\Bsb$) of the decaying particle; the hardware level trigger decision.
Each subset is fit separately, and no significant deviations are seen in the fit parameters.
To cross check the stability of the default amplitude model, a number of fits are performed with an additional resonance with fixed parameters included.
All values of mass, width and spin (up to 3), and all combinations of resonance daughters, are considered.
None of the additional resonances are found to contribute significantly.
\section{Results}
\label{sec:Results}

\begin{figure}[!tb]
\centering
\includegraphics[scale=0.50]{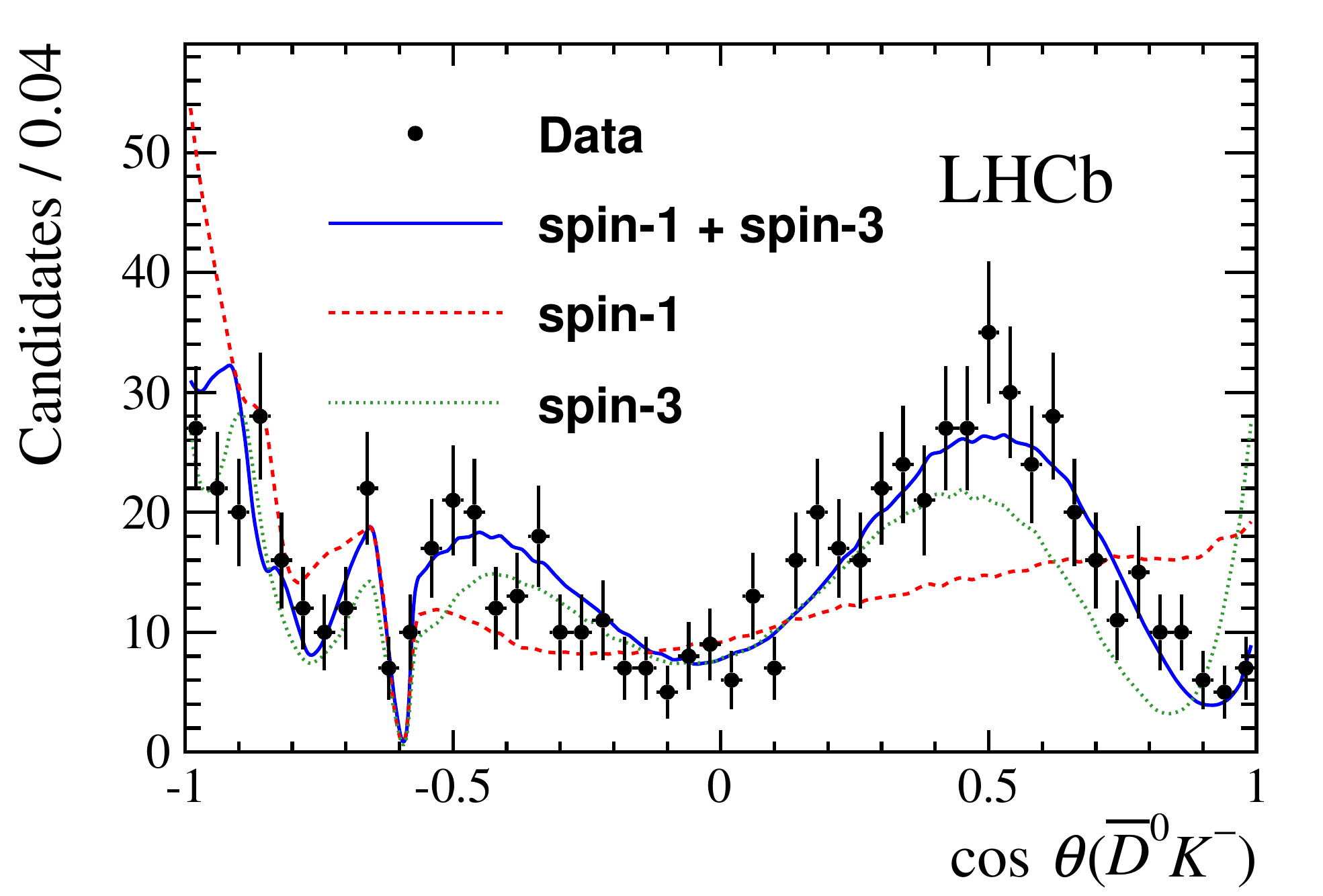}
\caption{\small
  Projections of the data and Dalitz plot fit results with alternative models
  onto the cosine of the helicity angle of the $\Dzb\Km$ system,
  $\cos\theta(\Dzb\Km)$, for $2.77 < m(\Dzb\Km) < 2.91\gevcc$.
  The data are shown as black points, the result of the baseline fit with both
  spin-1 and spin-3 resonances is given as a solid blue curve, and results of
  fits from the best models with only either a spin-1 or a spin-3 resonance
  are shown as dashed red and dotted green lines, respectively.
  The dip at $\cos\theta(\Dzb\Km) \approx -0.6$ is due to the $\Dzb$ veto.
  Comparison of the data and the different fit results in the 50 bins of this
  projection gives $\chisq$ values of 47.3, 214.0 and 150.0 for the default,
  spin-1 only and spin-3 only models, respectively.
}
\label{fig:dsjsep}
\end{figure}

As discussed in Sec.~\ref{sec:DalitzBaseline}, the data require both a spin-1 and a spin-3 resonance in the $m(\Dzb\Km) \approx 2.86 \gevcc$ region.
Figure~\ref{fig:dsjsep} shows the result of the baseline fit compared to
alternative models containing only a single resonance, either spin-1 or spin-3, in this region.
The expected angular distributions for different spin hypotheses are given in Eqs.~(\ref{eq:ZTFactors})-(\ref{eq:ZTFactors-end}).
As shown in Table~\ref{tab:DsJDNLL}, the changes in NLL relative to the baseline model are 156.8 and 136.5 for the spin-1 only and spin-3 only models, respectively.
The \chisq values in the 70 SDP bins discussed in Sec.~\ref{sec:DalitzBaseline} are 233 and 139 for the spin-1 only and spin-3 only, respectively.

\begin{figure}[!tb]
  \centering
  \includegraphics[width=0.45\textwidth]{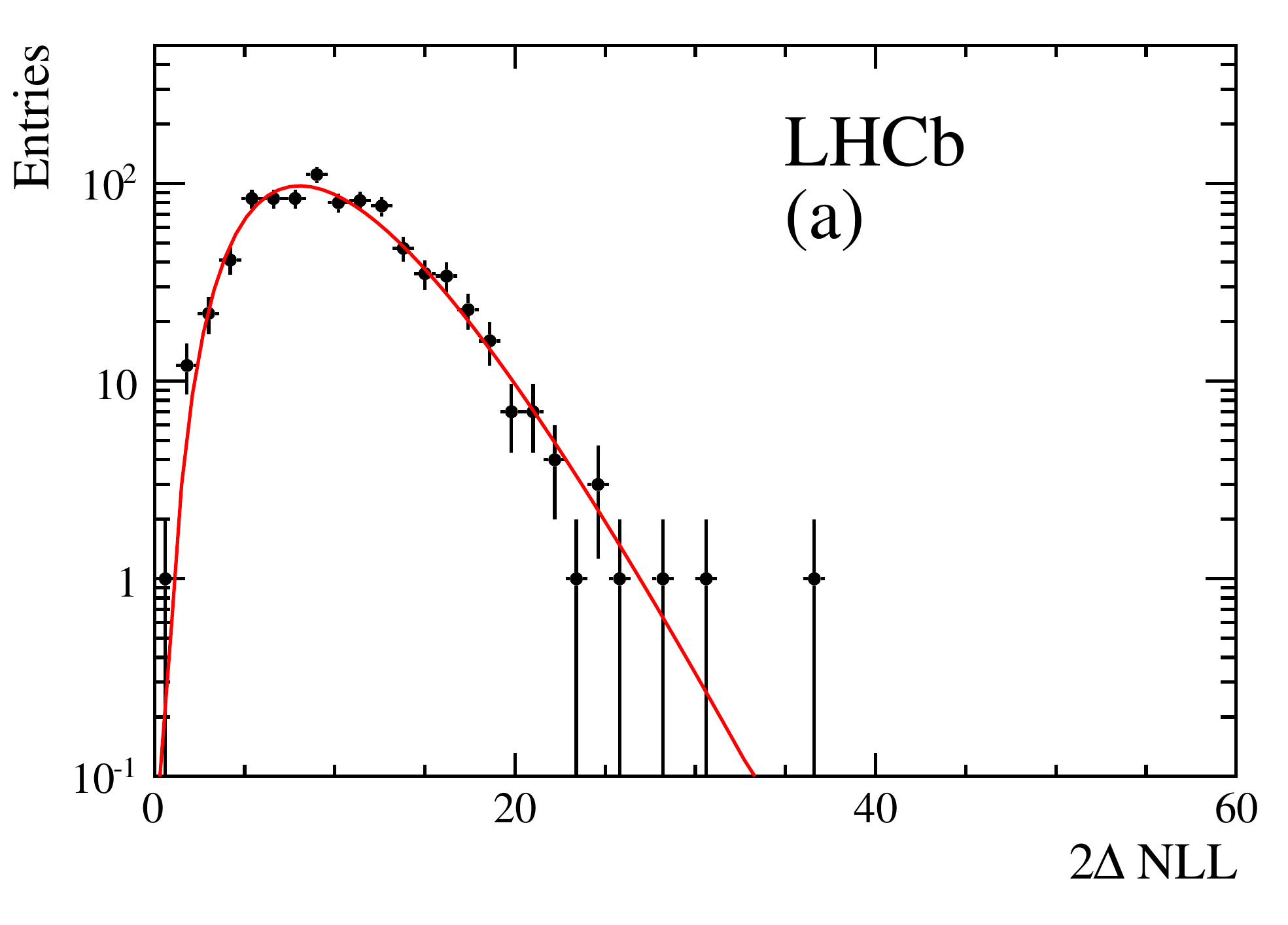}
  \includegraphics[width=0.45\textwidth]{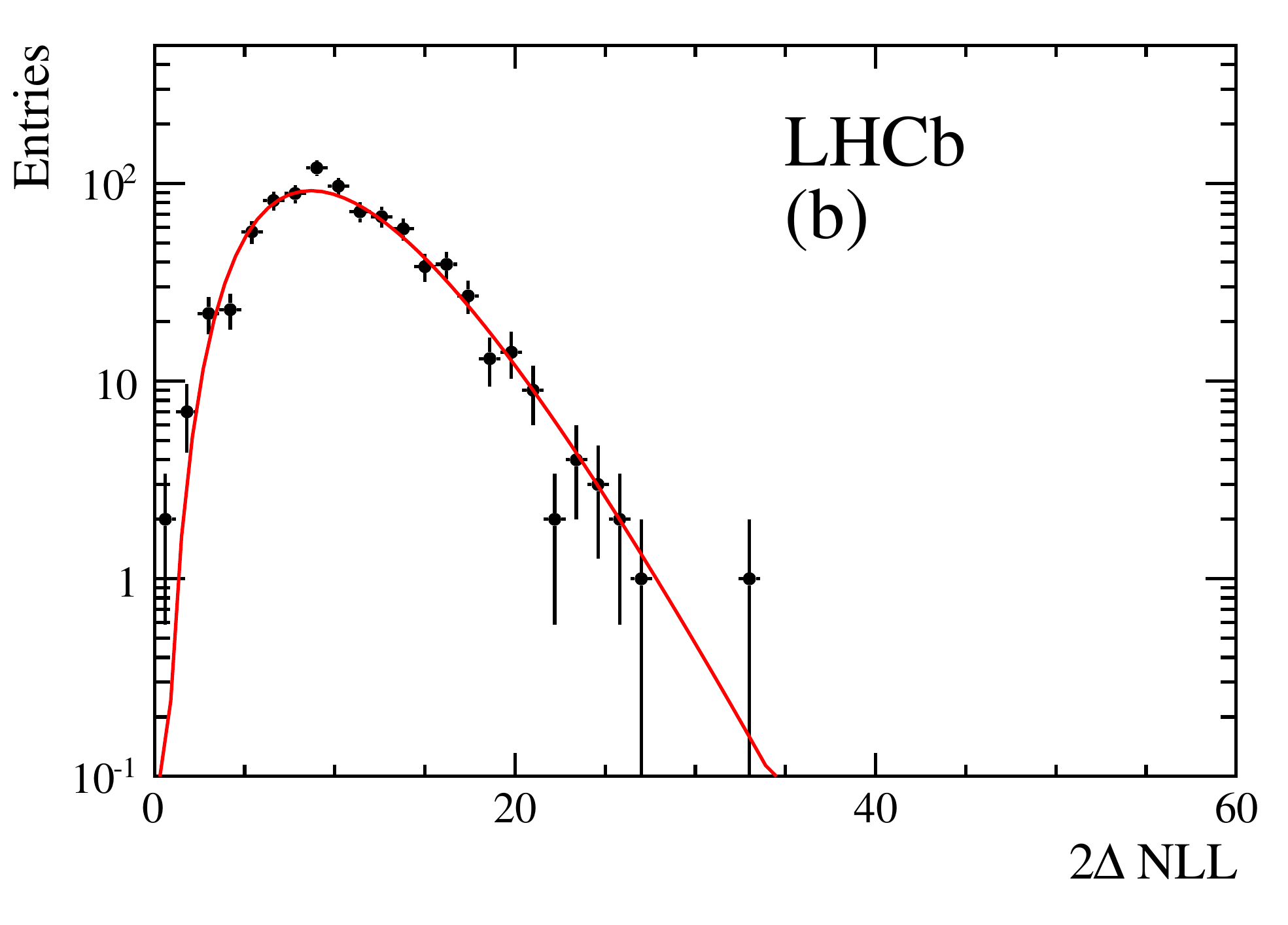}
  \caption{\small
    Fits of $\chi^2$ functions to the $2\Delta{\rm NLL}$ distributions obtained from
    fits to pseudoexperiments generated with (left) no $D_{s1}^*(2860)^-$ and
    (right) no $D_{s3}^*(2860)^-$ component.
    The corresponding $2\Delta{\rm NLL}$ values observed in data are 273 and
    314, respectively (see Table~\ref{tab:DsJDNLL}).
  }
  \label{fig:signifFits}
\end{figure}

To obtain a value for the significance of both states being present in the
data, ensembles of simulated pseudoexperiments are generated with parameters
corresponding to the best fit spin-1 only and spin-3 only models, and are
fitted with both resonances included.
The distributions of twice the difference in NLL ($2\Delta {\rm NLL}$) obtained from these ensembles, shown in Fig.~\ref{fig:signifFits}, are fitted with $\chi^2$ functions with the number of degrees of freedom floated.
The tails of the fitted functions are extrapolated to obtain the $p$-values to
find $2\Delta {\rm NLL}$ to be at least as large as the values seen in data.
These are found to correspond to $16$ and $15$ standard deviations for the spin-1 only and spin-3 only models, respectively.
Consistent values are obtained if only the tails of the distributions are fitted.
In addition $2\Delta {\rm NLL}$ distributions are constructed from an ensemble of simulated pseudoexperiments generated with the default model (containing both $D^{*}_{s1}(2860)^-$ and $D^{*}_{s3}(2860)^-$ resonances) fitted with either one or both resonances.
The values of $2\Delta {\rm NLL}$ observed in data are found to lie well within the bulk of the distributions with $p$-values of $24\,\%$ and $4\,\%$ for retaining the $D^{*}_{s1}(2860)^-$ and $D^{*}_{s3}(2860)^-$ resonances, respectively.

These significances include only statistical uncertainties, so the effect of the largest systematic uncertainties is tested by repeating the procedure with the variations in the models discussed in Sec.~\ref{sec:Systematics} that give the largest effects on the fit fractions, masses and widths of the $D_{sJ}^*(2860)^-$ states.
For the spin-1 only model, the effect of using the $\kappa$ model to describe the $\Km\pip$ S-wave is evaluated.
For the spin-3 only model, the $\kappa$ description of the $\Km\pip$ S-wave, the addition of the $\Kstarb_4(2045)^0$ state and the variation of the $\Dzb$ mass are considered.
The conclusion is that two states are required in this region with significance of at least $10$ standard deviations.

The masses and widths of these three states are determined to be
\begin{eqnarray*}
  m(D^{*}_{s2}(2573)^-)      & = & 2568.39 \pm 0.29 \pm 0.19 \pm 0.18 \mevcc \, ,\\
  \Gamma(D^{*}_{s2}(2573)^-) & = & 16.9    \pm 0.5  \pm 0.4  \pm 0.4  \mevcc \, ,\\
  m(D^{*}_{s1}(2860)^-)      & = &  2859   \pm 12   \pm 6    \pm 23   \mevcc \, ,\\
  \Gamma(D^{*}_{s1}(2860)^-) & = &  159    \pm 23   \pm 27   \pm 72   \mevcc \, ,\\
  m(D^{*}_{s3}(2860)^-)      & = &  2860.5 \pm 2.6  \pm 2.5  \pm 6.0  \mevcc \, ,\\
  \Gamma(D^{*}_{s3}(2860)^-) & = &  53     \pm 7    \pm 4    \pm 6    \mevcc \, ,
\end{eqnarray*}
where the first uncertainty is statistical, the second is due to experimental systematic effects and the third due to model variations.
The phase difference between the $D^{*}_{s1}(2860)^{-}$ and $D^{*}_{s3}(2860)^{-}$ amplitudes is consistent with $\pi$ within a large model uncertainty.
The results for the complex amplitudes, expressed both as real and imaginary
parts and as magnitudes and phases, are given in Table~\ref{tab:amplitude-results}.
The results for the fit fractions are given in Table~\ref{tab:fitfrac-results}, while results for the interference fit fractions are given in App.~\ref{app:IFF-results}.

\begin{table}[!tb]
\centering
\caption{\small
  Results for the complex amplitudes and their uncertainties.
  The three quoted errors are statistical, experimental systematic and
  model uncertainties, respectively.
  The central values and statistical uncertainties are as reported in Table~\ref{tab:cfitfrac}, while the experimental and model systematic uncertainties are as reported in Tables~\ref{tab:expt-syst} and~\ref{tab:model-syst}.
}
\label{tab:amplitude-results}
\resizebox{\textwidth}{!}{
\begin{tabular}{lc@{$\pm$}c@{$\pm$}c@{$\pm$}cc@{$\pm$}c@{$\pm$}c@{$\pm$}cc@{$\pm$}c@{$\pm$}c@{$\pm$}cc@{$\pm$}c@{$\pm$}c@{$\pm$}c}
\hline
Resonance & \multicolumn{4}{c}{Real part} & \multicolumn{4}{c}{Imaginary part} & \multicolumn{4}{c}{Magnitude} & \multicolumn{4}{c}{Phase (radians)} \\
\hline \\ [-2.5ex]
$\Kstarb(892)^{0}$       & $          -0.75$ & $0.08$ &0.16 &0.72 & $\phantom{-}0.74$ & $0.08$ &0.13 &0.33 & $1.06$ & 0.02 &0.03 &0.03 & $\phantom{-}2.36$ & 0.13 &0.20 &0.76 \\
$\Kstarb(1410)^{0}$      & $          -0.25$ & $0.03$ &0.02 &0.15 & $          -0.04$ & $0.05$ &0.12 &0.22 & $0.25$ & 0.04 &0.02 &0.14 & $          -2.96$ & 0.21 &0.50 &1.09 \\
LASS nonresonant         & $          -0.43$ & $0.09$ &0.16 &0.14 & $\phantom{-}0.59$ & $0.06$ &0.06 &0.18 & $0.73$ & 0.06 &0.05 &0.11 & $\phantom{-}2.19$ & 0.16 &0.26 &0.26 \\
$\Kstarbsubz(1430)^{0}$ & $          -0.49$ & $0.10$ &0.22 &0.14 & $\phantom{-}0.73$ & $0.07$ &0.07 &0.08 & $0.88$ & 0.04 &0.03 &0.07 & $\phantom{-}2.16$ & 0.20 &0.25 &0.16 \\
$\Kstarbsubt(1430)^{0}$ & $\phantom{-}0.09$ & $0.05$ &0.08 &0.26 & $          -0.37$ & $0.03$ &0.02 &0.03 & $0.38$ & 0.03 &0.02 &0.05 & $          -1.34$ & 0.10 &0.20 &0.65 \\
$\Kstarb(1680)^{0}$      & $          -0.08$ & $0.04$ &0.06 &0.14 & $\phantom{-}0.12$ & $0.04$ &0.02 &0.20 & $0.14$ & 0.06 &0.04 &0.11 & $\phantom{-}2.16$ & 0.26 &0.32 &2.66 \\
$\Kstarbsubz(1950)^{0}$ & $\phantom{-}0.11$ & $0.03$ &0.03 &0.21 & $          -0.01$ & $0.04$ &0.04 &0.23 & $0.11$ & 0.04 &0.03 &0.22 & $          -0.09$ & 0.41 &0.32 &1.71 \\
$D^{*}_{s2}(2573)^-$    & \multicolumn{4}{c}{1.00} &                 \multicolumn{4}{c}{0.00} &                 \multicolumn{4}{c}{1.00} &  \multicolumn{4}{c}{0.00} \\
$D^{*}_{s1}(2700)^-$    & $          -0.22$ & $0.04$ &0.02 &0.06 & $          -0.13$ & $0.04$ &0.06 &0.13 & $0.25$ & 0.04 &0.03 &0.04 & $          -2.61$ & 0.17 &0.18 &0.53 \\
$D^{*}_{s1}(2860)^-$    & $          -0.41$ & $0.05$ &0.05 &0.24 & $\phantom{-}0.16$ & $0.06$ &0.05 &0.09 & $0.44$ & 0.05 &0.03 &0.17 & $\phantom{-}2.78$ & 0.20 &0.12 &0.52 \\
$D^{*}_{s3}(2860)^-$    & $\phantom{-}0.27$ & $0.02$ &0.03 &0.05 & $          -0.12$ & $0.03$ &0.02 &0.04 & $0.29$ & 0.02 &0.02 &0.03 & $          -0.42$ & 0.07 &0.10 &0.18 \\
\hline
Nonresonant             & $\phantom{-}0.58$ & $0.07$ &0.25 &0.28 & $          -0.39$ & $0.06$ &0.04 &0.28 & $0.70$ & 0.08 &0.15 &0.19 & $          -0.59$ & 0.10 &0.36 &0.48 \\
\hline
$D^{*-}_{s\,v}$         & $\phantom{-}0.36$ & $0.04$ &0.04 &0.18 & $\phantom{-}0.23$ & $0.05$ &0.05 &0.17 & $0.43$ & 0.05 &0.05 &0.16 & $\phantom{-}0.57$ & 0.12 &0.08 &0.43 \\
$D^{*}_{s0\,v}(2317)^-$ & $\phantom{-}0.18$ & $0.08$ &0.22 &0.18 & $\phantom{-}0.24$ & $0.04$ &0.05 &0.09 & $0.30$ & 0.06 &0.16 &0.13 & $\phantom{-}0.91$ & 0.21 &0.72 &0.43 \\
$B^{*+}_{v}$            & $          -0.09$ & $0.10$ &0.08 &0.25 & $          -0.26$ & $0.05$ &0.11 &0.31 & $0.27$ & 0.09 &0.06 &0.13 & $          -1.90$ & 0.40 &0.34 &1.53 \\
\hline
\end{tabular}
}
\end{table}

For resonances without a significant signal, it is possible to set upper
limits on their fit fractions, and therefore on their branching fractions.
This is done for the $\Kstarb(1680)^{0}$, $\Kstarbsubz(1950)^{0}$, $D^{*}_{s0\,v}(2317)^{-}$ and $B^{*+}_{v}$ components of the default model, as well as for the $\Kstarb_3(1780)^0$ and $\Kstarb_4(2045)^0$ states.
The values of $2\,{\rm NLL}$ as functions of the fit fractions are obtained, and converted into likelihood functions.
The effect of systematic uncertainties is included by convolving the likelihood function with a Gaussian of width given by the systematic uncertainty.
These are then used to set 90\,\% and 95\,\% confidence level (CL) upper limits by integrating the likelihood.
The upper limits obtained with this procedure are included in Table~\ref{tab:fitfrac-results}.

\begin{table}[!tb]
\centering
\caption{\small
  Results for the fit fractions and their uncertainties (\%).
  The three quoted errors are statistical, experimental systematic and
  model uncertainties, respectively.
  Upper limits at both 90\,\% and 95\,\% confidence level (CL) are given for components that are not significant.
  The central values and statistical uncertainties are as reported in Table~\ref{tab:cfitfrac}, while the experimental and model systematic uncertainties are as reported in Tables~\ref{tab:expt-syst} and~\ref{tab:model-syst}.
}
\label{tab:fitfrac-results}
\begin{tabular}{lc@{$\,\pm\,$}c@{$\,\pm\,$}c@{$\,\pm\,$}ccc}
\hline
Resonance & \multicolumn{4}{c}{Fit fraction}  & \multicolumn{2}{c}{Upper limits} \\
          & \multicolumn{4}{c}{}              & 90\,\% CL & 95\,\% CL \\
\hline \\ [-2.5ex]
$\Kstarb(892)^{0}$         & 28.6 & 0.6 &0.7 &0.9 \\
$\Kstarb(1410)^{0}$        & \phani1.7 & 0.5 &0.2 &1.4 \\
LASS nonresonant          & 13.7 & 2.5 &1.5 &4.1 \\
$\Kstarbsubz(1430)^{0}$   & 20.0 & 1.6 &0.7 &3.3 \\
\ \ \ LASS total          & 21.4 & 1.4 &1.0 &4.7 \\
$\Kstarbsubt(1430)^{0}$    & \phani3.7 & 0.6 &0.4 &1.1 \\
$\Kstarb(1680)^{0}$        & \phani0.5 & 0.4 &0.3 &0.8 & $<2.0$ & $<2.4$ \\
$\Kstarbsubz(1950)^{0}$    & \phani0.3 & 0.2 &0.1 &2.4 & $<3.7$ & $<4.1$ \\
$\Kstarb_3(1780)^0$        & \multicolumn{4}{c}{---}   & $<0.33$ & $<0.38$ \\
$\Kstarb_4(2045)^0$        & \multicolumn{4}{c}{---}   & $<0.21$ & $<0.24$ \\
$D^{*}_{s2}(2573)^-$    & 25.7 & 0.7 &0.8 &1.1 \\
$D^{*}_{s1}(2700)^-$    & \phani1.6 & 0.4 &0.4 &0.5 \\
$D^{*}_{s1}(2860)^-$    & \phani5.0 & 1.2 &0.7 &3.3 \\
$D^{*}_{s3}(2860)^-$    & \phani2.2 & 0.1 &0.3 &0.4 \\
\hline
Nonresonant               & 12.4 & 2.7 &4.3 &7.6 \\
\hline
$D^{*-}_{s\,v}$           & \phani4.7 & 1.4 &1.1 &4.0 \\
$D^{*}_{s0\,v}(2317)^-$ & \phani2.3 & 1.1 &1.9 &2.3 & $<7.2$ & $<8.4$ \\
$B^{*+}_{v}$              & \phani1.9 & 1.2 &1.1 &1.8 & $<7.7$ & $<8.7$ \\
\hline
\end{tabular}
\end{table}

The fit fractions of the resonant components are converted into quasi-two-body branching fractions by multiplying by the previously measured value ${\cal B}(\Bs \to \Dzb\Km\pip) = \left(1.00\pm0.04\stat\pm0.10\syst\pm0.10\,({\cal B}) \right)\times10^{-3}$~\cite{LHCb-PAPER-2013-022}, where the third uncertainty is due to the knowledge of the branching fraction of the $\Bz \to \Dzb\pip\pim$ normalisation channel~\cite{PDG2012}.
For resonances where the subdecay branching fraction is known~\cite{PDG2012}
the product branching fraction can be converted into the $B$ decay branching fraction.
These results are given in Table~\ref{tab:BFresults}.

\begin{table}[!tb]
\centering
\caption{\small
  Results for the product branching fractions
  (top) ${\cal B}(\Bs \to \Dzb\Kstarzb)\times{\cal B}(\Kstarzb \to \Km\pip)$ and
  (bottom) ${\cal B}(\Bs \to D_s^{*-}\pip)\times{\cal B}(D_s^{*-} \to \Dzb\Km)$,
  for each $\Kstarzb$ and $D_s^{*-}$ resonance.
  For the $\Kstarzb$ resonances, where ${\cal B}(\Kstarzb \to \Km\pip)$ is known~\cite{PDG2012}, the $\Bs$ decay branching fraction is also given.
  The four quoted uncertainties are statistical, experimental systematic, model and PDG uncertainties, respectively.
  Upper limits are given at 90\,\% (95\,\%) confidence level.
}
\label{tab:BFresults}
\begin{tabular}{l r@{$\,\pm\,$}c@{$\,\pm\,$}c@{$\,\pm\,$}c@{$\,\pm\,$}l r@{$\,\pm\,$}c@{$\,\pm\,$}c@{$\,\pm\,$}c@{$\,\pm\,$}l}
\hline
Resonance & \multicolumn{5}{c}{Product branching fraction} & \multicolumn{5}{c}{Branching fraction} \\
              & \multicolumn{5}{c}{($10^{-5}$)}                & \multicolumn{5}{c}{ ($10^{-4}$)} \\
\hline \\ [-2.5ex]
$\Kstarb(892)^{0}$      & 28.6 & 0.6 &0.7 &0.9 & 4.2 & 4.29 & 0.09 & 0.11 & 0.14 & 0.63 \\
$\Kstarb(1410)^{0}$     &  1.7 & 0.5 &0.2 &1.4 & 0.2 & 3.86 & 1.14 & 0.45 & 3.18 & 0.89 \\
LASS nonresonant        & 13.7 & 2.5 &1.5 &4.1 & 2.0 & 2.06 & 0.38 & 0.23 & 0.62 & 0.30 \\
$\Kstarbsubz(1430)^{0}$ & 20.0 & 1.6 &0.7 &3.3 & 2.9 & 3.00 & 0.24 & 0.11 & 0.50 & 0.44 \\
\ \ \ LASS total        & 21.4 & 1.4 &1.0 &4.7 & 3.1 & 3.21 & 0.21 & 0.15 & 0.71 & 0.47 \\
$\Kstarbsubt(1430)^{0}$ &  3.7 & 0.6 &0.4 &1.1 & 0.5 & 1.11 & 0.18 & 0.12 & 0.33 & 0.15 \\
$\Kstarb(1680)^{0}$     & \multicolumn{5}{c}{$<2.0~(2.4)$}   & \multicolumn{5}{c}{$<0.78~(0.93)$} \\
$\Kstarbsubz(1950)^0$   & \multicolumn{5}{c}{$<3.7~(4.1)$}   & \multicolumn{5}{c}{$<1.1~(1.2)$} \\
$\Kstarb_3(1780)^0$     & \multicolumn{5}{c}{$<0.33~(0.38)$} & \multicolumn{5}{c}{$<0.26~(0.30)$} \\
$\Kstarb_4(2045)^0$     & \multicolumn{5}{c}{$<0.21~(0.24)$} & \multicolumn{5}{c}{$<0.31~(0.36)$} \\
\hline
$D^{*}_{s2}(2573)^-$ & 25.7 & 0.7 &0.8 &1.1 & 3.8 \\
$D^{*}_{s1}(2700)^-$ &  1.6 & 0.4 &0.4 &0.5 & 0.2 \\
$D^{*}_{s1}(2860)^-$ &  5.0 & 1.2 &0.7 &3.3 & 0.7 \\
$D^{*}_{s3}(2860)^-$ &  2.2 & 0.1 &0.3 &0.4 & 0.3 \\
\hline
\end{tabular}
\end{table}

\section{Summary}
\label{sec:Summary}

The first amplitude analysis of the $\Bs \to \Dzb \Km \pip$ decay has been presented.
The $\Bs \to \Dzb \Km \pip$ decay amplitude model contains a total of fourteen components: six $\Km\pip$ resonances, four $\Dzb\Km$ resonances, three virtual resonances and a nonresonant contribution.
The complex amplitudes of these are determined, and fit fractions and interference fit fractions are reported in addition, to enable convention-independent comparisons of the model.
The fit fraction results are converted into branching fraction measurements.
The result for ${\cal B}(\Bs\to\Dzb\Kstarb(892)^{0})$ is significantly more precise than the previous measurement~\cite{LHCb-PAPER-2011-008}, which was obtained from a much smaller and statistically independent data sample collected by LHCb during 2010.
All other branching fraction results are first reported measurements.

A structure at $m(\Dzb\Km) \approx 2.86 \gevcc$ is found to be an admixture of a spin-1 and a spin-3 resonance with a significance of at least 10 standard deviations.
Therefore the $D^{*}_{sJ}(2860)^{-}$ state previously observed by the BaBar collaboration in inclusive $e^+e^- \to \Dzb\Km X$ production~\cite{Aubert:2009ah} and by the LHCb collaboration in $pp \to \Dzb\Km X$ processes~\cite{LHCb-PAPER-2012-016} consists of at least these two resonances.
The properties of those states and of the $D^{*}_{s2}(2573)^{-}$ resonance are measured.

The spin of the $D^{*}_{s2}(2573)^{-}$ resonance is experimentally determined for the first time, and is confirmed to be 2.
The mass and width of this state are determined with significantly better precision than previous measurements~\cite{PDG2012}.
The result for the width is consistent with the previous world average.
The result for the mass, however, is somewhat below the previous average, which is dominated by a measurement by the BaBar collaboration~\cite{Aubert:2006mh} based on inclusive production in $\epem$ collisions.
The Dalitz plot analysis technique used in this paper ensures that the background under the $D^{*}_{s2}(2573)^{-}$ peak is small and does not contain large contributions from decays of higher \Dsm resonances, resulting in much lower systematic uncertainties on the measured parameters compared to the inclusive approach.

The masses of the $D_{s1}^*(2860)^-$ and $D_{s3}^*(2860)^-$ states are found to be consistent within uncertainties, while a larger width of the spin-1 state than of the spin-3 state is preferred.
These results appear to support an interpretation of these states being the $J^P = 1^-$ and $3^-$ members of the 1D family, though the $1^-$ state may be partially mixed with the vector member of the 2S family to give the physical $D_{s1}^*(2700)^-$ and $D_{s1}^*(2860)^-$ states.
The discovery of the $D_{s3}^*(2860)^-$ resonance represents the first observation of a heavy flavoured spin-3 particle, and the first time that a spin-3 state is seen to be produced in $B$ decays.
This discovery demonstrates that 1D charm resonances can be investigated experimentally, and therefore opens a new window for potential studies of the spectroscopy of heavy flavoured mesons.

\section*{Acknowledgements}

\noindent We express our gratitude to our colleagues in the CERN
accelerator departments for the excellent performance of the LHC. We
thank the technical and administrative staff at the LHCb
institutes. We acknowledge support from CERN and from the national
agencies: CAPES, CNPq, FAPERJ and FINEP (Brazil); NSFC (China);
CNRS/IN2P3 (France); BMBF, DFG, HGF and MPG (Germany); SFI (Ireland); INFN (Italy);
FOM and NWO (The Netherlands); MNiSW and NCN (Poland); MEN/IFA (Romania);
MinES and FANO (Russia); MinECo (Spain); SNSF and SER (Switzerland);
NASU (Ukraine); STFC (United Kingdom); NSF (USA).
The Tier1 computing centres are supported by IN2P3 (France), KIT and BMBF
(Germany), INFN (Italy), NWO and SURF (The Netherlands), PIC (Spain), GridPP
(United Kingdom).
We are indebted to the communities behind the multiple open
source software packages on which we depend. We are also thankful for the
computing resources and the access to software R\&D tools provided by Yandex LLC (Russia).
Individual groups or members have received support from
EPLANET, Marie Sk\l{}odowska-Curie Actions and ERC (European Union),
Conseil g\'{e}n\'{e}ral de Haute-Savoie, Labex ENIGMASS and OCEVU,
R\'{e}gion Auvergne (France), RFBR (Russia), XuntaGal and GENCAT (Spain), Royal Society and Royal
Commission for the Exhibition of 1851 (United Kingdom).

\ifx\mcitethebibliography\mciteundefinedmacro
\PackageError{LHCb.bst}{mciteplus.sty has not been loaded}
{This bibstyle requires the use of the mciteplus package.}\fi
\providecommand{\href}[2]{#2}

\newpage

\appendix
\section{Results for interference fit fractions}
\label{app:IFF-results}

The central values of the interference fit fractions are given in Table~\ref{tab:interferencefrac}.
The statistical, experimental systematic and model uncertainties on these quantities are given in Tables~\ref{tab:IFF-statErrs},~\ref{tab:expt-systIFF} and~\ref{tab:model-systIFF}, respectively.

\begin{table}[!hb]
\centering
\caption{\small
  Interference fit fractions (\%) from the nominal Dalitz plot fit. The amplitudes are: ($A_{0}$) $\Kstarb(892)^{0}$, ($A_{1}$) $\Kstarb(1410)^{0}$,
  ($A_{2}$) $\Kstarbsubz(1430)^{0}$, ($A_{3}$) LASS nonresonant, ($A_{4}$) $\Kstarbsubt(1430)^{0}$, ($A_{5}$) $\Kstarb(1680)^{0}$, ($A_{6}$) $\Kstarbsubz(1950)^{0}$,
  ($A_{7}$) $D^{*-}_{s\,v}$, ($A_{8}$) $D^{*}_{s0\,v}(2317)^-$, ($A_{9}$) $D^{*}_{s2}(2573)^-$, ($A_{10}$) $D^{*}_{s1}(2700)^-$, ($A_{11}$) $D^{*}_{s3}(2860)^-$,
  ($A_{12}$) $D^{*}_{s1}(2860)^-$, ($A_{13}$) $B^{*+}_{v}$, ($A_{14}$) Nonresonant.
  The diagonal elements correspond to the fit fractions shown in Table~\ref{tab:cfitfrac}.
}
\label{tab:interferencefrac}
\resizebox{\textwidth}{!}{
\begin{tabular}{lccccccccccccccc}
\hline
          &           $A_{0}$  &           $A_{1}$  &           $A_{2}$  &           $A_{3}$  &           $A_{4}$  &           $A_{5}$  &           $A_{6}$  &           $A_{7}$  &           $A_{8}$  &           $A_{9}$  &           $A_{10}$ &           $A_{11}$ &           $A_{12}$ &           $A_{13}$ &           $A_{14}$ \\
\hline
$A_{0}$   & $\phantom{-}28.6$  & $ \phantom{-}2.2$  & $           -0.0$  & $           -0.0$  & $           -0.0$  & $           -0.6$  & $ \phantom{-}0.0$  & $           -0.4$  & $           -0.3$  & $ \phantom{-}0.5$  & $           -0.3$  & $ \phantom{-}0.3$  & $           -1.2$  & $           -0.8$  & $           -3.5$  \\
$A_{1}$   &                    & $ \phantom{-}1.7$  & $ \phantom{-}0.0$  & $ \phantom{-}0.0$  & $           -0.0$  & $           -0.6$  & $ \phantom{-}0.0$  & $           -0.0$  & $ \phantom{-}0.2$  & $ \phantom{-}0.3$  & $           -0.3$  & $ \phantom{-}0.1$  & $           -0.6$  & $           -0.5$  & $           -2.2$  \\
$A_{2}$   &                    &                    & $\phantom{-}20.0$  & $          -12.3$  & $           -0.0$  & $           -0.0$  & $ \phantom{-}2.1$  & $           -2.9$  & $           -2.2$  & $           -1.4$  & $ \phantom{-}0.7$  & $           -0.4$  & $ \phantom{-}0.6$  & $           -3.2$  & $ \phantom{-}0.0$  \\
$A_{3}$   &                    &                    &                    & $\phantom{-}13.7$  & $ \phantom{-}0.0$  & $           -0.0$  & $           -1.5$  & $ \phantom{-}6.1$  & $ \phantom{-}1.8$  & $ \phantom{-}2.1$  & $           -1.5$  & $ \phantom{-}0.0$  & $           -0.5$  & $ \phantom{-}2.5$  & $           -2.5$  \\
$A_{4}$   &                    &                    &                    &                    & $ \phantom{-}3.7$  & $           -0.0$  & $           -0.0$  & $           -0.6$  & $           -0.6$  & $           -0.3$  & $           -0.1$  & $           -0.1$  & $           -0.5$  & $           -0.1$  & $           -0.3$  \\
$A_{5}$   &                    &                    &                    &                    &                    & $ \phantom{-}0.5$  & $           -0.0$  & $           -0.9$  & $           -0.5$  & $           -0.5$  & $ \phantom{-}0.1$  & $           -0.1$  & $           -0.2$  & $           -0.2$  & $           -0.3$  \\
$A_{6}$   &                    &                    &                    &                    &                    &                    & $ \phantom{-}0.3$  & $           -0.6$  & $           -0.2$  & $           -0.3$  & $ \phantom{-}0.1$  & $           -0.1$  & $           -0.1$  & $           -0.1$  & $           -0.7$  \\
$A_{7}$   &                    &                    &                    &                    &                    &                    &                    & $ \phantom{-}4.7$  & $           -0.0$  & $           -0.0$  & $           -0.6$  & $           -0.0$  & $ \phantom{-}3.0$  & $ \phantom{-}0.3$  & $ \phantom{-}0.0$  \\
$A_{8}$   &                    &                    &                    &                    &                    &                    &                    &                    & $ \phantom{-}2.3$  & $ \phantom{-}0.0$  & $           -0.0$  & $ \phantom{-}0.0$  & $           -0.0$  & $ \phantom{-}0.1$  & $           -0.7$  \\
$A_{9}$   &                    &                    &                    &                    &                    &                    &                    &                    &                    & $\phantom{-}25.7$  & $           -0.0$  & $ \phantom{-}0.0$  & $           -0.0$  & $ \phantom{-}0.4$  & $ \phantom{-}0.0$  \\
$A_{10}$  &                    &                    &                    &                    &                    &                    &                    &                    &                    &                    & $ \phantom{-}1.6$  & $           -0.0$  & $           -0.9$  & $           -0.6$  & $ \phantom{-}0.0$  \\
$A_{11}$  &                    &                    &                    &                    &                    &                    &                    &                    &                    &                    &                    & $ \phantom{-}2.2$  & $           -0.0$  & $ \phantom{-}0.0$  & $           -0.0$  \\
$A_{12}$  &                    &                    &                    &                    &                    &                    &                    &                    &                    &                    &                    &                    & $ \phantom{-}5.0$  & $           -1.6$  & $ \phantom{-}0.0$  \\
$A_{13}$  &                    &                    &                    &                    &                    &                    &                    &                    &                    &                    &                    &                    &                    & $ \phantom{-}1.9$  & $ \phantom{-}3.7$  \\
$A_{14}$  &                    &                    &                    &                    &                    &                    &                    &                    &                    &                    &                    &                    &                    &                    & $\phantom{-}12.4$  \\
\hline
\end{tabular}
}
\end{table}

\begin{table}[!tb]
\centering
\caption{\small
  Absolute statistical uncertainties on the interference fit fractions (\%) from the Dalitz plot fit.
  The amplitudes are: ($A_{0}$) $\Kstarb(892)^{0}$, ($A_{1}$) $\Kstarb(1410)^{0}$,
  ($A_{2}$) $\Kstarbsubz(1430)^{0}$, ($A_{3}$) LASS nonresonant, ($A_{4}$) $\Kstarbsubt(1430)^{0}$, ($A_{5}$) $\Kstarb(1680)^{0}$, ($A_{6}$) $\Kstarbsubz(1950)^{0}$,
  ($A_{7}$) $D^{*-}_{s\,v}$, ($A_{8}$) $D^{*}_{s0\,v}(2317)^-$, ($A_{9}$) $D^{*}_{s2}(2573)^-$, ($A_{10}$) $D^{*}_{s1}(2700)^-$, ($A_{11}$) $D^{*}_{s3}(2860)^-$,
  ($A_{12}$) $D^{*}_{s1}(2860)^-$, ($A_{13}$) $B^{*+}_{v}$, ($A_{14}$) Nonresonant.
  The diagonal elements correspond to the statistical uncertainties on the fit fractions shown in Table~\ref{tab:cfitfrac}.
}
\label{tab:IFF-statErrs}
\resizebox{\textwidth}{!}{
\begin{tabular}{lccccccccccccccc}
\hline
          & $A_{0}$  & $A_{1}$  & $A_{2}$  & $A_{3}$  & $A_{4}$  & $A_{5}$  & $A_{6}$  & $A_{7}$  & $A_{8}$  & $A_{9}$  & $A_{10}$ & $A_{11}$ & $A_{12}$ & $A_{13}$ & $A_{14}$ \\
\hline
$A_{0}$   & 0.6 & 0.4 & 0.0 & 0.0 & 0.0 & 0.3 & 0.0 & 0.5 & 0.3 & 0.2 & 0.1 & 0.1 & 0.1 & 0.3 & 0.4  \\
$A_{1}$   &     & 0.5 & 0.0 & 0.0 & 0.0 & 0.4 & 0.0 & 0.4 & 0.2 & 0.1 & 0.1 & 0.0 & 0.1 & 0.2 & 0.4  \\
$A_{2}$   &     &     & 1.6 & 1.2 & 0.0 & 0.0 & 0.6 & 0.5 & 0.6 & 0.2 & 0.1 & 0.1 & 0.2 & 1.3 & 0.8  \\
$A_{3}$   &     &     &     & 2.5 & 0.0 & 0.0 & 0.6 & 1.0 & 0.6 & 0.3 & 0.3 & 0.0 & 0.5 & 1.5 & 0.7  \\
$A_{4}$   &     &     &     &     & 0.6 & 0.0 & 0.0 & 0.1 & 0.1 & 0.1 & 0.1 & 0.0 & 0.1 & 0.1 & 0.3  \\
$A_{5}$   &     &     &     &     &     & 0.4 & 0.0 & 0.3 & 0.1 & 0.2 & 0.1 & 0.1 & 0.2 & 0.2 & 0.3  \\
$A_{6}$   &     &     &     &     &     &     & 0.2 & 0.3 & 0.1 & 0.1 & 0.0 & 0.0 & 0.1 & 0.2 & 0.3  \\
$A_{7}$   &     &     &     &     &     &     &     & 1.4 & 0.0 & 0.0 & 0.6 & 0.0 & 0.8 & 1.5 & 0.0  \\
$A_{8}$   &     &     &     &     &     &     &     &     & 1.1 & 0.0 & 0.0 & 0.0 & 0.0 & 0.5 & 2.9  \\
$A_{9}$   &     &     &     &     &     &     &     &     &     & 0.6 & 0.0 & 0.0 & 0.0 & 0.2 & 0.0  \\
$A_{10}$  &     &     &     &     &     &     &     &     &     &     & 0.4 & 0.0 & 0.5 & 0.3 & 0.0  \\
$A_{11}$  &     &     &     &     &     &     &     &     &     &     &     & 0.1 & 0.0 & 0.0 & 0.0  \\
$A_{12}$  &     &     &     &     &     &     &     &     &     &     &     &     & 1.2 & 0.6 & 0.0  \\
$A_{13}$  &     &     &     &     &     &     &     &     &     &     &     &     &     & 1.2 & 1.0  \\
$A_{14}$  &     &     &     &     &     &     &     &     &     &     &     &     &     &     & 2.7  \\
\hline
\end{tabular}
}
\end{table}

\begin{table}[!tb]
\centering
\caption{\small
  Absolute experimental systematic uncertainties on the interference fit fractions (\%).
  The amplitudes are: ($A_{0}$) $\Kstarb(892)^{0}$, ($A_{1}$) $\Kstarb(1410)^{0}$,
  ($A_{2}$) $\Kstarbsubz(1430)^{0}$, ($A_{3}$) LASS nonresonant, ($A_{4}$) $\Kstarbsubt(1430)^{0}$, ($A_{5}$) $\Kstarb(1680)^{0}$, ($A_{6}$) $\Kstarbsubz(1950)^{0}$,
  ($A_{7}$) $D^{*-}_{s\,v}$, ($A_{8}$) $D^{*}_{s0\,v}(2317)^-$, ($A_{9}$) $D^{*}_{s2}(2573)^-$, ($A_{10}$) $D^{*}_{s1}(2700)^-$, ($A_{11}$) $D^{*}_{s3}(2860)^-$,
  ($A_{12}$) $D^{*}_{s1}(2860)^-$, ($A_{13}$) $B^{*+}_{v}$, ($A_{14}$) Nonresonant.
  The diagonal elements correspond to the experimental systematic
  uncertainties on the fit fractions shown in Table~\ref{tab:expt-syst}.
}
\label{tab:expt-systIFF}
\resizebox{\textwidth}{!}{
\begin{tabular}{lccccccccccccccc}
\hline
          &           $A_{0}$  &           $A_{1}$  &           $A_{2}$  &           $A_{3}$  &           $A_{4}$  &           $A_{5}$  &           $A_{6}$  &           $A_{7}$  &           $A_{8}$  &           $A_{9}$  &           $A_{10}$ &           $A_{11}$ &           $A_{12}$ &           $A_{13}$ &           $A_{14}$ \\
\hline
$A_{0}$   & 0.74 & 0.60 & 0.00 & 0.00 & 0.00 & 0.12 & 0.00 & 0.61 & 0.40 & 0.21 & 0.08 & 0.04 & 0.08 & 0.20 & 0.67  \\
$A_{1}$   &      & 0.16 & 0.00 & 0.00 & 0.00 & 0.34 & 0.00 & 0.61 & 0.54 & 0.30 & 0.11 & 0.08 & 0.07 & 0.22 & 0.39  \\
$A_{2}$   &      &      & 0.72 & 0.65 & 0.00 & 0.00 & 0.45 & 0.67 & 1.61 & 0.30 & 0.11 & 0.07 & 0.16 & 0.72 & 0.74  \\
$A_{3}$   &      &      &      & 1.52 & 0.00 & 0.00 & 0.37 & 0.62 & 1.57 & 0.26 & 0.19 & 0.06 & 0.43 & 0.78 & 0.70  \\
$A_{4}$   &      &      &      &      & 0.39 & 0.00 & 0.00 & 0.08 & 0.19 & 0.14 & 0.03 & 0.05 & 0.06 & 0.13 & 0.17  \\
$A_{5}$   &      &      &      &      &      & 0.26 & 0.00 & 0.23 & 0.33 & 0.12 & 0.05 & 0.03 & 0.17 & 0.11 & 0.30  \\
$A_{6}$   &      &      &      &      &      &      & 0.13 & 0.18 & 0.08 & 0.06 & 0.02 & 0.02 & 0.08 & 0.15 & 0.25  \\
$A_{7}$   &      &      &      &      &      &      &      & 1.09 & 0.00 & 0.00 & 0.52 & 0.00 & 0.86 & 1.25 & 0.00  \\
$A_{8}$   &      &      &      &      &      &      &      &      & 1.94 & 0.00 & 0.00 & 0.00 & 0.00 & 1.97 & 4.63  \\
$A_{9}$   &      &      &      &      &      &      &      &      &      & 0.78 & 0.00 & 0.00 & 0.00 & 0.17 & 0.00  \\
$A_{10}$  &      &      &      &      &      &      &      &      &      &      & 0.44 & 0.00 & 0.54 & 0.28 & 0.00  \\
$A_{11}$  &      &      &      &      &      &      &      &      &      &      &      & 0.28 & 0.00 & 0.06 & 0.00  \\
$A_{12}$  &      &      &      &      &      &      &      &      &      &      &      &      & 0.65 & 0.63 & 0.00  \\
$A_{13}$  &      &      &      &      &      &      &      &      &      &      &      &      &      & 1.07 & 1.29  \\
$A_{14}$  &      &      &      &      &      &      &      &      &      &      &      &      &      &      & 4.30  \\
\hline
\end{tabular}
}
\end{table}

\begin{table}[!tb]
\centering
\caption{\small
  Absolute model uncertainties on the interference fit fractions (\%).
  The amplitudes are: ($A_{0}$) $\Kstarb(892)^{0}$, ($A_{1}$) $\Kstarb(1410)^{0}$,
  ($A_{2}$) $\Kstarbsubz(1430)^{0}$, ($A_{3}$) LASS nonresonant, ($A_{4}$) $\Kstarbsubt(1430)^{0}$, ($A_{5}$) $\Kstarb(1680)^{0}$, ($A_{6}$) $\Kstarbsubz(1950)^{0}$,
  ($A_{7}$) $D^{*-}_{s\,v}$, ($A_{8}$) $D^{*}_{s0\,v}(2317)^-$, ($A_{9}$) $D^{*}_{s2}(2573)^-$, ($A_{10}$) $D^{*}_{s1}(2700)^-$, ($A_{11}$) $D^{*}_{s3}(2860)^-$,
  ($A_{12}$) $D^{*}_{s1}(2860)^-$, ($A_{13}$) $B^{*+}_{v}$, ($A_{14}$) Nonresonant.
  The diagonal elements correspond to the model uncertainties on the fit fractions shown in Table~\ref{tab:model-syst}.
}
\label{tab:model-systIFF}
\resizebox{\textwidth}{!}{
\begin{tabular}{lccccccccccccccc}
\hline
          &           $A_{0}$  &           $A_{1}$  &           $A_{2}$  &           $A_{3}$  &           $A_{4}$  &           $A_{5}$  &           $A_{6}$  &           $A_{7}$  &           $A_{8}$  &           $A_{9}$  &           $A_{10}$ &           $A_{11}$ &           $A_{12}$ &           $A_{13}$ &           $A_{14}$ \\
\hline
$A_{0}$   & 0.88 & 1.58 & 0.00 & 0.00 & 0.00 & 1.18 & 0.00 & 1.67 & 0.96 & 0.88 & 0.35 & 0.36 & 0.66 & 0.81 & 2.66  \\
$A_{1}$   &      & 1.37 & 0.00 & 0.00 & 0.00 & 1.11 & 0.00 & 0.79 & 0.68 & 0.62 & 0.29 & 0.24 & 0.45 & 0.48 & 1.58  \\
$A_{2}$   &      &      & 3.32 & 3.63 & 0.00 & 0.00 & 1.33 & 1.21 & 0.85 & 0.42 & 0.17 & 0.11 & 0.43 & 1.22 & 2.16  \\
$A_{3}$   &      &      &      & 4.09 & 0.00 & 0.00 & 0.95 & 1.96 & 1.28 & 0.77 & 0.54 & 0.11 & 0.99 & 3.28 & 2.82  \\
$A_{4}$   &      &      &      &      & 1.06 & 0.00 & 0.00 & 0.20 & 0.28 & 0.46 & 0.16 & 0.22 & 0.26 & 0.43 & 1.84  \\
$A_{5}$   &      &      &      &      &      & 0.80 & 0.00 & 1.50 & 0.76 & 0.78 & 0.27 & 0.23 & 0.40 & 0.43 & 1.15  \\
$A_{6}$   &      &      &      &      &      &      & 2.42 & 1.45 & 0.68 & 0.60 & 0.20 & 0.16 & 0.30 & 1.29 & 2.49  \\
$A_{7}$   &      &      &      &      &      &      &      & 4.02 & 0.00 & 0.00 & 1.24 & 0.00 & 1.47 & 4.30 & 0.00  \\
$A_{8}$   &      &      &      &      &      &      &      &      & 2.30 & 0.00 & 0.00 & 0.00 & 0.00 & 1.69 & 7.43  \\
$A_{9}$   &      &      &      &      &      &      &      &      &      & 1.05 & 0.00 & 0.00 & 0.00 & 0.35 & 0.00  \\
$A_{10}$  &      &      &      &      &      &      &      &      &      &      & 0.54 & 0.00 & 0.75 & 1.12 & 0.00  \\
$A_{11}$  &      &      &      &      &      &      &      &      &      &      &      & 0.42 & 0.00 & 0.03 & 0.00  \\
$A_{12}$  &      &      &      &      &      &      &      &      &      &      &      &      & 3.28 & 1.27 & 0.00  \\
$A_{13}$  &      &      &      &      &      &      &      &      &      &      &      &      &      & 1.83 & 4.68  \\
$A_{14}$  &      &      &      &      &      &      &      &      &      &      &      &      &      &      & 7.64  \\
\hline
\end{tabular}
}
\end{table}

\clearpage

\centerline{\large\bf LHCb collaboration}
\begin{flushleft}
\small
R.~Aaij$^{41}$,
B.~Adeva$^{37}$,
M.~Adinolfi$^{46}$,
A.~Affolder$^{52}$,
Z.~Ajaltouni$^{5}$,
S.~Akar$^{6}$,
J.~Albrecht$^{9}$,
F.~Alessio$^{38}$,
M.~Alexander$^{51}$,
S.~Ali$^{41}$
% ...

{\footnotesize \it
% ...
$ ^{5}$Clermont Universit\'{e}, Universit\'{e} Blaise Pascal, CNRS/IN2P3, LPC, Clermont-Ferrand, France\\
$ ^{6}$CPPM, Aix-Marseille Universit\'{e}, CNRS/IN2P3, Marseille, France\\
% ...
$ ^{9}$Fakult\"{a}t Physik, Technische Universit\"{a}t Dortmund, Dortmund, Germany\\
% ...
$ ^{37}$Universidad de Santiago de Compostela, Santiago de Compostela, Spain\\
$ ^{38}$European Organization for Nuclear Research (CERN), Geneva, Switzerland\\
% ...
$ ^{41}$Nikhef National Institute for Subatomic Physics, Amsterdam, The Netherlands\\
% ...
$ ^{46}$H.H. Wills Physics Laboratory, University of Bristol, Bristol, United Kingdom\\
% ...
$ ^{51}$School of Physics and Astronomy, University of Glasgow, Glasgow, United Kingdom\\
$ ^{52}$Oliver Lodge Laboratory, University of Liverpool, Liverpool, United Kingdom\\
% ...
}
\end{flushleft}

\end{document}